%% file: hewawaduge_zare_PRF2021_v3_arXiv.tex
\documentclass[aps,prf,superscriptaddress,longbibliography]{revtex4-1}

\usepackage{amsmath,rotating}
\usepackage{dashrule}

\usepackage{psfrag}
\usepackage{latexsym}
\usepackage{mathrsfs}
\usepackage{graphicx}
\usepackage{pifont}
\usepackage{amsmath}
\usepackage{subfigure}
\usepackage{mathtools}
\usepackage{mathcomp}
\usepackage{mathdots}
\usepackage{rotating}
\usepackage{array}
\usepackage{color}
\usepackage{algorithm}
\usepackage{algorithmic}
\usepackage[normalem]{ulem}
\usepackage{bm}
\usepackage{amssymb}
\usepackage[dvipsnames]{xcolor}

\usepackage{caption}
\usepackage{tikz}

\usepackage{dsfont}

\usetikzlibrary{arrows,shapes,trees,calc,positioning}
\usetikzlibrary{matrix}

\newcommand{\cM}{{\cal M}}

\newcommand{\bvarphi}{{\bf \varphi}}

\newtheorem{theorem}{Theorem}

\newtheorem{remark}{Remark}

\newcommand{\enma}[1]   {\ensuremath{#1}}

\newcommand{\beq}{\begin{equation}}
\newcommand{\eeq}{\end{equation}}
\newcommand{\bseq}{\begin{subequations}}
\newcommand{\eseq}{\end{subequations}}
\newcommand{\beqn}{\begin{eqnarray}}
\newcommand{\eeqn}{\end{eqnarray}}
\newcommand{\ba}{\begin{array}}
\newcommand{\ea}{\end{array}}
\newcommand{\bct}{\begin{center}}
\newcommand{\ect}{\end{center}}
\newcommand{\btmz}{\begin{itemize}}
\newcommand{\etmz}{\end{itemize}}
\newcommand{\benum}{\begin{enumerate}}
\newcommand{\eenum}{\end{enumerate}}

\newcommand{\norm}[1]{\| #1 \|}                 

\newcommand{\diag}      {\enma{\mathrm{diag}}}

\newcommand{\trace}     {\enma{\mathrm{trace}}}

\newcommand{\inner}[2]{\left\langle #1,#2 \right\rangle}

\newcommand{\bv}{{\bf v}}

\newcommand{\matbegin}{
        \left[
}
\newcommand{\matend}{
        \right]
}

\newcommand{\tbo}[2]{
  \matbegin \begin{array}{c}
       #1 \\ #2
       \end{array} \matend }
\newcommand{\thbo}[3]{
  \matbegin \begin{array}{c}
       #1 \\ #2 \\ #3
       \end{array} \matend }
\newcommand{\obt}[2]{
  \matbegin \begin{array}{cc}
       #1 & #2
       \end{array} \matend }

\newcommand{\tbt}[4]{
  \matbegin \begin{array}{cc}
       #1 & #2 \\ #3 & #4
       \end{array} \matend }
\newcommand{\thbt}[6]{
  \matbegin \begin{array}{cc}
       #1 & #2 \\ #3 & #4 \\ #5 & #6
       \end{array} \matend }
\newcommand{\tbth}[6]{
  \matbegin \begin{array}{ccc}
       #1 & #2 & #3\\ #4 & #5 & #6
       \end{array} \matend }

\newcommand{\be}{\begin{equation}}
\newcommand{\ee}{\end{equation}}

\newcommand{\cplxs}{ C\kern -.35em \rule{0.03 em}{.7 ex}~   }

\def\complex{\hbox{C\kern -.45em \rule{0.03 em}{1.5 ex}}~}

\newcommand{\bi}{\begin{itemize}}
\newcommand{\ei}{\end{itemize}}

\newcommand{\bU}{\mathbf{U}}

\newcommand{\bu}{{\bf u}}
\newcommand{\bz}{{\bf z}}

\newcommand{\tc}{\textcolor}
\newcommand{\btab}{\begin{tabular}}
\newcommand{\etab}{\end{tabular}}

\newcommand{\bx}{{\bf x}}

\newcommand{\bpsi}{\mbox{\boldmath$\psi$}}

\newcommand{\non}{\nonumber}
\newcommand{\mrd}{\mathrm{d}}
\newcommand{\mre}{\mathrm{e}}
\newcommand{\mri}{\mathrm{i}}

\newcommand{\ds}{\displaystyle}

\newcommand{\br}{{\bf r}}

\newcommand{\bA}{\mathbf{A}}
\newcommand{\bB}{\mathbf{B}}
\newcommand{\bC}{\mathbf{C}}

\newcommand{\bk}{\mathbf{k}}
\newcommand{\DefinedAs}[0]{\mathrel{\mathop:}=}
\newcommand{\vsp}{\vspace*{0.15cm}}

\definecolor{bgblue}{rgb}{0.04,0.19,0.53}
\definecolor{dblue1}{rgb}{0,0.3,0.7}
\definecolor{dred}{rgb}{0.4,0.2,0}

\begin{document}

\title{\LARGE \bf
Input-output analysis of stochastic base flow uncertainty
}

\author{Dhanushki Hewawaduge}
\email{dhanushki.hewawaduge@utdallas.edu}
\author{Armin Zare}
\email{armin.zare@utdallas.edu}
\affiliation{Department of Mechanical Engineering, University of Texas at Dallas, Richardson, TX 75080, USA.}

    \begin{abstract}
    We adopt an input-output approach to analyze the effect of persistent white-in-time {structured} stochastic base flow perturbations on the mean-square properties of the linearized Navier-Stokes equations. Such base flow variations enter the linearized dynamics as multiplicative sources of uncertainty that can alter the stability of the linearized dynamics and their receptivity to exogenous excitations. {Our approach does not rely on costly stochastic simulations or adjoint-based sensitivity analysis. We provide verifiable conditions for mean-square stability and study the frequency response of the flow subject to additive and multiplicative sources of uncertainty using the solution to the generalized Lyapunov equation. For small-amplitude base flow perturbations, we bypass the need to solve large generalized Lyapunov equations by adopting a perturbation analysis.} We use our framework to study the destabilizing effects of stochastic base flow variations in transitional parallel flows, and the reliability of numerically estimated mean velocity profiles in turbulent channel flows. We uncover the Reynolds number scaling of critically destabilizing perturbation variances and demonstrate how the wall-normal shape of base flow modulations can influence the amplification of various length scales. Furthermore, we explain the robust amplification of streamwise streaks in the presence of streamwise base flow variations by analyzing the dynamical structure of the governing equations as well as the Reynolds number dependence of the energy spectrum.
    \end{abstract}

\maketitle

\section{Introduction}
\label{sec.intro}

The linearized Navier-Stokes (NS) equations have been used to capture the early stages of transition and identify key mechanisms for subcritical transition in wall-bounded shear flows. Even in the absence of transition, the non-normality of the linearized dynamical generator induces interactions of the exponentially decaying normal modes~\cite{tretrereddri93,schhen01}, which in turn result in the high sensitivity of velocity fluctuations to different sources of perturbation. This feature of the linearized dynamics has played a critical role in explaining the large transient growth of velocity fluctuations~\cite{gus91,butfar92,redhen93,henred94,schhen94} and the amplification of deterministic and stochastic disturbances in transitional and turbulent wall-bounded flows~\cite{tretrereddri93,farioa93b,farioa98,bamdah01,jovbamJFM05,hwacosJFM10a,hwacosJFM10b,mcksha10,ranzarhacjovPRF19b}. The success of this approach has also paved the way for the model-based design of active and passive flow control strategies for suppressing turbulence or reducing skin-friction drag~\cite{kimbew07,jovPOF08,moajovJFM10,moajovJFM12,luhshamck14,ranzarjovJFM21}. In these studies, additive stochastic excitation is used to model the effect of background disturbances and exogenous perturbations, or model the uncertainty caused by excluding the nonlinear terms in the NS equations. While most studies consider stochastic excitations to be white-in-time, efforts have also been made to shape the spectra of colored-in-time stochastic forcing to match the second-order statistics of turbulent flows~\cite{zarjovgeoJFM17,morsemhencos19,towlozyan20,zargeojovARC20}, which highlights the dynamical significance of such additive stochastic excitations in augmenting the linearized dynamics~\cite{zarjovgeoCDC16,zarjovgeoJFM17,zarmohdhigeojovTAC20}. An important, but rather less studied aspect of the linearized NS equations, however, arises from the uncertainty surrounding the choice of a base flow state and its implications for stability analysis, turbulence modeling, and the performance of model-based flow control.

Depending on the flow configuration and its characteristic regime, a base flow profile can either be obtained as the solution to the NS equations in steady state, or as a long-time averaged mean of a simulation-based flow field or experimental dataset. Due to insufficient data or imprecise measurements, it is often the case that the time-averaged mean can only be poorly approximated, resulting in uncertainties that prevail over the statistical averaging process (small data issues). For example, experimental constraints may confine reliable measurement collection and subsequent data acquisition procedures to certain parts of the flow domain, and in numerical simulations, {segments} of the computational domain may be poorly resolved. Furthermore, analytical or numerical approximations may have been made outside their range of validity implying a degree of uncertainty in the expressions for base flow profiles. This calls for the development of techniques that account for {such} sources of uncertainty and evaluate the validity and robustness of linearized models around uncertain base flow profiles. 

Previous studies have examined the sensitivity of the eigenvalues of the Orr–Sommerfeld operator to deterministic variations in the base flow. In~\cite{botcorluc03}, an adjoint-based variational procedure was used to identify worst-case perturbations with the most destabilizing effect on the eigenspectrum. Similar tools were later used in a locally temporal framework for identifying the optimal modification to the base flow for stabilizing a bluff-body wake~\cite{hwacho06} and were extended to global stability analysis~\cite{marsipjac08,prabragia10}. 
{While it has been shown that minute perturbations of the dynamic generator can cause significant displacement of eigenvalues~\cite{redschhen93,schhenkhomal93,treemb05}, it is generally accepted that the disturbance behavior of the linearized NS equations would be robust. Nevertheless, efforts have been made in quantifying the flow response to deterministic and stochastic base flow variations.}
In~\cite{brasippramar11}, an analytical expression was found for the gradient of singular values of the resolvent operator with respect to base flow modifications thereby accounting for variations in the non-modal behavior of wall-bounded shear flows. Besides adjoint-based methods for analyzing the sensitivity to deterministic modifications, there has also been efforts in quantifying the effect of random spatial base flow variations using stochastic spectral projection based on generalized polynomial chaos theory~\cite{kolucsag11}.

Additive sources of uncertainty in the base flow enter the linearized dynamics multiplicatively and in a structured manner. For deterministic and set-valued uncertainties, the structured singular value can be used to provide a robust stability theory for the uncertain dynamics~\cite{skopos07}. However, this approach is based on a worst-case analysis and may not provide a realistic model for experimental/numerical imperfections and measurement noise that are unlikely to bear an optimal shape. {On the other hand, the dynamical equations for the second moments of stochastically perturbed linear systems can be used to determine the effect of perturbations on optimal finite-time energy growth~\cite{farioa02}. Application of similar analysis techniques to stochastically perturbed Poiseuille flow uncovers the effect of multiplicative uncertainty on optimal energy growth as well as the robust amplification of streaks~\cite{sch07}. As highlighted in these studies, persistent multiplicative uncertainty increases the sensitivity of non-normal linear dynamical systems by influencing their asymptotic and transient mean-square response. In contrast to its additive counterpart, however, white-in-time multiplicative uncertainty can compromise the mean-square stability (MSS) properties of linear systems. Mean-square stability is a strong form of stability that implies stability of the mean and convergence of all trajectories of the stochastic dynamical system (in the absence of exogenous excitation) to zero with probability one~\cite{kus67,wil73}.
}

\begin{figure}
	\begin{tabular}{cccccc}
\subfigure[]{\label{fig.couette}}
        &&
        \hspace{.2cm}
       \subfigure[]{\label{fig.poiseuille}}
        &&
        \hspace{-.6cm}
       \subfigure[]{\label{fig.turbulent}}
       &
        \\[-.15cm]
        &
\includegraphics[width=5.1cm]{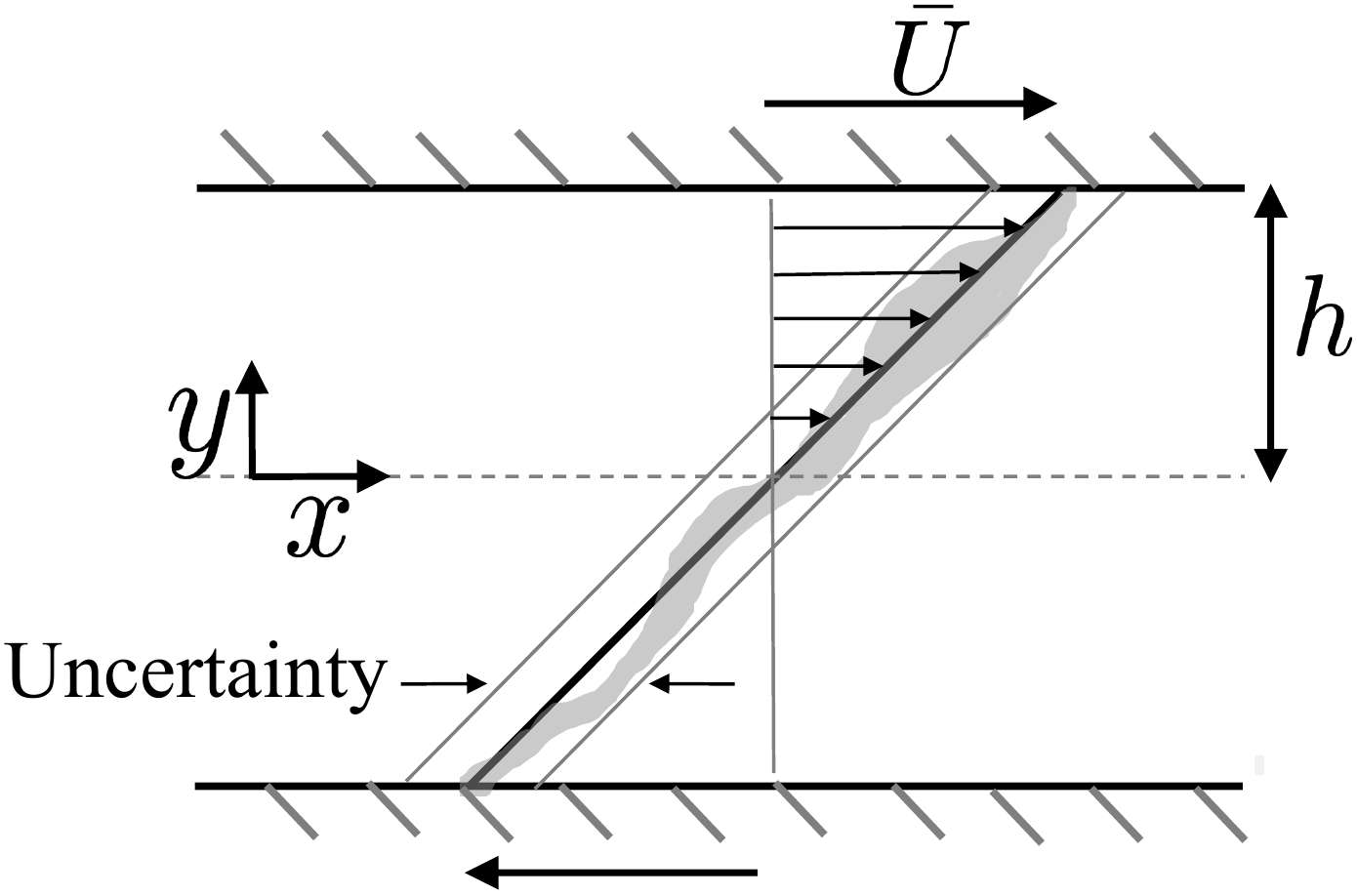}
        &&
        \hspace{-.1cm}
        \begin{tabular}{c}
        \vspace{3.2cm}
\includegraphics[width=5.1cm]{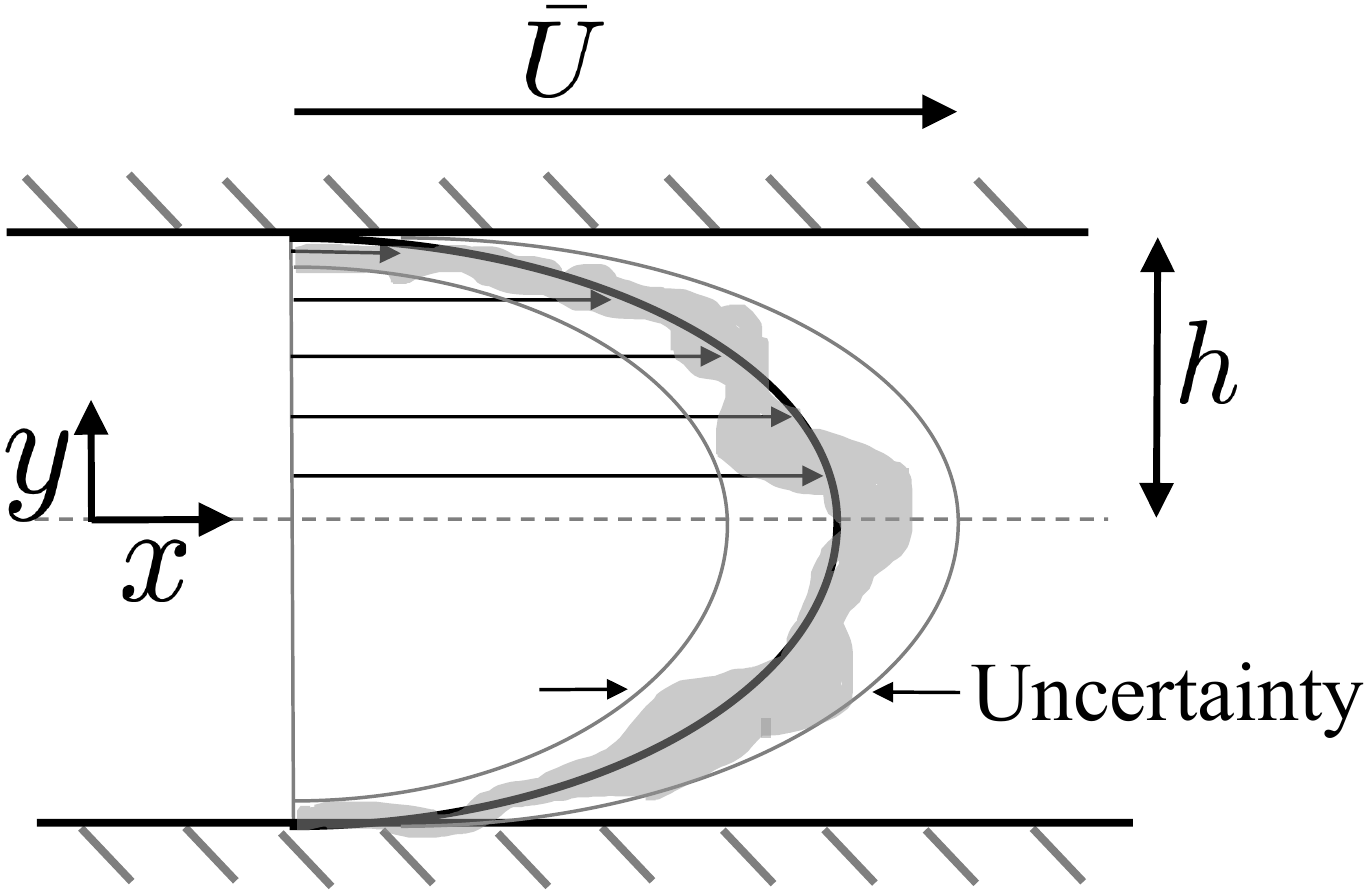}
	\end{tabular}
        &&
        \hspace{-.3cm}
        \begin{tabular}{c}
        \vspace{3.2cm}
	\includegraphics[width=5.8cm]{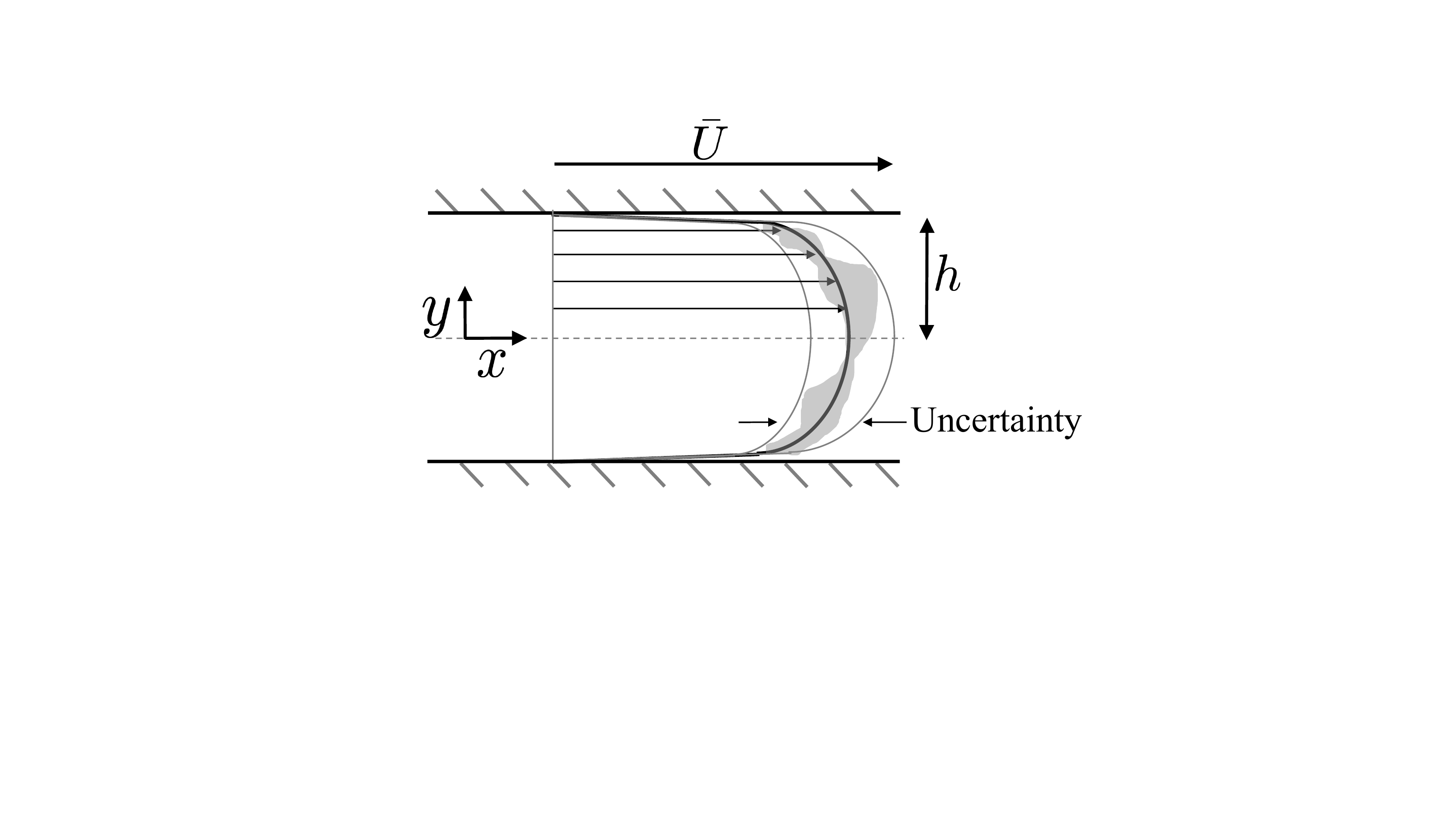}
	\end{tabular}
        \end{tabular}
        \vspace{-3cm}
        \caption{Side view of the three-dimensional canonical flows considered in this study along with various realizations of stochastic base flow perturbations $\gamma_u(t)$ represented by the shaded area surrounding the base flow profiles. (a) Couette flow; (b) Poiseuille flow; and (c) turbulent channel flow.}
        \label{fig.uncertain-flows}
\end{figure}

In this paper, we revisit the problem of analyzing said internal stochastic uncertainties by modeling structured perturbations to the base flow as white-in-time stochastic processes. In addition to persistent stochastic uncertainty in the base, we model the effect of exogenous excitations as a persistent white-in-time stochastic forcing. The dynamics of velocity fluctuations around the uncertain base state are governed by a set of stochastic differential equations (SDEs). We provide an input-output treatment by rewriting the SDEs as a feedback interconnection of the linearized dynamics and structured stochastic uncertainties. This allows us to separate the nominal (known) dynamics from the sources of uncertainty and facilitates both stability and receptivity analyses of the fluctuation dynamics in the presence of persistent additive and multiplicative stochastic excitation. Building on the recent developments of~\cite{filbam18}, we provide specialized conditions for the MSS of the uncertain dynamics. Furthermore, we analyze the energy spectrum of the linearized NS equations subject to additive and multiplicative sources of excitation. To this end, we compute the second-order statistics of the velocity field from the solution to generalized Lyapunov equations. For small-amplitude perturbations of the base state, we adopt a perturbation analysis to compute the energy spectrum of velocity fluctuations using a computationally efficient method that by-passes the need to solve the associated generalized Lyapunov equations. We demonstrate the utility of our approach by studying the stability and receptivity of the three-dimensional channel flow around canonical Couette and Poiseuille profiles as well as a turbulent mean velocity profile resulting from direct numerical simulations (DNS) all of which are contaminated with persistent stochastic perturbations; see Fig.~\ref{fig.uncertain-flows} for an illustration. We also uncover the Reynolds number scaling of the critical variance of stochastic base flow uncertainty that guarantees MSS and identify length scales that are most influenced by such perturbations.

The rest of our presentation is organized as follows. In Sec.~\ref{sec.dynamics}, we describe our model of stochastic base flow perturbation, introduce the stochastically forced linearized NS equations around the uncertain base flow, and demonstrate how base flow perturbations enter the dynamics as multiplicative sources of uncertainty. In Sec.~\ref{sec.InOut-MSS}, we rewrite the linearized dynamics as a feedback interconnection between nominal dynamics and sources of stochastic uncertainty. We then use this input-output representation to provide MSS conditions for our model, characterize its frequency response, and describe the generalized Lyapunov equation that we use to compute the second-order statistics and energy spectrum of velocity fluctuations. In Sec.~\ref{sec.laminar-flows}, we examine the MSS and energy amplification of velocity fluctuations around Couette and Poiseuille profiles, {and study the influence of base flow perturbations on flow structures. In Sec.~\ref{sec.turbulent-flows}, we extend this analysis to} the linearized NS equations around the DNS-based mean velocity profiles of turbulent channel flow at various Reynolds numbers. In Sec.~\ref{sec.Reynolds-number-dependence}, we study the Reynolds number dependence of variance amplification for streamwise elongated flow structures. 
We provide concluding remarks in Sec.~\ref{sec.conclusion}.

\section{Dynamics of fluctuations around uncertain base flow}
\label{sec.dynamics}
{The dynamics of incompressible Newtonian fluids is governed by the NS equations, 
\begin{align}
	\label{eq.NS-eqn}
	\ba{rcl}
		\tilde{\bu}_t 
		&\;=\;& 
		-\left( \tilde{\bu} \cdot \nabla \right) \tilde{\bu} \,-\, \nabla \tilde{P} \,+\,\dfrac{1}{R} \, \Delta \tilde{\bu} 
		\\[.15cm]
		0 
		&\;=\;& 
		\nabla \cdot \tilde{\bu}
	\ea
\end{align}
where $\tilde{\bu}$ is the velocity vector, $\tilde{P}$ is the pressure, $\nabla$ is the gradient, $\Delta = \nabla \cdot \nabla$ is the Laplacian, and $t$ is time. Here, the Reynolds number $R$ is defined in terms of appropriate length and velocity scales, e.g., for a laminar channel flow configuration, $R = \bar{\bU} h/\nu$, where $\bar{\bU}$ is the maximum nominal velocity, $h$ is the channel half-height, and $\nu$ denotes the kinematic viscosity. Linearization of the NS equations around an arbitrary, parallel base flow $\bu = [\,U(y)\,~0\,~W(y)\,]^T$ and pressure $P$ yields the equations that govern the dynamics of velocity, $\bv$, and pressure, $p$, fluctuations,
\begin{align}
	\label{eq.dyn-fluc}
	\ba{rcl}
		\bv_t 
		&\;=\;& 
		-\left( \nabla \cdot \bu  \right) \bv \,-\, \left( \nabla \cdot \bv  \right) \bu \,-\, \nabla p \,+\,\dfrac{1}{R} \, \Delta \bv \,+\, {\bf f}
		\\[.15cm]
		0 
		&\;=\;& 
		\nabla \cdot \bv.
	\ea
\end{align}
Here, $\bv = [\,u\,~v\,~w\,]^T$, with $u$, $v$, and $w$ representing the fluctuating components in the streamwise, $x$, wall-normal, $y$, and spanwise, $z$ directions, and ${\bf f}$ denoting a three-dimensional zero-mean white-in-time additive stochastic forcing.}

{We assume the base flow $\bu$ to be contaminated with an additive source of uncertainty, i.e.,}
\begin{align}
\label{eq.base-flow-form}
{\bu(y,t) \;=\; \bar{\bu}(y) \,+\, \gamma(y,t).}
\end{align}
Here, $\bar{\bu}(y) = [\,\bar{U}(y)\,~0\,~\bar{W}(y)\,]^T$ is the nominal base flow in the absence of uncertainty and $\gamma$ is a zero-mean white-in-time stochastic process that can enter both streamwise and spanwise components of $\bar{\bu}$, i.e., $\gamma(y,t) = [\,\gamma_u(y,t)\,~0\,~\gamma_w(y,t)\,]^T$.
The uncertain base flow $\bu$ enters the linearized Eqs.~\eqref{eq.dyn-fluc} as a coefficient that multiplies the vector of velocity fluctuations $\bv$. While $\bu$ includes the sources of uncertainty $\gamma$, it remains constant in $x$ and $z$. Elimination of pressure and application of the Fourier transform in the spatially invariant wall-parallel directions brings Eqs.~\eqref{eq.dyn-fluc} into the evolution form
\begin{align}
	\label{eq.lnse}
	\ba{rcl}
	\bvarphi_t(y,\bk,t)
	&\;=\;&
	\left[\bA(\bk, t)\, \bvarphi(\cdot,\bk,t)\right](y) \,+\, \left[\bB(\bk)\,{\bf f}(\cdot,\bk,t)\right](y)
	\\[.15cm]
	\bv(y,\bk,t)
	&\;=\;&
	\left[\bC(\bk)\, \bvarphi(\cdot,\bk,t)\right](y)
	\ea
\end{align}
where the state variable $\bvarphi = [\,v\,~ \eta\,]^T$ contains the wall-normal velocity $v$ and vorticity $\eta = \partial_z u - \partial_x w$, and $\bk = [\,k_x\,~k_z\,]^T$ is the vector of streamwise and spanwise wavenumbers. These SDEs involve multiplicative sources of stochastic uncertainty $\gamma_u$ and $\gamma_w$ in addition to the additive source of stochastic uncertainty ${\bf f}$. In~\eqref{eq.lnse}, operators $\bA$, $\bB$, and $\bC$ are given by
\begin{align}
	\label{eq.Aform}
        \bA(\bk,t)
        &\;\DefinedAs\;
        \tbt{\bA_{11}}{0}{\bA_{21}}{\bA_{22}}
        \\[.25cm]
        \non
        \bA_{11}(\bk,t)
        &\;\DefinedAs\;
        \Delta^{-1}\,\Big(\dfrac{1}{R}\,\Delta^2 \,+\, \mri k_x \left(\bar{U}''(y) + \gamma_u''(y,t) \,-\, (\bar{U}(y) + \gamma_u(y,t))\,\Delta \right) \,+\,
        \\[.05cm]
        \non
        & \hspace{3cm} 
        \mri k_z \left(\bar{W}''(y) + \gamma_w''(y,t) \,-\, (\bar{W}(y) + \gamma_w(y,t))\,\Delta\right) \Big)
         \\[.15cm]
         \non
        \bA_{21}(\bk,t)
        &\;\DefinedAs\;
        -\mri k_z \left(\bar{U}'(y) + \gamma'_u(y,t)\right) \,+\,  \mri k_x \left( \bar{W}'(y) + \gamma'_w(y,t)\right)
         \\[.15cm]
         \non
        \bA_{22}(\bk,t)
        &\;\DefinedAs\
        \dfrac{1}{R}\, \Delta \,-\, \mri k_x\, \left(\bar{U}(y) +\gamma_u(y,t) \right) \,-\, \mri k_z \left(\bar{W}(y) + \gamma_w(y,t) \right)
	\\[.15cm]
        \bB(\bk)
        &\;\DefinedAs\;
        \tbth{-\mri k_x \Delta^{-1} \partial_y}{-k^2 \Delta^{-1}}{-\mri k_z \Delta^{-1} \partial_y}{\mri k_z}{0}{-\mri k_x},
        \quad\quad
        \bC(\bk)
        \;\DefinedAs\;
        \thbo{\bC_u}{\bC_v}{\bC_w}
        \;=\;
        \dfrac{1}{k^2}
        \thbt{\mri k_x\partial_y}{-\mri k_z}{k^2}{0}{\mri k_z\partial_y}{\mri k_x}
        \non
\end{align}
where {prime denotes differentiation with respect to the wall-normal coordinate, $\mri$ is the imaginary unit,}  $k^2 = k_x^2 \,+\, k_z^2$, $\Delta = \partial_y^2 \,-\, k^2$ is the Laplacian, $\Delta^2 = \partial_y^4 \,-\, 2k^2\partial_y^2 \,+\, k^4$, and $v(\,\pm1,\,\bk,\,t\,) = v_y(\,\pm1,\,\bk,\,t\,)=\eta(\,\pm1,\,\bk,\,t\,)=0$, which can be derived from the original no-slip {and no-penetration} boundary conditions on $u$, $v$, and $w$.

\begin{figure}
	\begin{tabular}{cccccc}
	 \hspace{-.6cm}
	\subfigure[]{\label{fig.fshape1}}
        &&
        \hspace{.6cm}
      \subfigure[]{\label{fig.fvec}}
      &
        \\[-.25cm]
        \begin{tabular}{c}
                \vspace{5.5cm}
                {\normalsize \rotatebox{90}{$y$}}
        \end{tabular}
        &
	\includegraphics[width=6cm]{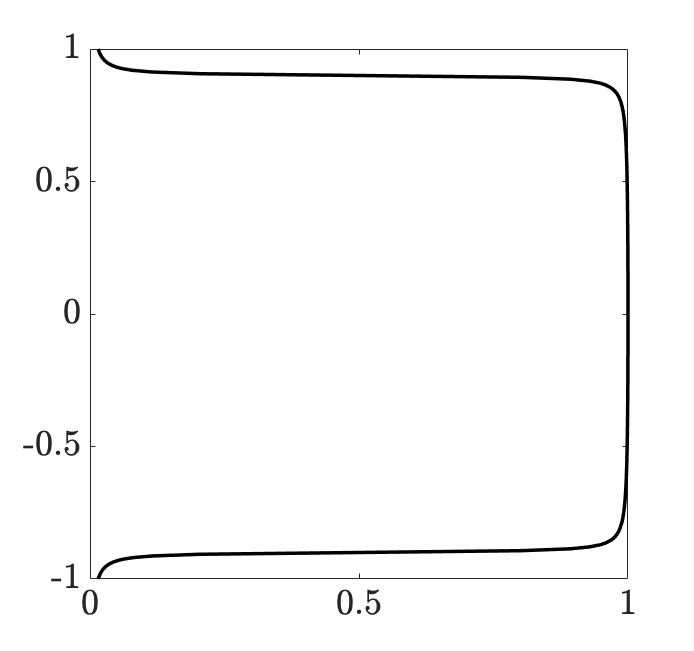}
        &&
        \hspace{.2cm}
	 \includegraphics[width=6cm]{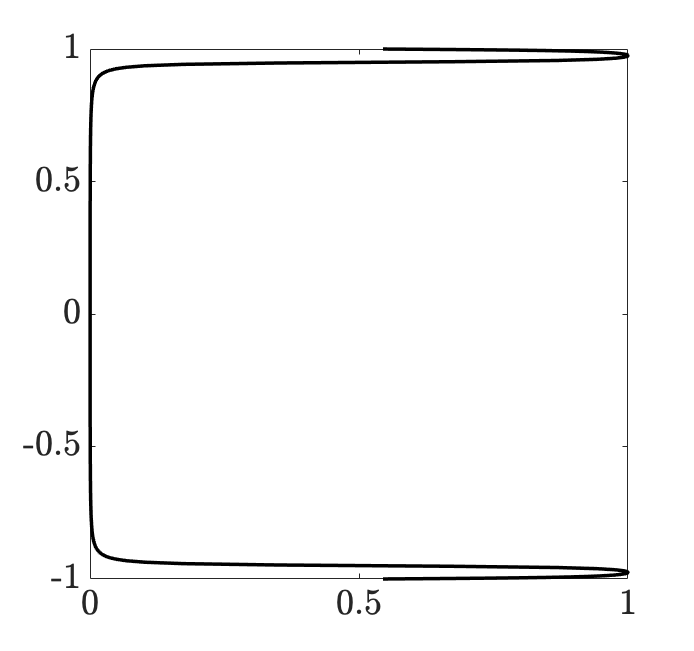}
        \\[-2.8cm]
        &
        \normalsize {$f(y)$}
	&&
        \normalsize {$f(y)$}
        \end{tabular}
             \caption{{(a) A shape function $f(y)$ determined using Eq.~\eqref{eq.f} with $\{y_1, y_2\} =\{-0.9,0.9\}$ and $a=200$; and (b) a two-sided shape function $f(y) = f_1(y) + f_2(y)$ in which $f_1(y)$ and $f_2(y)$ are determined using Eq.~\eqref{eq.f} with $\{y_1, y_2\} = \{-1,-0.95\}$ for $f_1(y)$, $\{y_1, y_2\} = \{0.95,1\}$ for $f_2(y)$, and $a=200$.}}
        \label{fig.shape functions}
\end{figure}

We confine the class of stochastic base flow perturbations to the form {$\gamma(y,t) = \alpha \,\bar{\gamma}(t) f(y)$}, in which {$\alpha>0$ is the constant amplitude}, $\bar{\gamma}(t)$ is a zero-mean stochastic parameter {of unit amplitude,} and $f(y)$ is a smooth filter function that determines the wall-normal region of influence and is defined as
\begin{align}
	\label{eq.f}
	 f(y) \DefinedAs  \dfrac{1}{\pi} [ \,\arctan{(a(y\,-\,y_1))} \,-\, \arctan{(a(y\,-\,y_2))}\, ].
\end{align}
Here, $y_1$ and $y_2$ determine the wall-normal extent of $f (y)$ and $a$ specifies the roll-off rate. In Secs.~\ref{sec.laminar-flows} and~\ref{sec.turbulent-flows}, we study the influence of stochastic base flow perturbations {with wall-normal dependence corresponding to the shape functions shown in Fig.~\ref{fig.shape functions}, as well as a normalized variant of the associated nominal base flow profile $\bar{U}(y)$, i.e.,} 
\begin{align}
\label{eq.fshape-Ubar}
    {f(y) \;=\; \dfrac{\bar{U}(y)}{\max\left(|\bar{U}(y)|\right)}}
\end{align}
{where normalization ensures that the multiplicative source of uncertainty amplifies the state only via its amplitude $\alpha$ and the variance of its stochastic temporal component $\bar{\gamma}(t)$. The variance of $\bar{\gamma}(t)$ will be typically chosen to represent critically stable conditions beyond which the state dynamics lose stability.} While the shape function in Fig.~\ref{fig.fshape1} does not restrict the wall-normal extent of the perturbations (besides a roll-off at the walls in accordance with the boundary conditions), Fig.~\ref{fig.fvec} represents an extreme case corresponding to base flow perturbations that may result from active/passive boundary actuation (e.g., blowing and suction), surface roughness, or the secondary effects of perturbation growth in transition mechanisms.
Beyond application specificities, these extreme cases allow us to study the dependence of our results on the wall-normal extent of base flow perturbations.

\begin{remark}
In Eq.~\eqref{eq.base-flow-form}, $\gamma(y,t)$ accounts for the effect of temporal sources of uncertainty on the base flow $\bar{\bu}$. We note that spatially random effects can also be translated into temporal uncertainty using the local convection velocity together with a Taylor-like transformation, i.e., $\gamma(y,t) = r(\bx,y)/\bu_c$, where $\bu_c$ is the local convection velocity and $r(\bx,y)$ is a spatially random process in $\bx$. The specification of $\bu_c$, i.e., its directivity and magnitude, is problem dependent, but the result of numerical studies can provide guidelines for its determination; see, e.g.,~\cite[Sec.~5.2]{zarjovgeoJFM17}.
\end{remark}


Based on the class of stochastic perturbations $\gamma(y,t)$ considered in this paper, the operator-valued matrix $\bA$ in evolution model~\eqref{eq.lnse} can be decomposed {into nominal and perturbed components} as
\begin{align}
\label{eq.A-decomp-original}
	\bA(\bk,t)
	~=~
	\bar{\bA}(\bk) \;+\; {\alpha \left(\bar{\gamma}_u(t)\,\bA_u(\bk)\, \,+\, \bar{\gamma}_w(t)\, \bA_w(\bk) \right)}
\end{align}
where expressions for $\bar{\bA}$, $\bA_u$, and $\bA_w$ are given in Appendix~\ref{app.A0-Au-Aw}. The nominal base flow profile $\bar{\bu}(y)$, and shape functions $f_u(y)$ and $f_w(y)$ enter operators $\bar{\bA}$, $\bA_u$, and $\bA_w$ as deterministic parameters, respectively. {Note that while we have assumed the streamwise and spanwise components of the base flow uncertainty $\gamma(y,t)$ to be of equal amplitude, all mathematical developments can be easily extended to scenarios where the streamwise and spanwise components have different amplitudes.}

In this study, we use a pseudospectral scheme with $N$ Chebyshev collocation points in the wall-normal direction~\cite{weired00} to discretize the operators in the linearized equations~\eqref{eq.lnse}. In addition, we employ a change of variables to obtain a state-space representation in which the kinetic energy is determined by the Euclidean norm of the state vector~\cite[Appendix A]{zarjovgeoJFM17}. This yields the state-space model
\begin{align}
	\label{eq.lnse1}
	\ba{rcl}
	\dot{\bpsi}(\bk,t)
	&\;=\;&
	A(\bk,t)\,\bpsi(\bk,t) \,+\, B(\bk)\,{\bf f}(\bk,t)
	\\[.15cm]
	\bv(\bk,t)
	&\;=\;&
	C(\bk)\, \bpsi(\bk,t)
	\ea
\end{align}
where vectors $\bpsi$ and $\bv$ are vectors with complex-valued entries and $2N$ and $3N$ components, respectively, and matrices $A$, $B$, and $C$ are discretized versions of the corresponding operators that incorporate the aforementioned change of coordinates. We next provide an input-output reformulation of SDE~\eqref{eq.lnse1} to analyze the influence of stochastic sources of uncertainty on the mean-square asymptotic stability and second-order statistics of velocity fluctuations.

\section{Mean-square stability and Input-output analysis}
\label{sec.InOut-MSS}

The evolution of $\bpsi$ in SDE~\eqref{eq.lnse1} is affected by the presence of both stochastic base flow perturbations $\gamma(y,t)$ and additive forcing ${\bf f}(t)$. While there is no ambiguity in the treatment of additive noise in continuous-time systems, multiplicative noise is not generally well-defined and its treatment calls for the adoption of a suitable stochastic calculus (e.g., It\={o}~\cite{ito79} or Stratonovich~\cite{str66}). In this section, we provide an appropriate interpretation for the multiplicative uncertainty, extract these sources using a linear fractional transformation, and establish an input-output relation between stochastic sources and the output velocity fluctuations of system~\eqref{eq.lnse1}. Building on this representation, we examine conditions for MSS and analyze the frequency response of the system in the presence of multiplicative stochastic uncertainty.

\subsection{Stochastic feedback interconnection}
\label{sec.feedback-interconnection}

\begin{figure}
\centering
        \input{Fig3.tex}
\caption{Linear fractional transformation of an LTI system subject to both additive and multiplicative stochastic disturbances (Eqs.~\eqref{eq.ss-differential}). Here, $\mrd \tilde{\bf f}$ and the $\mrd \tilde{\gamma}_{i}$ represent differentials of Wiener processes that model additive and multiplicative sources of stochastic uncertainty, respectively.}
\label{fig.M-Delta-diagram}
\end{figure}
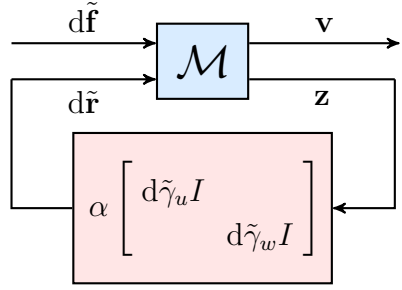

In input-output form, SDE~\eqref{eq.lnse1} can be rewritten as
\begin{align}
	\non
	\tbo{\bv}{\bz}
	\;=\;
	\cM \tbo{{\bf f}}{\br} ~~
	&\,\Leftrightarrow\,~~ \tbo{\bv(\bk,t)}{\bz(\bk,t)} \;=\; \ds{\int^t_0 M(\bk,t \,-\, \tau)} \tbo{{\bf f}(\bk,\tau)}{\br(\bk,\tau)} \mrd \tau
	\\[.35cm]
	\label{eq.impulse-resp1}
	\br(\bk,t)
	&\;=\;
	{\alpha\,}{\cal D} (\bar{\gamma}(t))\, \bz(\bk,t)
\end{align}
which extracts the role of multiplicative uncertainties by rearranging the dynamics as a feedback connection between the nominal (known) dynamics (captured by the impulse response operator $\cM$) and the structured uncertainty ${\cal D}(\bar{\gamma}(t)) \DefinedAs \diag\{\bar{\gamma}_u(t)I,\bar{\gamma}_w(t)I\}$. In Eqs.~\eqref{eq.impulse-resp1}, $M$ denotes the finite-dimensional approximation to the impulse response operator $\cM$, $\bv$ is the output velocity vector (cf.~Eqs.~\eqref{eq.lnse1}), and $\bz$ is computed from the state $\bpsi$. Moreover, the exogenous stochastic input ${\bf f}$, the uncertain feedback signal $\br$, and the sources of uncertainty $\bar{\gamma}_u$ and $\bar{\gamma}_w$ are white processes that are all defined as derivatives of Wiener processes (or Brownian motion)~\cite{oks03}, i.e.,
\begin{align*}
	\bar{\gamma}_u(t) \;\DefinedAs\; \dfrac{\mrd \tilde{\gamma}_u(t)}{\mrd t} ~;~~~~~
	\bar{\gamma}_w(t) \;\DefinedAs\; \dfrac{\mrd \tilde{\gamma}_w(t)}{\mrd t} ~;~~~~~
	{\bf f}(\bk,t)\;\DefinedAs\; \dfrac{\mrd \tilde{\bf f}(\bk,t)}{\mrd t} ~;~~~~~
	\br(\bk,t)\;\DefinedAs\; \dfrac{\mrd \tilde\br(\bk,t)}{\mrd t}.
\end{align*}
Here, $\tilde{\gamma}_i$ are zero-mean Wiener processes with variance $\sigma_i^2$ and 
$\tilde{\bf f}$ is a zero-mean vector-valued Wiener process with instantaneous covariance
\begin{align*}
	\left< \tilde{\bf f} (\bk,t)\, \tilde{\bf f}^* (\bk,t) \right> 
	\;=\;
	\Omega(\bk)\, t
\end{align*}
in which $\Omega(\bk) = \Omega^*(\bk) \succeq 0$ is the spatial covariance matrix. We assume that $\tilde{\gamma}_i$ and $\tilde{\bf f}$ are uncorrelated at all times, adopt the It\={o} interpretation, and assume that $\br$ has temporally independent increments, i.e., its differentials $(\mrd \br(\bk,t_1), \mrd \br (\bk,t_2))$ are independent when $t_1\neq t_2$. Given this mathematical interpretation, the differential form of Eqs.~\eqref{eq.impulse-resp1} is given by 
\begin{align}
	\non
	\tbo{\bv}{\bz}
	\;=\;
	\cM \tbo{\mrd \tilde{\bf f}}{\mrd \tilde{\br}} ~~
	&\,\Leftrightarrow\,~~ \tbo{\bv(\bk,t)}{\bz(\bk,t)} \;=\; \ds{\int^t_0 M(\bk,t \,-\, \tau)} \tbo{\mrd\tilde{\bf f}(\bk,\tau)}{\mrd\tilde{\br}(\bk,\tau)}
	\\[.35cm]
	\label{eq.impulse-resp2}
	\mrd\tilde{\br}(\bk,t)
	&\;=\;
	{\alpha\,}{\cal D} (\mrd\tilde{\gamma}(t))\, \bz(\bk,t)
\end{align}
and is described by the block diagram in Fig.~\ref{fig.M-Delta-diagram}. A corresponding state-space model is given by
\begin{align}
\label{eq.ss-differential}
	\cM
	: &\,\left\{ 
	\ba{rcl}\!\!
		\mrd \bpsi(\bk,t)
		&=\,&
		\bar{A}(\bk)\, \bpsi(\bk,t) \mrd t \,+\, B_0(\bk)\,\mrd \tilde{\br}(\bk,t) \,+\, B(\bk)\, \mrd\tilde{\bf f}(\bk,t)
		\\[.15cm]
		\bz(\bk,t)
		&=\,&
		C_0(\bk)\, \bpsi(\bk,t)
		\\[.15cm]
		\bv(\bk,t)
		&=\,&
		C(\bk)\,\bpsi(\bk,t)
	\ea
	\right.
	\non
\\[.35cm]
	\mrd \tilde{\br}(\bk,t) &\;=\; {\alpha\,}{\cal D} (\mrd\tilde{\gamma}(t))\, \bz(\bk,t)
\end{align}
with
\begin{align}
	\label{eq.B0-C0}
	B_0(\bk)
	\,\DefinedAs\,
	\obt{I\, ~}{I \,},
	\quad
	C_0(\bk)
	\,\DefinedAs\,
	\tbo{A_u(\bk)}{A_w(\bk)}.
\end{align}
Here, $\bpsi$, $\bz$, and $\bv$ are complex-valued vectors of appropriate dimension, $B$ and $C$ are finite-dimensional approximations of the input and output operators in~\eqref{eq.lnse1}, and $\bar{A}$, $A_u$ and $A_w$ are finite-dimensional approximations of the nominal dynamics and their perturbations in~\eqref{eq.A-decomp-original}. 

\subsection{Mean-square stability conditions}
\label{sec.MSS}

For the causal LTI system~\eqref{eq.ss-differential}, MSS certifies that for all differential inputs, $[\,\mrd \tilde{\bf f}\,~ \mrd \tilde{\br}\,]^T$, with independent increments and uniformly bounded variances, the output process
\begin{align*}
	\ba{rccl}
	\tbo{\bv}{\bz}
	&=& 
	\underbrace{\tbt{\cM_{11}}{\cM_{12}}{\cM_{21}}{\cM_{22}}} 
	&
	\tbo{\mrd \tilde{\bf f}}{\mrd \tilde{\br}}
	\\
	&&
	\cM&
	\ea
\end{align*}
has a uniformly bounded variance; see, e.g.,~\cite{sam59}. {Following~\cite[Theorem 3.2]{filbam20}, the necessary and sufficient conditions for MSS can be generalized for the continuous-time scenario, i.e., the output $\bv$ in~\eqref{eq.ss-differential} has a finite covariance if and only if the feedback subsystem ($\cM_{22},\Gamma$) is MSS. Based on this, the exact} necessary and sufficient conditions for the MSS of~\eqref{eq.ss-differential} are: (i) $\bar{A}$ is Hurwitz; and (ii) the spectral radius of the loop gain operator
\begin{align}
 \label{eq.Loop gain operator}
	\mathbb{L}({\bf R}) \;\DefinedAs\; \Gamma \circ \left( \ds{\int_0^\infty }M_{22}(\tau)\, {\bf R}\,M_{22}^*(\tau) \mrd \tau\right)
\end{align}
is strictly less than {$1/\alpha^2$}, i.e., $\rho(\mathbb{L})<{1/\alpha^2}$. Here, $\circ$ is the Hadamard product, $M_{22}$ is the impulse response of the subsystem $\cM_{22}: \mrd\tilde{\br} \to \bz$, which is given by
\begin{align*}
	M_{22}(\bk,t) 
	\;=\; 
	C_0(\bk)\, \mre^{\,\bar{A}(\bk,t)t}\, B_0(\bk) 
\end{align*}
and $*$ denotes complex-conjugate-transpose. The matrix $\Gamma$ denotes the mutual correlation of {uncertainties} $\tilde{\gamma}_i$, i.e., $\Gamma \DefinedAs \left<\tilde{\gamma}_i(t)\,\tilde{\gamma}^*_j(t)\right>$. {For example, for mutually independent multiplicative uncertainties in the streamwise and spanwise directions that are spatially uncorrelated, $\Gamma= \diag\{\sigma_u^2\, I,\,\sigma_w^2\, I\}$, where $\sigma^2_u$ and $\sigma^2_w$ are variances of $\bar{\gamma}_u$ and $\bar{\gamma}_w$, respectively. In this paper, we consider $\bar{\gamma}_u$ and $\bar{\gamma}_w$ to be mutually independent, but repeated throughout the spatial domain, i.e., $\Gamma = \diag\{\sigma_u^2\, \mathds{1}\mathds{1}^T,\,\sigma_w^2\, \mathds{1}\mathds{1}^T\}$, where $\mathds{1}$ represents the vector of $2N$ ones. As explained in Sec.~\ref{sec.dynamics}, the wall-normal support of each multiplicative uncertainty $\bar{\gamma}_i$ will be captured by its associated shape function $f_i(y)$ within operators $A_i$ in Eq.~\eqref{eq.B0-C0}.} 

\begin{remark}
We note that a similar condition for global mean-square asymptotic stability was established in~\cite{buckel14}. This condition was based on the stability of the differential generalized Lyapunov equation and amounts to the eigenvalue stability of the mean-square stability matrix, which takes a similar form as the loop gain operator $\mathbb{L}(\cdot)$. The differential generalized Lyapunov equation governs the evolution of the state covariance matrix $X$. In Sec.~\ref{sec.FreqResp}, we use the steady-state solution of this equation to compute the second-order statistics and energy spectrum of velocity fluctuations in the presence of both additive and multiplicative stochastic excitation.
\end{remark}

The loop gain operator propagates the steady-state covariance of $\mrd \tilde{\br}$ denoted by ${\bf R}$ through the feedback configuration in Fig.~\ref{fig.M-Delta-diagram}. Equivalently, we have
\begin{align*}
	\mathbb{L}({\bf R})
	\;=\;
	\Gamma \circ (C_0\, X\, C_0^* )	
\end{align*}
where $X$ is the solution to the algebraic Lyapunov equation
\begin{align*}
	\bar{A}\,X \,+\, X\,\bar{A}^*
	\;=\;
	-B_0\, {\bf R}\, B_0^*.
\end{align*}
In practice, the spectral radius of $\mathbb{L}$ can be numerically computed using the power iteration algorithm; see, e.g.,~\cite[Section VI.A]{filbam20}. Starting from an initial ${\bf R}_0 \succeq 0$ an estimate for the spectral radius is updated via a sequence of steps:
\begin{align*}
	\ba{rcl}
	\bar{A}\,X_{k+1} \,+\, X_{k+1}\,\bar{A}^*
	&=&
	-B_0\, {\bf R}_k\, B_0^*
	\\[.15cm]
	{\bf R}_{k+1}
	&\DefinedAs&
	\dfrac{1}{\norm{{\bf R}_k}_F} \left( \Gamma \circ (C_0\, X_{k+1}\, C_0^* ) \right)
	\\[.45cm]
	\rho_{k+1}
	&\DefinedAs&
	\inner{{\bf R}_k}{{\bf R}_{k+1}}
	\ea
\end{align*}
until the residual 
$
	\left({\bf R}_{k+1} - \rho_{k+1} {\bf R}_{k} \right)/ \norm{{\bf R}_{k+1}}_F
$
is smaller than a desirable tolerance.

\subsection{Frequency response of uncertain dynamics}
\label{sec.FreqResp}

We build on the input-output representation provided in Sec.~\ref{sec.MSS} and characterize the frequency response of  the system subject to both additive and multiplicative sources of uncertainty. We show that the second-order statistics of the uncertain system and the energy spectrum of velocity fluctuations can be obtained from the solution of a generalized Lyapunov equation. 

The impulse response $M$ in~\eqref{eq.impulse-resp2} corresponding to the state-space representation~\eqref{eq.ss-differential} takes the form
\begin{align*}
	M(\bk,t) \;\DefinedAs\; \tbo{C(\bk)\,}{\,C_0(\bk)}\, \mre^{\,\bar{A}(\bk,t)t}\, \obt{B(\bk)\,}{\,B_0(\bk)}
\end{align*}
When ${\bf f}$, $\bar{\gamma}_u$, and $\bar{\gamma}_w$ are zero-mean white-in-time processes with covariance matrix $\Omega$, and variances $\sigma_{u}^2$ and $\sigma_{w}^2$, the steady-state covariance of the state,
\begin{align}
\label{eq.X-def}
	X(\bk) \;=\; \lim_{t\to\infty} \left< \bpsi(\bk,t)\, \bpsi^*(\bk,t)\right>
\end{align}
can be determined as the solution to the generalized Lyapunov equation
\begin{align}
	\label{eq.gen-lyap-original}
	\bar{A} \, X \;+\; X\,\bar{A}^* \;+\; {\alpha^2\,} B_0 \left(\Gamma \circ (C_0\, X\, C_0^* )\right) B_0^*
	~=\;
	- B\,\Omega\, B^*
\end{align}
which is parameterized over wavenumber pairs $\bk$. The generalized Lyapunov equation relates the statistics of white-in-time forcing ${\bf f}$ and multiplicative sources of excitation ${\alpha\,}\bar{\gamma}_u$ and ${\alpha\,}\bar{\gamma}_w$ {with wall-normal support $f_u(y)$ and $f_w(y)$} to the steady-state covariance $X$ via system matrices {$\bar{A}$} and $B$ and perturbation matrices $A_u$ and $A_w$. It can also be used to compute the energy spectrum of velocity fluctuations $\bv$, 
\begin{align}
	\label{eq.E}	
	E(\bk) \;=\; \trace \left(\Phi(\bk)\right) \;=\; \trace \left(C(\bk) X(\bk) C^*(\bk)\right)
\end{align}
where $\Phi$ is the covariance matrix of $\bv$. We can then capture the influence of multiplicative uncertainty on the energy spectrum using the discounted spectrum
\begin{align}
	\label{eq.Ec}	
	E_c(\bk) \;=\; E(\bk)\,-\, E_0(\bk)
\end{align}
where $E_0$ denotes the nominal energy spectrum in the absence of uncertainties $\gamma_u$ and $\gamma_w$.

{Following~\eqref{eq.B0-C0} and the assumption of repeated base flow perturbations, which yield $\Gamma = \diag\{\sigma_u^2\, \mathds{1}\mathds{1}^T,\,\sigma_w^2\, \mathds{1}\mathds{1}^T\}$, Eq.~\eqref{eq.gen-lyap-original} can be expanded to reflect contributions from uncertainties affecting the streamwise and spanwise components of the base flow as
\begin{align}
	\label{eq.gen-lyap1}
	\bar{A} \, X \;+\; X\,\bar{A}^* \;+\; {\alpha^2 \left( \sigma_u^2 \left( A_u\, X \,A_u^* \right) \;+\; \sigma_w^2 \left(A_w\, X \,A_w^* \right) \right)}
	~=\;
	- B\,\Omega\, B^*
\end{align}
A direct approach to solving~\eqref{eq.gen-lyap1}} as a linear system of equations yields
\begin{align}
    \label{eq.genlyap-vectorized}
 \left( I\otimes \bar{A} \;+\; \bar{A}\otimes I \;+\; {\alpha^2 \left(\sigma_u^2\left(A_u \otimes A_u\right) \;+\; \sigma_w^2\left(A_w \otimes A_w \right) \right)} \right) \mathrm{vec}(X)
\;=\;
-\, \mathrm{vec}(B\,\Omega\, B^*)
\end{align}
where $\otimes$ is the Kronecker product and $\mathrm{vec}(\cdot)$ denotes vectorization. However, in the absence of sparse matrix structures, solving for $X$ can be challenging even for medium-size problems. {Other existing methods for solving~\eqref{eq.gen-lyap1} explore solutions to surrogate equations and utilize iterative methods to improve computational complexity~\cite{ben04,benlipen08,dam08,bendam11}. In what follows, we consider small-amplitude perturbations ($\alpha \ll 1$) and pursue an alternative approach by utilizing a perturbation analysis to achieve a computationally efficient way of obtaining the energy spectrum. As shown in Appendix~\ref{app.pert-anal}, this approach allows us to compute the second-order statistics of the uncertain model by solving a sequence of standard algebraic Lyapunov equations instead of the generalized Lyapunov equation~\eqref{eq.gen-lyap1}. In addition to the computational benefit, the choice of small perturbation amplitude is motivated by the desire to account for uncertainties arising from measurement imperfections, small-data issues in the statistical averaging process, or the effect of random active/passive boundary actuation strategies that influence the base flow. Based on this, up to a second order in the perturbation amplitude $\alpha$, the state covariance $X$ in~\eqref{eq.X-def} is given by
\begin{align}
	\label{eq.X-pert-anal}
	X(\bk)
	\;=\;
	X_0(\bk) \,+\, \alpha^2\,X_2(\bk) \,+\, O(\alpha^4)
\end{align}
where $X_0$ and $X_2$ are obtained from a set of decoupled Lyapunov equations; see Appendix~\ref{app.pert-anal} for details. Note that $X_0$ represents the steady-state covariance of $\bpsi$, i.e., the state of the nominal dynamics in the absence of base flow perturbations, and $X_2$ represents the second-order correction induced by the random base flow uncertainty. The energy spectrum of velocity fluctuations $\bv$ (Eq.~\eqref{eq.E}) follows a similar perturbation series as~\eqref{eq.X-pert-anal}:
\be
	\label{eq.E-pert}	
	E(\bk) \;=\; E_0(\bk) \,+\, \alpha^2 E_2(\bk) \,+\, O(\alpha^4)
\ee
where {$E_0(\bk) = \trace (C(\bk) X_0(\bk) C^*(\bk))$} is the nominal energy spectrum in the absence of base flow perturbations, and {$E_2(\bk) = \trace (C(\bk) X_2(\bk) C^*(\bk))$} captures the effect of base flow perturbations at the level of $\alpha^2$. When $\alpha \ll 1$, the correction $\alpha^2 E_2(\bk)$ provides a good approximation of the discounted spectrum $E_c(\bk)$ in Eq.~\eqref{eq.Ec}, and as $\alpha$ grows, higher-order terms may be needed to better approximate $E(\bk)$.
}

\section{Effect of base flow variations on transitional flows}
\label{sec.laminar-flows}

{
In this section, we examine the dynamics of stochastically forced Couette and Poiseuille flows in the presence of zero-mean white-in-time stochastic base flow uncertainty $\bar{\gamma}_u$ in the streamwise direction. The nominal dynamics are obtained by linearizing the NS equations around $\bar{\bu} = [\,\bar{U}(y)\,~0\,~0\,]^T$ with $\bar{U}(y) = y$ for Couette flow (Fig.~\ref{fig.couette}) and $\bar{U}(y) = 1-y^2$ for Poiseuille flow (Fig.~\ref{fig.poiseuille}). Throughout this section, we use $N=101$ Chebyshev collocation points to discretize the operators involved in the linearized equations. Grid convergence is ensured by doubling the number of collocation points. We first examine the MSS of the flow in the presence of streamwise base flow perturbations. Our analysis identifies critically destabilizing perturbation variances over a range of Reynolds numbers. Using these critical variance levels, we examine the effect of base flow perturbations of various amplitude on the energy spectrum and dominant flow structures.
}

\subsection{{Stability analysis}}
\label{sec.stability-laminar}

{
For both Couette and Poiseuille flows, we use the stability condition presented in Sec.~\ref{sec.MSS} to examine the MSS of the horizontal wavenumber pair $\bk= (1,1)$, which corresponds to an oblique flow structure. Both DNS~\cite{redschbaghen98} and nonlinear optimal perturbation analysis~\cite{rabcauker12} have demonstrated the fragility of such flow structures in transition mechanisms, i.e., oblique modes require less energy to induce transition than streamwise elongated modes. The high sensitivity of such three-dimensional flow structures to additive streamwise excitations was also demonstrated using frequency response analysis of the linearized NS equations~\cite{jovbamJFM05}.}

\begin{figure}
                \begin{tabular}{cccc}
      \subfigure[]{\label{fig.sigma2vsR_couette}}
        &
        &
        \hspace{.4cm}
       \subfigure[]{\label{fig.sigma2vsR_poiseuille}}
        &
        \\[-.6cm]
        \hspace{.2cm}
        \begin{tabular}{c}
                \vspace{.1cm}
                {\normalsize \rotatebox{90}{$\sigma^2_u$}}
        \end{tabular}
        &
        \begin{tabular}{c}
               \includegraphics[width=6.5cm]{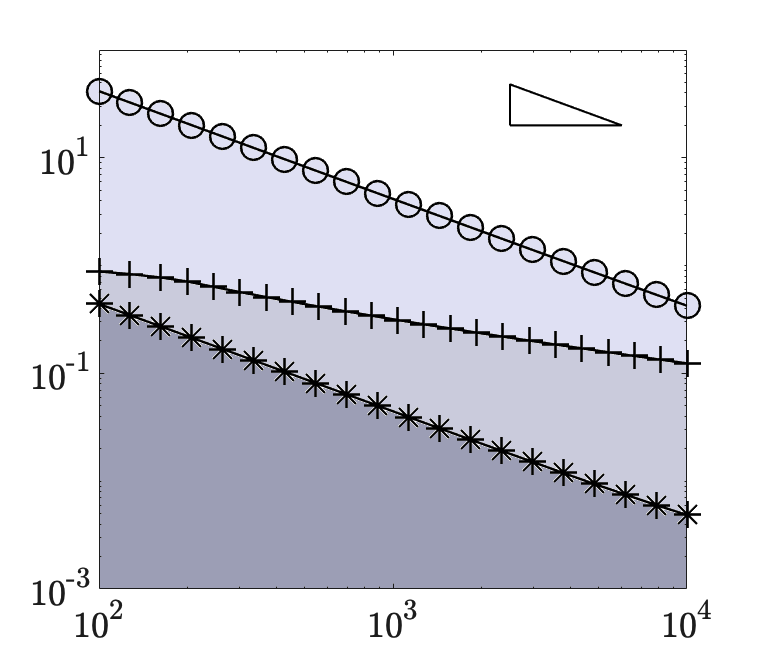}
        \end{tabular}
        &
        \hspace{.6cm}
        \begin{tabular}{c}
        		\vspace{.4cm}
        		{\normalsize \rotatebox{90}{$\sigma^2_u$}}
        \end{tabular}
        &
        \begin{tabular}{c}
                \includegraphics[width=6.5cm]{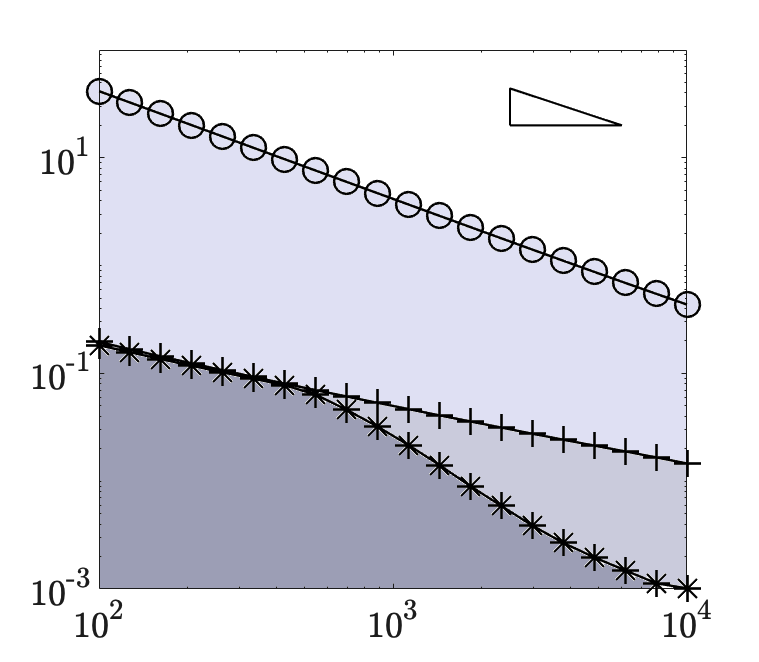}
        \end{tabular}
        \\[.2cm]
        &
        {\normalsize $R$}
        &&
        {\normalsize $R$}
        \end{tabular}
             \caption{Stability curves for fluctuation dynamics with $\bk = (1,1)$ in (a) Couette flow; and (b) Poiseuille flow. The curves demonstrate the Reynolds number dependence of the maximum tolerable variance for stochastic base flow perturbations entering the dynamics through {the shape functions $f(y)$ corresponding to Eq.~\eqref{eq.fshape-Ubar} ($+$) or} Figs.~\ref{fig.fshape1} ({\large$*$}) and~\ref{fig.fvec} ({\large$\circ$}). For a given Reynolds number, the shaded areas under the curves denote the variances of stochastic base flow uncertainty that do not violate MSS ($\rho(\mathbb{L})<1$ {with $\alpha = 1$}). {The triangles in the upper right corners demonstrate an $R^{-1}$ slope}.}
        \label{fig.sigma2vsR}
\end{figure}

{
Figure~\ref{fig.sigma2vsR} shows the minimum destabilizing variance $\sigma^2_u$ over a range of Reynolds numbers for $\bk= (1,1)$ when the base flow is perturbed at different wall-normal regions.} 
The shaded areas under the curves denote the Reynolds numbers and perturbation variances for which the flows remain asymptotically mean-square stable. 
In both flows, higher variances $\sigma_u^2$ could be tolerated when stochastic perturbations were confined to the wall-normal regions close to the walls, i.e., when $f(y)$ {corresponds to the shape function shown in Fig.~\ref{fig.fvec}. As expected, both flows become less robust to base flow perturbations as the region of influence grows in the wall-normal dimension. We also observe that while the stability curves corresponding to the oblique mode in Couette and Poiseuille flows are similar for near-wall perturbations (Fig.~\ref{fig.fvec}), the oblique mode in Poiseuille flow is, generally, more sensitive to channel-wide base flow perturbations ($f(y)$ given by Eq.~\eqref{eq.fshape-Ubar} and Fig.~\ref{fig.fshape1}).}

While it generally becomes easier to destabilize the flow at higher Reynolds numbers, critical variance levels demonstrate different Reynolds number scaling when base flow perturbations are confined to different wall-normal regions. In both Couette and Poiseuille flows the critical variance of near-wall base flow perturbations (Fig.~\ref{fig.fvec}) are found to scale as $R^{-1}$. In other words, it is reasonable to expect larger persistent stochastic perturbations with variances of the same order ($R^{-1}$) to induce $O(R^{-1})$ growth rates that can instigate transition. On the other hand, if perturbations follow the shape of the corresponding base flows ($f(y)$ given by Eq.~\eqref{eq.fshape-Ubar}), the critical variance levels decrease at a slower rate ($R^{-0.5}$). When the base flow perturbations are allowed to enter through the entire wall-normal extent of the channel (Fig.~\ref{fig.fshape1}), the critical variance levels are found to scale as $O(R^{-1})$ in Couette flow and at an increasing rate in Poiseuille flow ($O(R^{-0.5})$ for $R<600$ and $O(R^{-1.2})$ for $R>600$). {The elevated sensitivity of higher Reynolds number Poiseuille flow to base flow perturbations entering the entire wall-normal extent (Fig.~\ref{fig.fshape1}) may be explained by an increase in the energy and vertical reach of near-wall motions, which are less affected when $f(y)=\bar{U}(y)/\max(|\bar{U}(y)|)$.} It is noteworthy that the Reynolds number scaling obtained using our stochastic approach is in agreement with the scaling observed in~\cite{botcorluc03} for the magnitude of deterministic (worst-case) base flow perturbations. 

\begin{figure}
                \begin{tabular}{cccc}
      \subfigure[]{\label{fig.sigma2vsR_couette_Ubar}}
        &
        &
        \hspace{.4cm}
      \subfigure[]{\label{fig.sigma2vsR_poiseuille_Ubar}}
        &
        \\[-.2cm]
        \hspace{.2cm}
        \begin{tabular}{c}
                \vspace{.1cm}
                {\normalsize {$k_x$}}
        \end{tabular}
        &
        \begin{tabular}{c}
               \includegraphics[width=6.5cm]{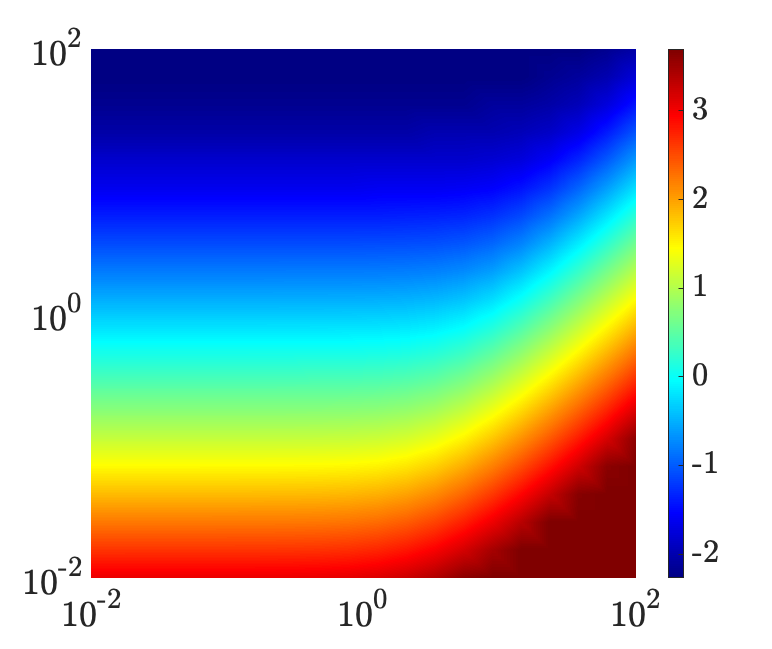}
        \end{tabular}
        &
        \hspace{.6cm}
        \begin{tabular}{c}
        		\vspace{.4cm}
        		{\normalsize {$k_x$}}
        \end{tabular}
        &
        \begin{tabular}{c}
                \includegraphics[width=6.5cm]{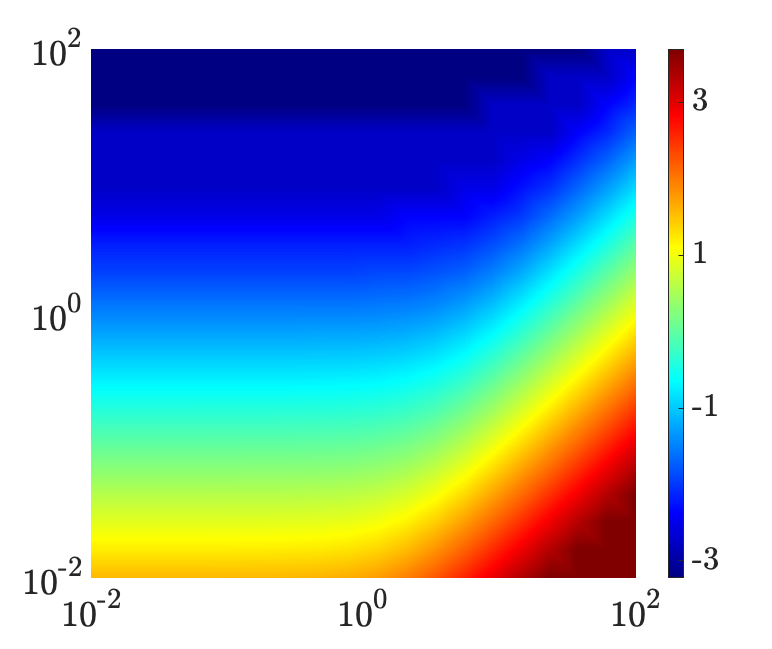}
        \end{tabular}
        \\[.2cm]
        &
        {\normalsize $k_z$}
        &&
        {\normalsize $k_z$}
        \end{tabular}
             \caption{{Logarithmically scaled critical variance levels for stochastic multiplicative uncertainty $\bar{\gamma}_u$ with $\alpha=1$ and {$f(y)=\bar{U}(y)/\max(|\bar{U}(y)|)$} over the horizontal wavenumber spectrum in (a) Couette flow with $R=500$; and (b) Poiseuille flow with $R=2000$.}}
        \label{fig.sigma2_critical_kxkzgrd}
\end{figure}

We note that besides $\bk=(1,1)$, a similar Reynolds number dependence can be observed for the critical variance of base flow perturbations at other horizontal wavenumber pairs. For Couette flow at $R=500$ and Poiseuille flow at $R=2000$ subject to base flow perturbations with {$f(y)=\bar{U}(y)/\max(|\bar{U}(y)|)$}, Fig.~\ref{fig.sigma2_critical_kxkzgrd} shows the critical variance levels $\sigma^2_u$ for flow fluctuations with different spanwise and streamwise wavenumbers. In both flows, streamwise elongated structures (smaller $k_x$) are more robust toward streamwise base flow perturbations. On the other hand, the sensitivity to such perturbations is largely invariant to the width of flow structures and only decreases for longer flow structures when $\lambda_z\lesssim 1$. Based on Fig.~\ref{fig.sigma2_critical_kxkzgrd}, streamwise elongated structures (streaks) that are thin in the spanwise dimension exhibit the least sensitivity to such base flow uncertainty. Finally, we note that similar trends can be observed in both flows when base flow perturbations are allowed to enter a larger extent of the wall-normal domain (when $f(y)$ follows Fig.~\ref{fig.fshape1}).

\subsection{{Energy spectrum of velocity fluctuations}}
\label{sec.energy-plots-laminar}

{
We now use the maximum tolerable variance over all horizontal wavenumber pairs to study the effect of base flow perturbations on the energy spectrum of velocity fluctuations. In both flows, the most sensitive modes that have been considered in Fig.~\ref{fig.sigma2_critical_kxkzgrd} correspond to $\bk = (100,0.01)$. In Couette flow with $R=500$, the critical variance levels $\sigma_u^2$ for streamwise base flow perturbations that enter through wall-normal regions corresponding to  {Eq.~\eqref{eq.fshape-Ubar}}, Fig.~\ref{fig.fshape1}, and Fig.~\ref{fig.fvec} to destabilize this mode are $0.44$, $0.09$, and $8.3$, respectively. These values change to $0.03$, $0.008$, and $2.09$ in Poiseuille flow with $R=2000$. In the numerical experiments of this section, we consider slightly lower variance levels than these critical values to ensure MSS (and thereby the validity of the covariance computed from the steady-state generalized Lyapunov equation~\eqref{eq.gen-lyap1}) over all length-scales. Moreover, we assume the stochastic input {\bf f} to be white-in-time with trivial covariance $\Omega=I$.}

\begin{figure}
		\begin{tabular}{cccc}
	\hspace{-.6cm} \subfigure[]{\label{fig.Enom_couette}}
	&&
	\hspace{.6cm} 
	\subfigure[]{\label{fig.Enom_poiseuille}}
	&
	\\[-.4cm]
	\begin{tabular}{c}
		\vspace{.4cm}
		\hspace{-.2cm}
		\normalsize{$k_x$}
	\end{tabular}
	&
	\hspace{-.3cm}
	\begin{tabular}{c}
		\includegraphics[width=6.5cm]{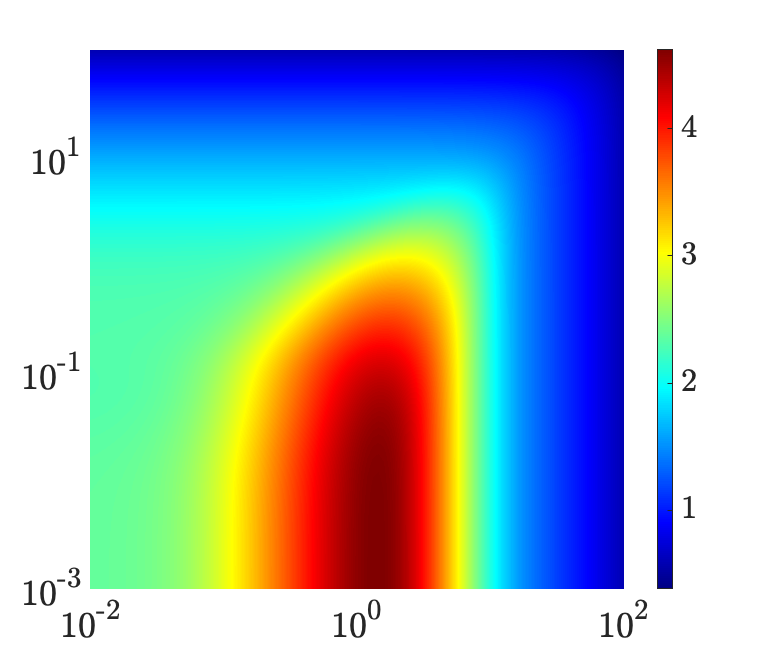}
		\\
		{\normalsize $k_z$}
	\end{tabular}
	&&
	\hspace{.1cm}
	\begin{tabular}{c}
		\includegraphics[width=6.5cm]{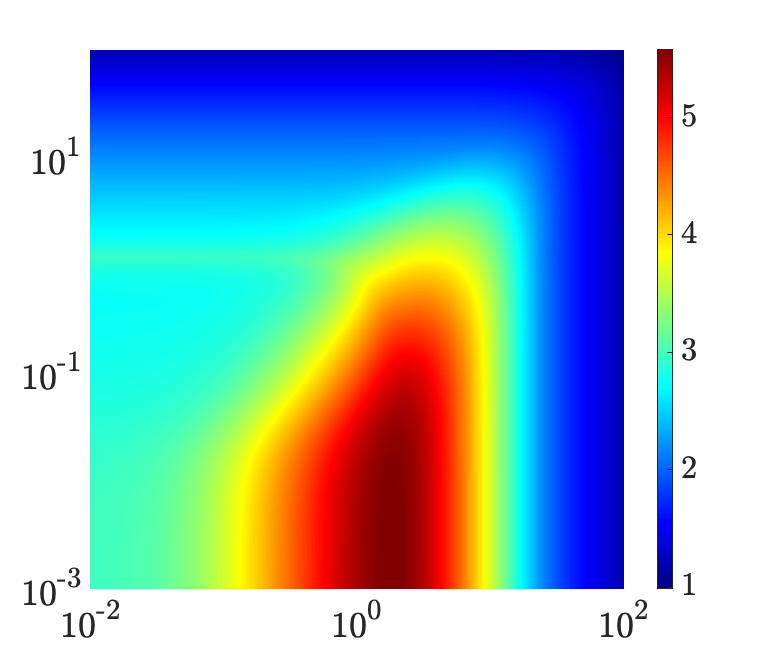}
		\\
		{\normalsize $k_z$}
	\end{tabular}
	\end{tabular}
		\caption{Energy spectra of (a) plane Couette flow with $R=500$ and (b) plane Poiseuille flow with $R=2000$. Color plots show $\log_{10}(E_0(\bk))$.}
        \label{fig.Energy_plots_laminar_nominal}
\end{figure}

\begin{figure}
\centering
	\begin{tabular}{cc}
        \begin{tabular}{c}
        		\vspace{.1cm}
        		{\normalsize \rotatebox{90}{$E_c$}}
        \end{tabular}
        &
        \hspace{-.2cm}
        \begin{tabular}{c}
         \includegraphics[width=6.5cm]{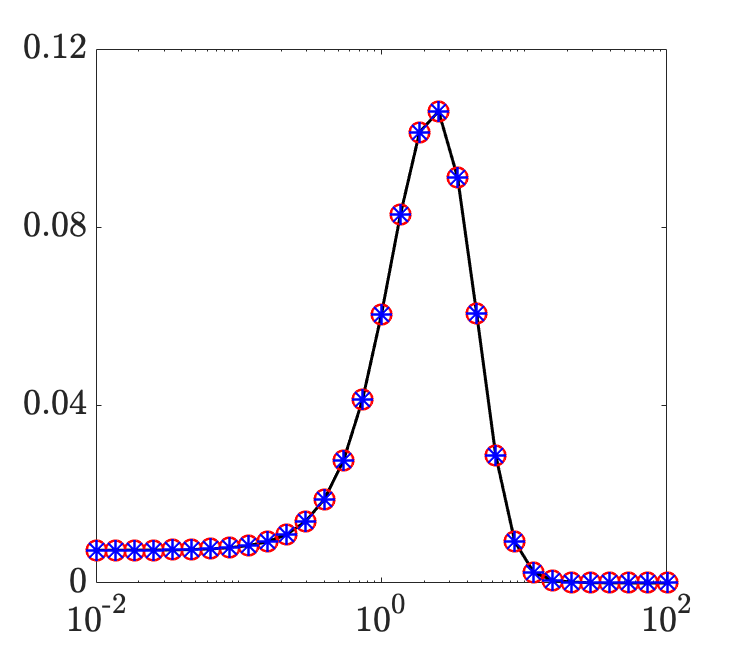}
	\end{tabular}
        \\[-.1cm]
        &
        {\normalsize $k_z$}
        \vspace{-.2cm}
        \end{tabular}
                \caption{{The discounted energy spectrum $E_c$ in Couette flow with $R=500$ and $k_x = 1$ subject to base flow perturbations with {$f(y) = \bar{U}(y)/\max(|\bar{U}(y)|)$} with $\sigma^2_u=0.43$ and $\alpha=0.01$. Direct solution from solving Eq.~\eqref{eq.genlyap-vectorized} ($-$); and approximate solutions from perturbation analysis: $E_c = \alpha^2 E_2(\bk)$ ({\large $*$}) and $E_c = \alpha^2 E_2(\bk) + \alpha^4 E_4(\bk)$ (\tc{red}{\large$\circ$}).}}
        \label{fig.vector-pert}
\end{figure}

\begin{figure}[h!]
		\begin{tabular}{cccc}
	\hspace{-.6cm} \subfigure[]{\label{fig.E2_couette_withUbar}}
	&&
	\hspace{.6cm} 
	\subfigure[]{\label{fig.E2_poiseuille_withUbar}}
	&
	\\[-.4cm]
	\begin{tabular}{c}
		\vspace{.4cm}
		\hspace{-.2cm}
		\normalsize{$k_x$}
	\end{tabular}
	&
	\hspace{-.3cm}
	\begin{tabular}{c}
		\includegraphics[width=6.5cm]{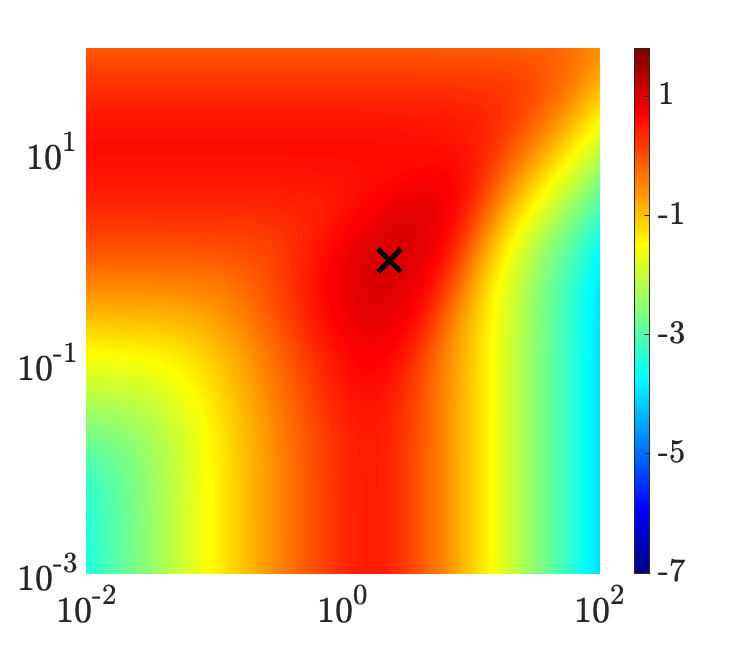}
	\end{tabular}
	&&
	\hspace{.1cm}
	\begin{tabular}{c}
		\includegraphics[width=6.5cm]{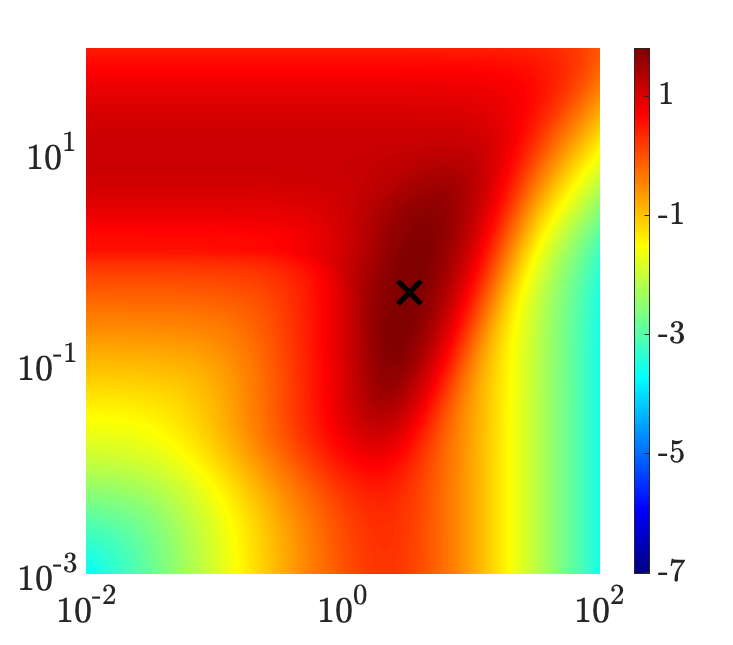}
	\end{tabular}
		\\[2cm]
	\hspace{-.6cm} \subfigure[]{\label{fig.E2_couette_fshape1}}
	&&
	\hspace{.6cm} 
	\subfigure[]{\label{fig.E2_poiseuille_fshape1}}
	&
	\\[-.4cm]
	\begin{tabular}{c}
		\vspace{.4cm}
		\hspace{-.2cm}
		\normalsize{$k_x$}
	\end{tabular}
	&
	\hspace{-.3cm}
	\begin{tabular}{c}
		\includegraphics[width=6.5cm]{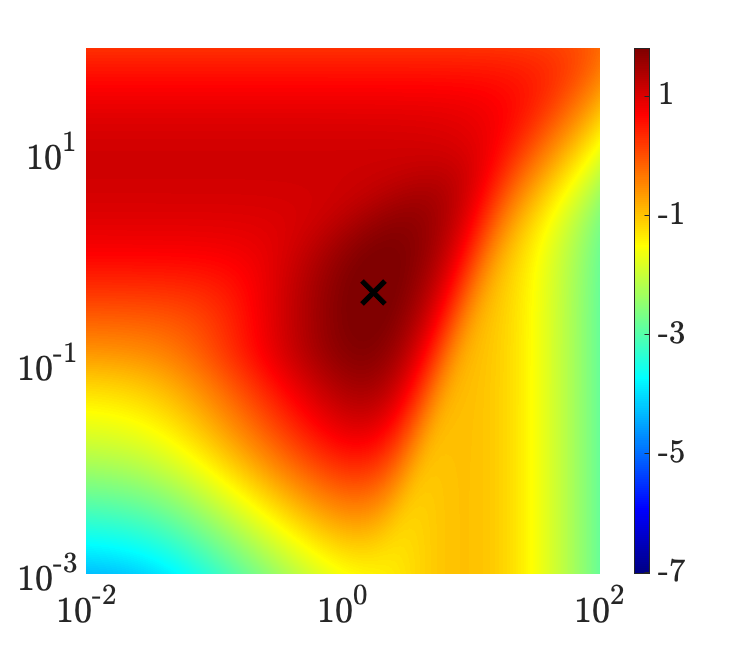}
	\end{tabular}
	&&
	\hspace{.1cm}
	\begin{tabular}{c}
		\includegraphics[width=6.5cm]{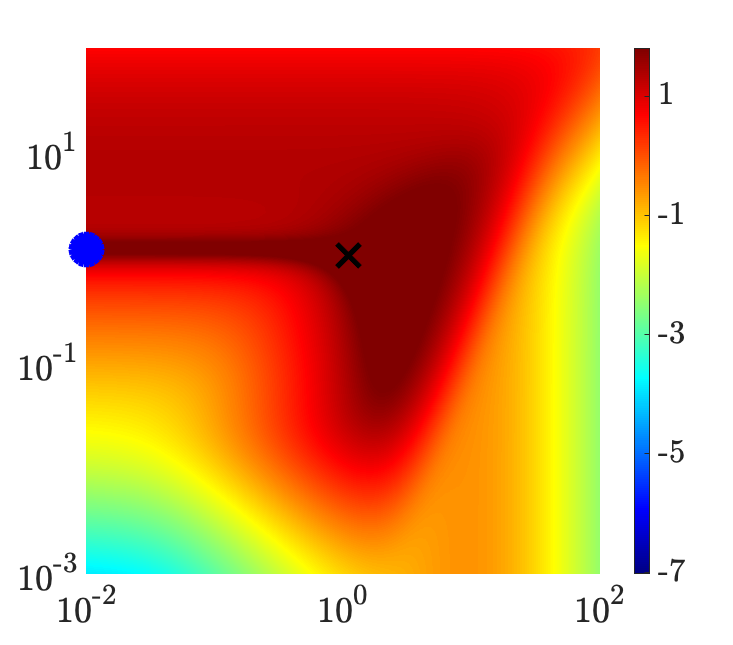}
	\end{tabular}
	\\[2cm]
	\hspace{-.6cm} \subfigure[]{\label{fig.E2_couette_withfvec}}
	&&
	\hspace{.6cm} 
	\subfigure[]{\label{fig.E2_poiseuille_withfvec}}
	&
	\\[-.4cm]
	\begin{tabular}{c}
		\vspace{.4cm}
		\hspace{-.2cm}
		\normalsize{$k_x$}
	\end{tabular}
	&
	\hspace{-.3cm}
	\begin{tabular}{c}
		\includegraphics[width=6.5cm]{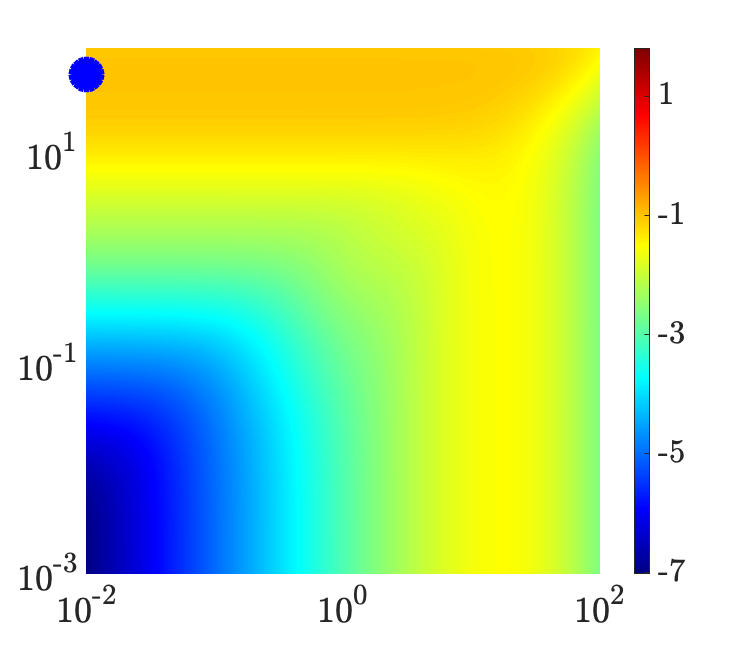}
		\\
		{\normalsize{$k_z$}}
	\end{tabular}
	&&
	\hspace{.1cm}
	\begin{tabular}{c}
		\includegraphics[width=6.5cm]{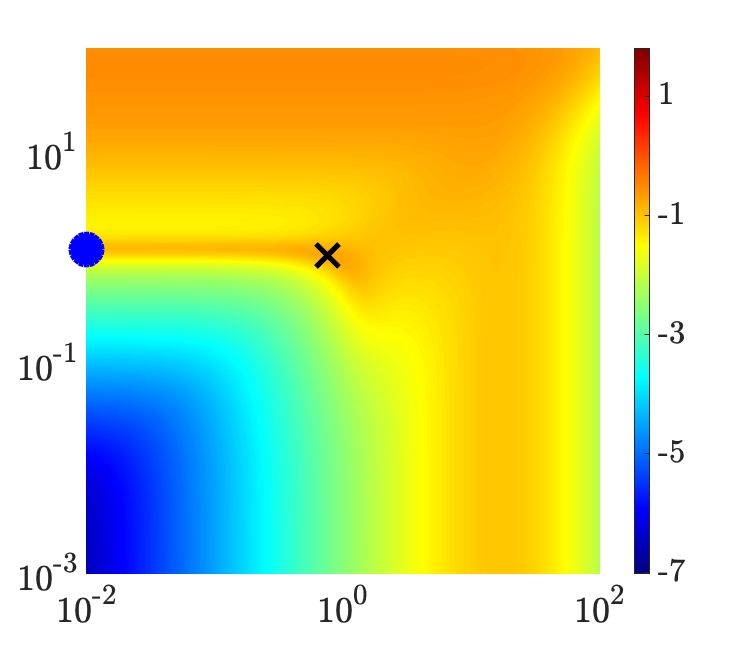}
		\\
		{\normalsize $k_z$}
	\end{tabular}
	\end{tabular}
		\caption{{The second-order correction to the energy spectrum $E_2(\bk)$ due to multiplicative uncertainty $\bar{\gamma}_u$ in Couette flow with $R=500$ (left); and Poiseuille flow with $R=2000$ (right). Shape functions {$f(y)$ correspond to (a,b) Eq.~\eqref{eq.fshape-Ubar}}; (c,d) Fig.~\ref{fig.fshape1}; (e,f) Fig.~\ref{fig.fvec}. Variances $\sigma_u^2$: (a) $4.38\times10^{-3}$; (b) $6.25\times10^{-4}$; (c) $0.004$; (d) $6.25\times10^{-4}$; (e) $0.005$; (f) $0.001$.  Color plots show $\log_{10}(E_2(\bk))$. The symbols (\tc{black}{$\times$}) and ({\large $\bullet$}) mark the wavenumber pairs associated with oblique waves and TS waves, respectively.}}
        \label{fig.Energy_plots_laminar}
\end{figure}

In the absence of multiplicative uncertainty ($\gamma_u = 0$), Eq.~\eqref{eq.gen-lyap-original} reduces to a standard algebraic Lyapunov equation. The nominal energy spectra $E_0$ of plane Couette and Poiseuille flows, which can be computed from the solution of this Lyapunov equation, are shown in Figs.~\ref{fig.Enom_couette} and~\ref{fig.Enom_poiseuille}, respectively. {In the perturbed case, we use the perturbation analysis presented in Sec.~\ref{sec.FreqResp} to compute the effect of base flow perturbations on the energy of velocity fluctuations by solving a sequence of standard algebraic Lyapunov equations (Eqs.~\eqref{eq.lyap-pert}) instead of the generalized Lyapunov equation~\eqref{eq.gen-lyap1}. Figure~\ref{fig.vector-pert} validates this approach
in predicting the discounted energy spectrum $E_c$ (Eq.~\eqref{eq.Ec}) of plane Couette flow at $k_x=1$ due to small amplitude base flow perturbations with {$f(y)=\bar{U}(y)/\max(|\bar{U}(y)|)$}. This streamwise wavenumber will be shown to contain the most sensitive region of the spectrum (Fig.~\ref{fig.Energy_plots_laminar}). It is evident that for small perturbation amplitudes $\alpha$, even the second-order correction (at the level of $\alpha^2$) is in excellent agreement with the direct solution whose computational cost is significantly higher.}

Figure~\ref{fig.Energy_plots_laminar} shows the second-order correction to the energy spectrum ($E_2(\bk)$) of Couette and Poiseuille flow induced by base flow perturbations of various shape $f(y)$. Clearly, base flow perturbations have resulted in the amplification of all spatial scales. As shown in Fig.~\ref{fig.Energy_plots_laminar_nominal}, the amplification of streamwise elongated flow structures (streaks) dominates the energy spectra of nominal (unperturbed) flows. In contrast, small-amplitude channel-wide base flow perturbations (when $f(y)$ is given by Eq.~\eqref{eq.fshape-Ubar} or Fig.~\ref{fig.fshape1}) predominantly influence the oblique modes with $k_x\approx 1$ and $k_z\sim O(1)$ (marked by (\tc{black}{$\times$})), and near-wall perturbations (when $f(y)$ corresponds to Fig.~\ref{fig.fvec}) result in the dominant amplification of Tollmien–Schlichting (TS) waves (marked by ({\large $\bullet$})). This is to be expected as the three-dimensional oblique modes predominantly reside farther way from the channel walls. We remark that even though the amplification of TS waves is overcome by that of oblique modes when channel-wide base flow perturbations are applied, their local signature at $k_z \approx 0$ prevails in all cases. Figures~\ref{fig.Energy_plots_laminar}(c,d) demonstrate that the amplification of streaks is quite robust to base flow perturbations that are not confined in the wall-normal direction (Fig.~\ref{fig.fshape1}). This is because at $k_x=0$ stochastic base flow perturbations have no way to influence the solution of Eq.~\eqref{eq.gen-lyap-original} as the main diagonal blocks of $A_u$ would be zero and the off-diagonal (coupling) term, which includes the wall-normal derivative of $f(y)$, is predominantly zero (apart from the immediate vicinity of the walls); see Appendix~\ref{app.A0-Au-Aw}. This is in agreement with the findings of the worst-case adjoint-based analysis conducted for a zero-pressure-gradient boundary layer in~\cite{brasippramar11}, where the lack of influence on streamwise streaks is evident from the structure of the analytically derived gradient of resolvent singular values. {It follows from the form of perturbation matrices $A_u$ and $A_w$ in Appendix~\ref{app.A0-Au-Aw} that streaks would be susceptible to multiplicative sources of uncertainty that enter the dynamics through other components of the base state or involve significant wall-normal variations (cf.\ Figs.~\ref{fig.Energy_plots_laminar}(a,b) and~\ref{fig.Energy_plots_laminar}(e,f))}.

\begin{figure}[h!]
		\begin{tabular}{cccc}
	\hspace{-.6cm} \subfigure[]{\label{fig.Ec_couette_withUbar}}
	&&
	\hspace{.6cm} 
	\subfigure[]{\label{fig.Ec_poiseuille_withUbar}}
	&
	\\[-.4cm]
	\begin{tabular}{c}
		\vspace{.4cm}
		\hspace{-.2cm}
		\normalsize{$k_x$}
	\end{tabular}
	&
	\hspace{-.3cm}
	\begin{tabular}{c}
		\includegraphics[width=6.5cm]{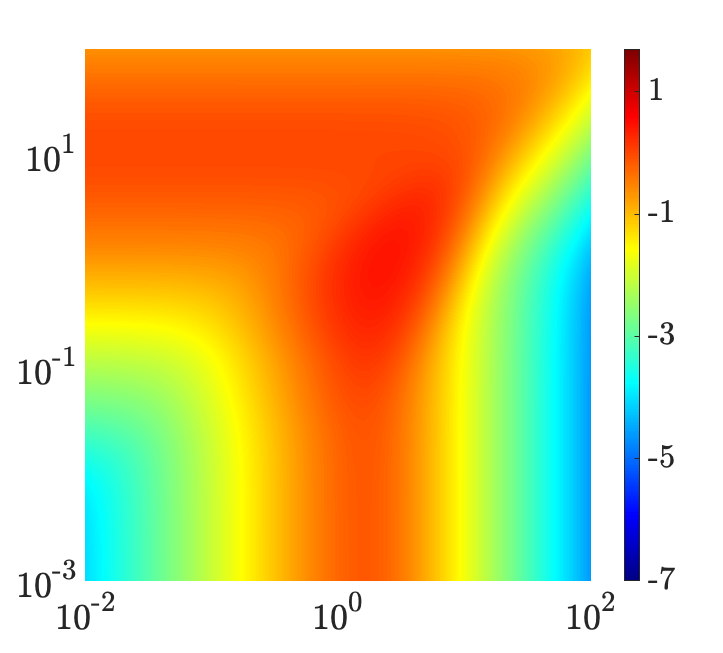}
	\end{tabular}
	&&
	\hspace{.1cm}
	\begin{tabular}{c}
		\includegraphics[width=6.5cm]{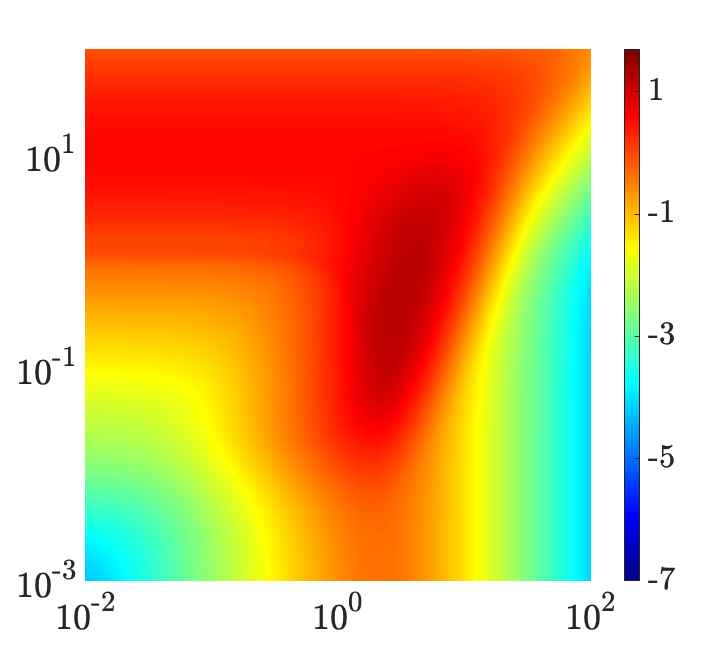}
	\end{tabular}
		\\[2cm]
	\hspace{-.6cm} \subfigure[]{\label{fig.Ec_couette_fshape1}}
	&&
	\hspace{.6cm} 
	\subfigure[]{\label{fig.Ec_poiseuille_fshape1}}
	&
	\\[-.4cm]
	\begin{tabular}{c}
		\vspace{.4cm}
		\hspace{-.2cm}
		\normalsize{$k_x$}
	\end{tabular}
	&
	\hspace{-.3cm}
	\begin{tabular}{c}
		\includegraphics[width=6.5cm]{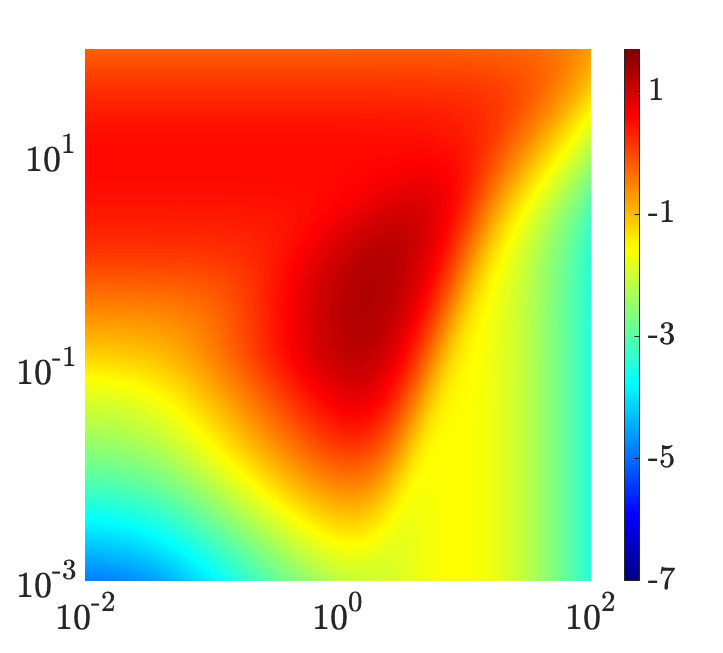}
	\end{tabular}
	&&
	\hspace{.1cm}
	\begin{tabular}{c}
		\includegraphics[width=6.5cm]{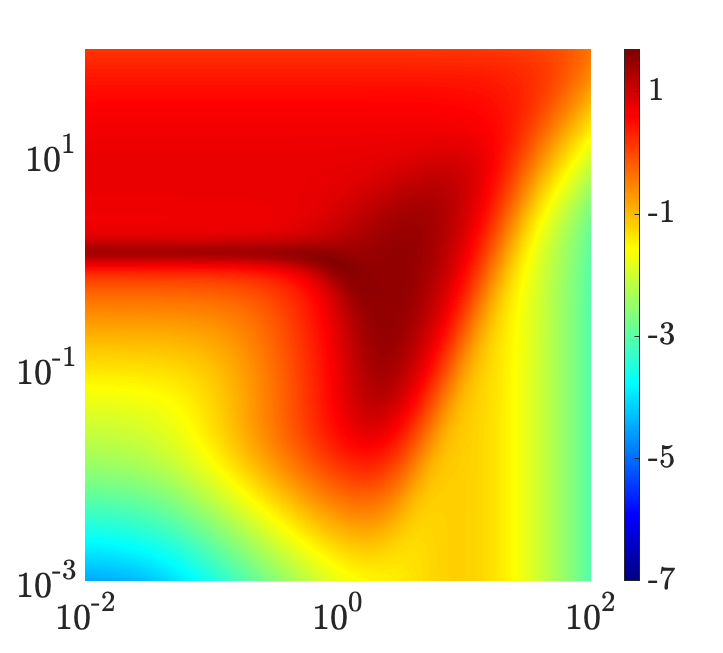}
	\end{tabular}
	\\[2cm]
	\hspace{-.6cm} \subfigure[]{\label{fig.Ec_couette_withfvec}}
	&&
	\hspace{.6cm} 
	\subfigure[]{\label{fig.Ec_poiseuille_withfvec}}
	&
	\\[-.4cm]
	\begin{tabular}{c}
		\vspace{.4cm}
		\hspace{-.2cm}
		\normalsize{$k_x$}
	\end{tabular}
	&
	\hspace{-.3cm}
	\begin{tabular}{c}
		\includegraphics[width=6.5cm]{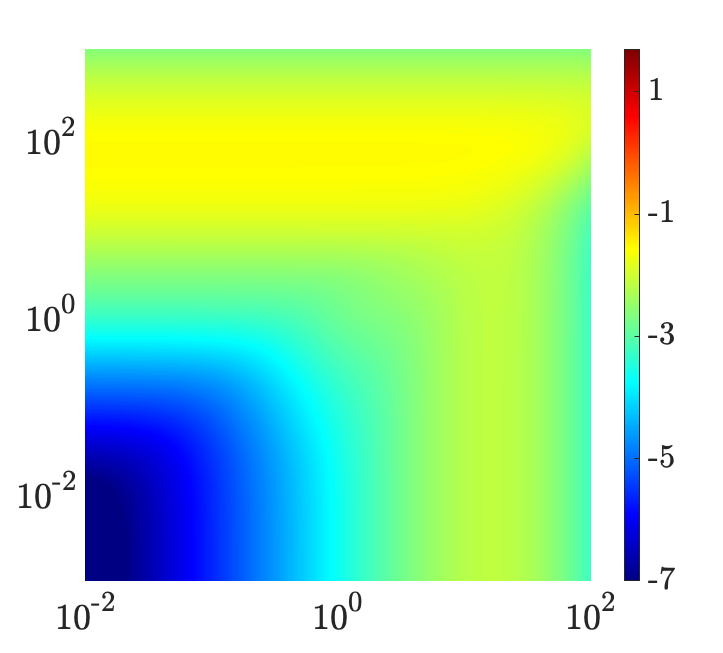}
		\\
		{\normalsize{$k_z$}}
	\end{tabular}
	&&
	\hspace{.1cm}
	\begin{tabular}{c}
		\includegraphics[width=6.5cm]{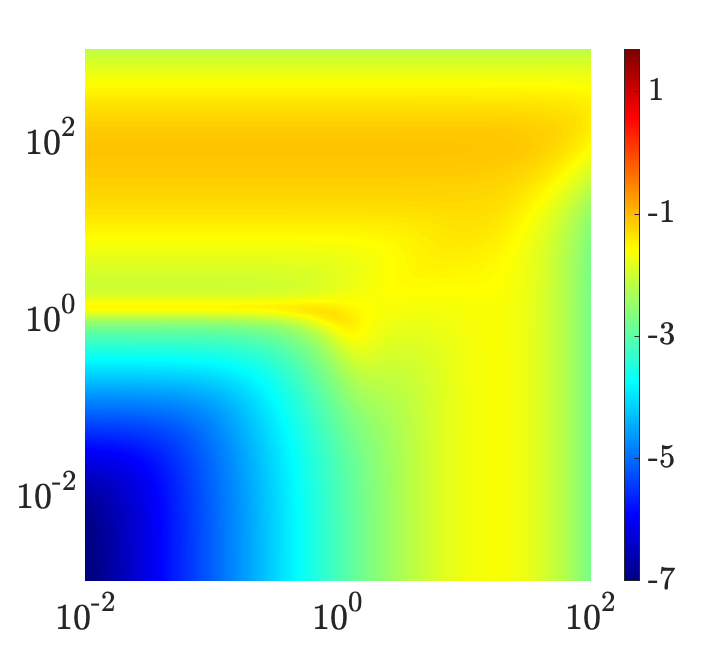}
		\\
		{\normalsize $k_z$}
	\end{tabular}
	\end{tabular}
		\caption{{The correction to the energy spectrum $E_c(\bk)$ due to multiplicative uncertainty $\bar{\gamma}_u$ with $\alpha = 0.5$ in Couette flow with $R=500$ (left); and Poiseuille flow with $R=2000$ (right). Shape functions {$f(y)$ correspond to (a,b) Eq.~\eqref{eq.fshape-Ubar}}; (c,d) Fig.~\ref{fig.fshape1}; (e,f) Fig.~\ref{fig.fvec}. Variances $\sigma_u^2$: (a) $4.38\times10^{-3}$; (b) $6.25\times10^{-4}$; (c) $0.004$; (d) $6.25\times10^{-4}$; (e) $0.005$; (f) $0.001$.  Color plots show $\log_{10}(E_c(\bk))$.}}
        \label{fig.Energy_plots_laminar_alpha0p5}
\end{figure}

\begin{figure}
		\begin{tabular}{cccc}
	\hspace{-.6cm} \subfigure[]{\label{fig.Ec_vs_alpha_couette}}
	&&
	\hspace{.6cm} 
	\subfigure[]{\label{fig.Ec_vs_alpha_poiseuille}}
	&
	\\[-.6cm]
	\begin{tabular}{c}
		\vspace{.4cm}
		\hspace{.2cm}
		\normalsize \rotatebox{90}{$\int_{\bk} E_c(\bk) \,d\bk / \int_{\bk} E_0(\bk) \,d\bk$}
	\end{tabular}
	&
	\hspace{-.3cm}
	\begin{tabular}{c}
		\includegraphics[width=6.5cm]{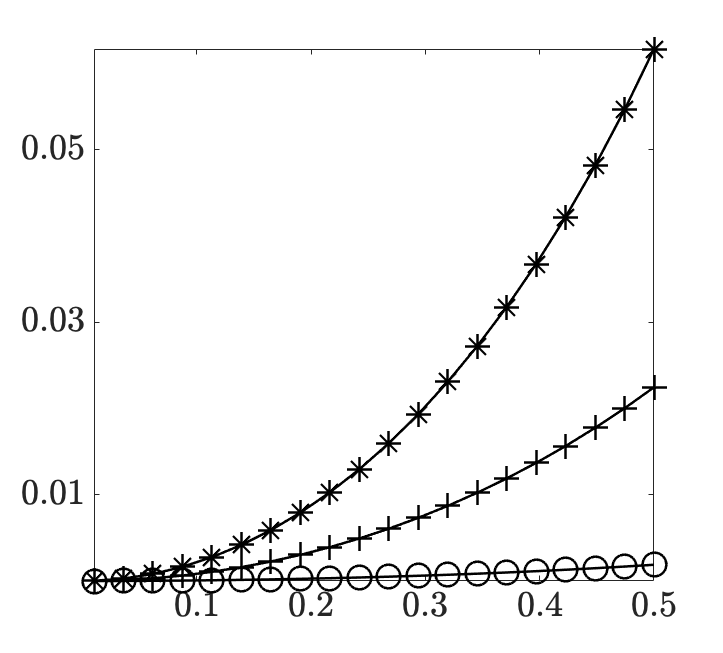}
		\\
		{\normalsize $\alpha$}
	\end{tabular}
	&&
	\hspace{.1cm}
	\begin{tabular}{c}
		\includegraphics[width=6.5cm]{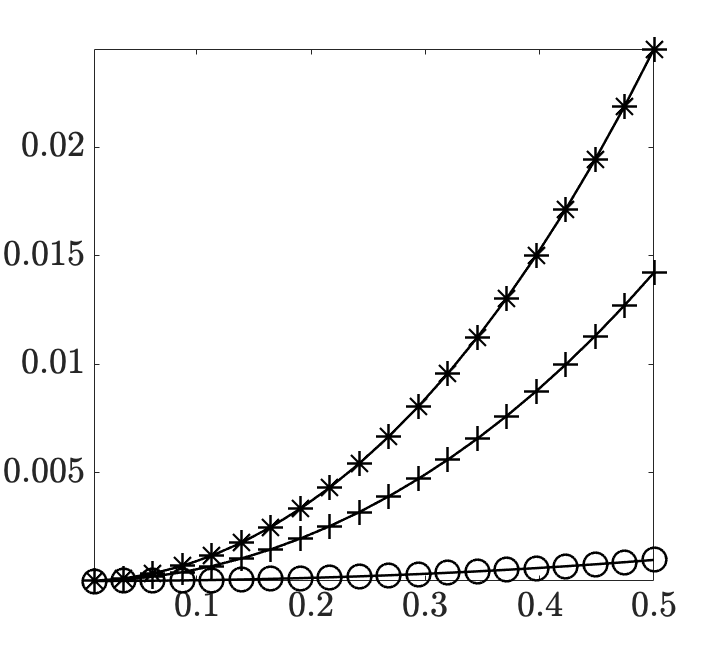}
		\\
		{\normalsize $\alpha$}
	\end{tabular}
	\end{tabular}
		\caption{{The relative correction to the turbulent kinetic energy $\int_{\bk} E_c(\bk) \,d\bk / \int_{\bk} E_0(\bk) \,d\bk$ in (a) Couette flow with $R=500$; and (b) Poiseuille flow with $R=2000$ subject to base flow variations with amplitude $\alpha$. The curves demonstrate the $\alpha$ dependence of the energy correction due to stochastic base flow perturbations entering the dynamics through {$f(y) = \bar{U}(y)/\max(|\bar{U}(y)|)$} (+), Fig.~\ref{fig.fshape1} ({\large$*$}), and Fig.~\ref{fig.fvec} ({\large$\circ$}). (a) Base flow perturbations are introduced with variances of $\sigma_u^2=4.38\times10^{-3}$ ($+$), $0.004$ ({\large$*$}); and $0.005$ ({\large$\circ$}) into Couette flow, and (b) $\sigma_u^2=6.25\times10^{-4}$ ($+$), $6.25\times10^{-4}$ ({\large$*$}); and $0.001$ ({\large$\circ$}) into Poiseuille flow.}}
        \label{fig.Ec_vs_alpha_laminar}
\end{figure}

Figure~\ref{fig.Energy_plots_laminar_alpha0p5} shows the correction to the energy spectrum ($E_c(\bk)$) of Couette and Poiseuille flows in the presence of higher-amplitude base flow perturbations. The perturbation amplitude $\alpha=0.5$ considered in this figure corresponds to the maximum perturbation amplitude for which the result of perturbation analysis is in agreement with the direct solution of Eq.~\eqref{eq.genlyap-vectorized}. In obtaining these plots the limit of the perturbation series~\eqref{eq.E-pert} was obtained using 2 terms (up to $4$th order in $\alpha$) in the perturbation series, i.e., $Ec = \alpha^2 E_2 + \alpha^4 E_4$, and verified using the Shanks transformation. This transformation provides the means to improve the convergence rate of slowly convergent series and to even achieve convergence when the original series is divergent; see~\cite{sha55,van64,wyn66,sid03} for additional details. Figure~\ref{fig.Energy_plots_laminar_alpha0p5} shows that for high amplitude base flow perturbations, apart from a uniform increase in the energy correction over all scales (by approximately $O(3)$), the amplification trends predominantly follow the predictions of Fig.~\ref{fig.Energy_plots_laminar} for small perturbations. For both Couette and Poiseuille flows, Fig.~\ref{fig.Ec_vs_alpha_laminar} examines the dependence of the normalized correction to the total kinetic energy on the amplitude of base flow perturbations that enter the dynamics through various shape functions $f(y)$ and with uncertainty variances $\sigma_u^2$ that correspond to the maximum tolerable values identified in Fig.~\ref{fig.sigma2vsR}. These figures show that the correction to kinetic energy increases as the wall-normal extent of base flow perturbations becomes larger. We also observe that as the amplitude of base flow perturbations increases the exponential growth rate approaches that of higher powers of $\alpha$ (cf.\ Eq.~\eqref{eq.E-pert}).

\begin{figure}
    \begin{tabular}{cccc}
      \subfigure[]{\label{fig.eig_couette}}
        &
        &
        \hspace{.4cm}
       \subfigure[]{\label{fig.eig_poiseuille}}
        &
        \\[-.6cm]
        \hspace{.4cm}
        \begin{tabular}{c}
                \vspace{.1cm}
                {\normalsize \rotatebox{90}{$\lambda_j/\Sigma_i \lambda_i$}}
        \end{tabular}
        &
        \begin{tabular}{c}
              \includegraphics[width=6.5cm]{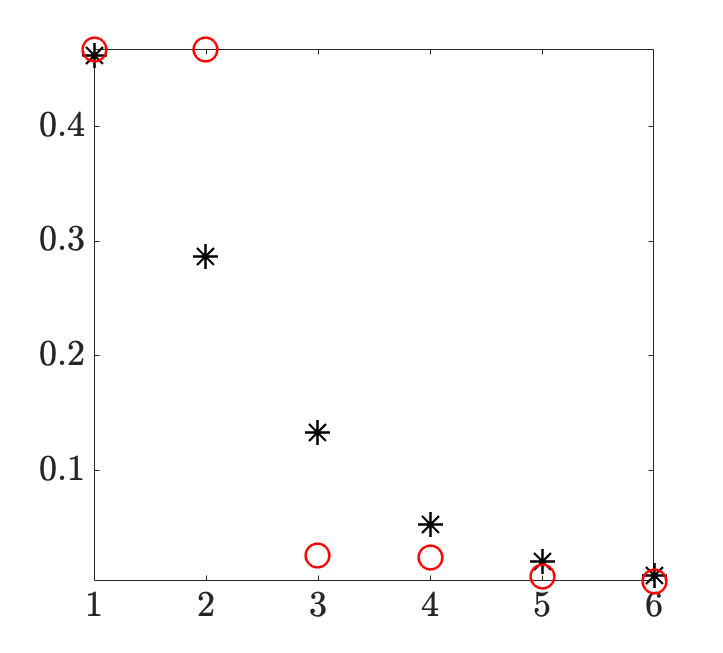}
        \end{tabular}
        &
        \hspace{1cm}
        \begin{tabular}{c}
        		\vspace{.4cm}
        		{\normalsize \rotatebox{90}{$\lambda_j/\Sigma_i \lambda_i$}}
        \end{tabular}
        &
        \begin{tabular}{c}
             \includegraphics[width=6.5cm]{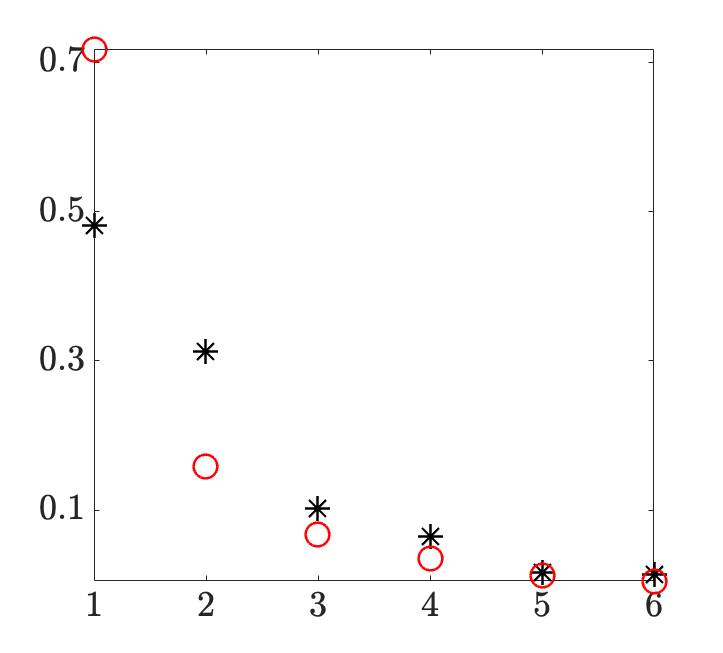}
        \end{tabular}
        \\[.2cm]
        &
        {\normalsize $j$}
        &&
        {\normalsize $j$}
        \end{tabular}
             \caption{{Contribution of the first six eigenvalues of the velocity covariance matrix $\Phi$ of channel flow in the absence (\tc{black}{\large $*$}) and presence (\tc{red}{\large $\circ$}) of base flow perturbations with {$f(y)=\bar{U}(y)/\max(|\bar{U}(y)|)$} and amplitude $\alpha = 1$. (a) Couette flow with $R = 500$ at $\bk=(0.95, 2.29)$; and (b) Poiseuille flow with $R = 2000$ at $\bk=(0.38, 3.02)$. The variance of base flow uncertainties: (a) $\sigma_u^2=0.50$; (b) $\sigma_u^2=0.21$.}}
      \label{fig.principal-modes_laminar}
\end{figure}

\subsection{{Maximally affected flow structures}}
\label{sec.principal flow structures}

{
Following the proper orthogonal decomposition of~\cite{baklum67,moimos89}, we extract the dominant flow structures that result from the steady-state stochastic analysis of transitional flow in the presence of base flow uncertainty. These flow structures can be formed from the energetically dominant eigenvectors of the velocity covariance matrix $\Phi(\bk) = C(\bk)X(\bk)C^*(\bk)$, where $X(\bk)$ represents the solution of the generalized Lyapunov equation~\eqref{eq.genlyap-vectorized}. Following the eigenvalue decomposition of $\Phi$, the symmetries in the wall-parallel directions can be used to construct the velocity components of flow structures~as 
\begin{align}
\label{eq.vel-components}
	\ba{rcl}
	u(x,y,z)
	&=&
	\phantom{-}4\,\ds{\int_{k_x, k_z >0} }\,\cos{(\,k_z z\,) \, \operatorname{Re}\,\left( \tilde{u} \left(\,y,\bk\,\right)\mre^{\mri (k_xx)}\right)\,  \mrd \bk}
	\\[.45cm]
	v(x,y,z)
	&=&
	\phantom{-}4\,\ds{\int_{k_x, k_z >0} }\,\cos{(\,k_z z\,)\,\operatorname{Re}\,\left( \tilde{v} \left(\,y,\bk\,\right)\mre^{\mri (k_xx)}\right)\, \mrd \bk}
	\\[.45cm]
	w(x,y,z)
	&=&
	-4\,\ds{\int_{k_x, k_z >0} }\,\sin{(\,k_z z\,)\,\operatorname{Im}\,\left( \tilde{w} \left (\,y,\bk\,\right)\mre^{\mri (k_xx)}\right)\, \mrd \bk}.
	\ea
\end{align}
Here, $\operatorname{Re}$ and $\operatorname{Im}$ denote real and imaginary parts, and $\tilde{u}$, $\tilde{v}$ and $\tilde{w}$ correspond to the streamwise, wall-normal, and spanwise velocity components of an eigenvector of $\Phi(\bk)$.
}

\begin{figure}
	\begin{tabular}{cccccc}
	 \hspace{-.4cm}
	\subfigure[]{\label{fig.Couette_Ubar_first}}
      &
        \\[-.5cm]
        \hspace{-.2cm}
        \begin{tabular}{c}
                \vspace{4.5cm}
                {\normalsize \rotatebox{90}{$y$}}
        \end{tabular}
        &
        \hspace{-.2cm}
	\includegraphics[width=5.4cm]{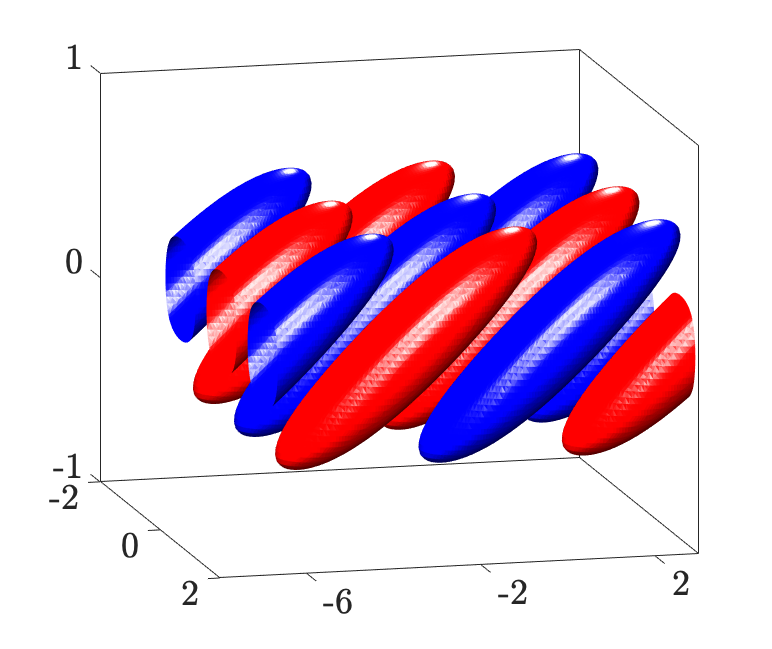}
        &&
	\includegraphics[width=5.4cm]{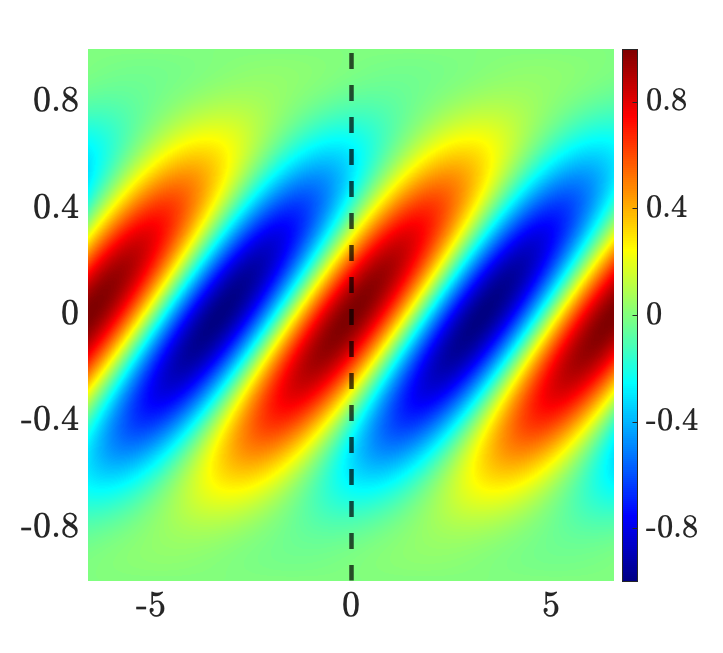}
        &&
	 \includegraphics[width=5.4cm]{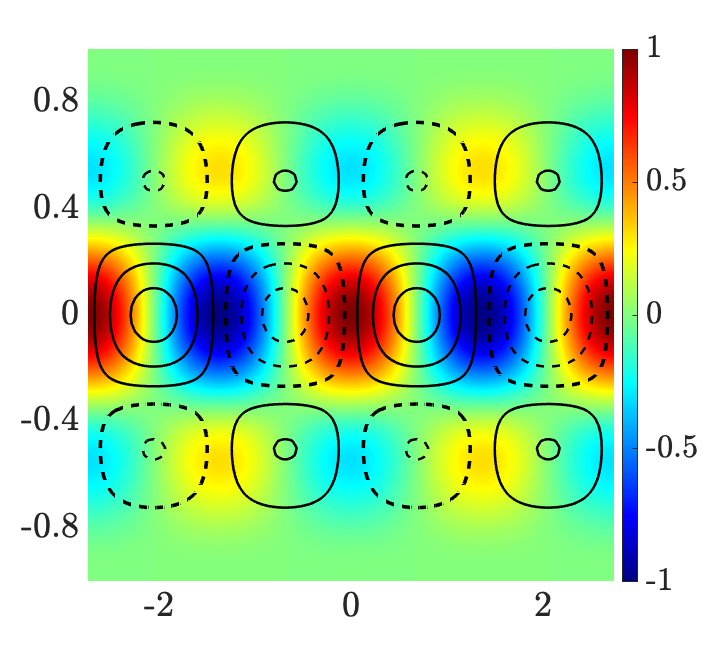}
        \\[-2.4cm]
      &
        \\[-.15cm]
        \hspace{-.2cm}
        \begin{tabular}{c}
                \vspace{4.5cm}
                {\normalsize \rotatebox{90}{$y$}}
        \end{tabular}
        &
        \hspace{-.2cm}
	\includegraphics[width=5.4cm]{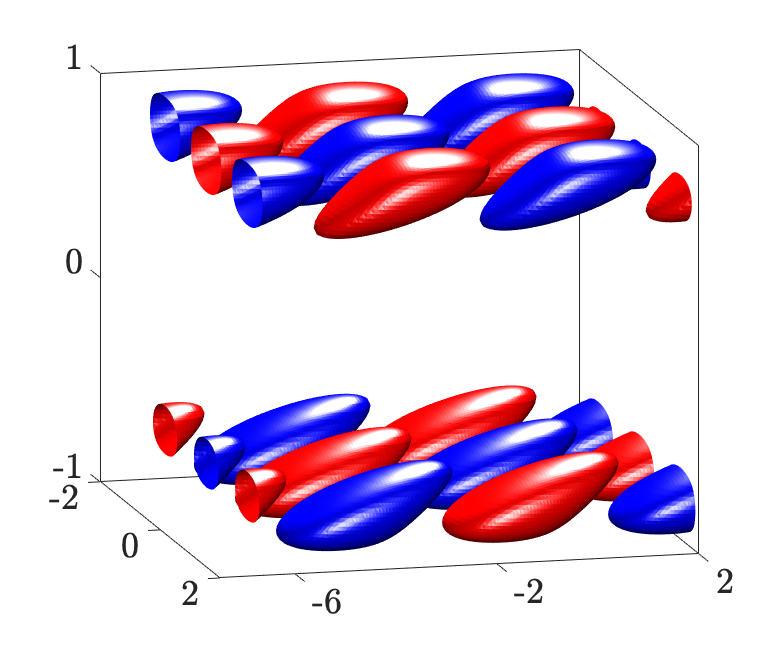}
        &&
	\includegraphics[width=5.4cm]{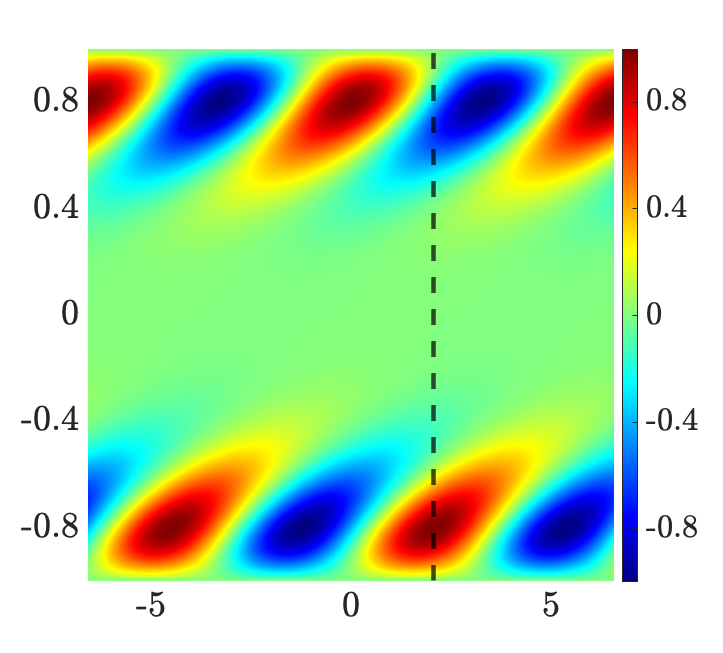}
        &&
	 \includegraphics[width=5.4cm]{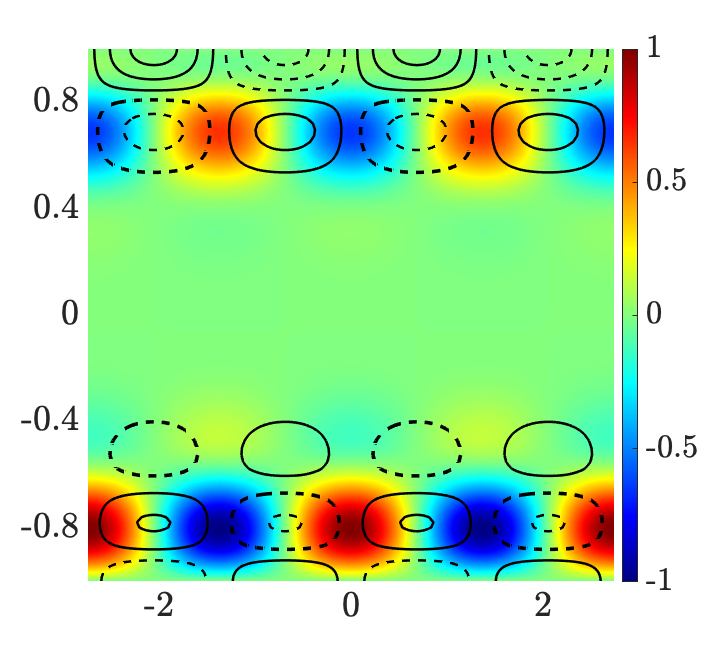}
        \\[-2.2cm]
         \hspace{-.4cm}
	\subfigure[]{\label{fig.Couette_Ubar_second}}
      &
        \\[-.5cm]
        \hspace{-.2cm}
        \begin{tabular}{c}
                \vspace{4.5cm}
                {\normalsize \rotatebox{90}{$y$}}
        \end{tabular}
        &
        \hspace{-.2cm}
	\includegraphics[width=5.4cm]{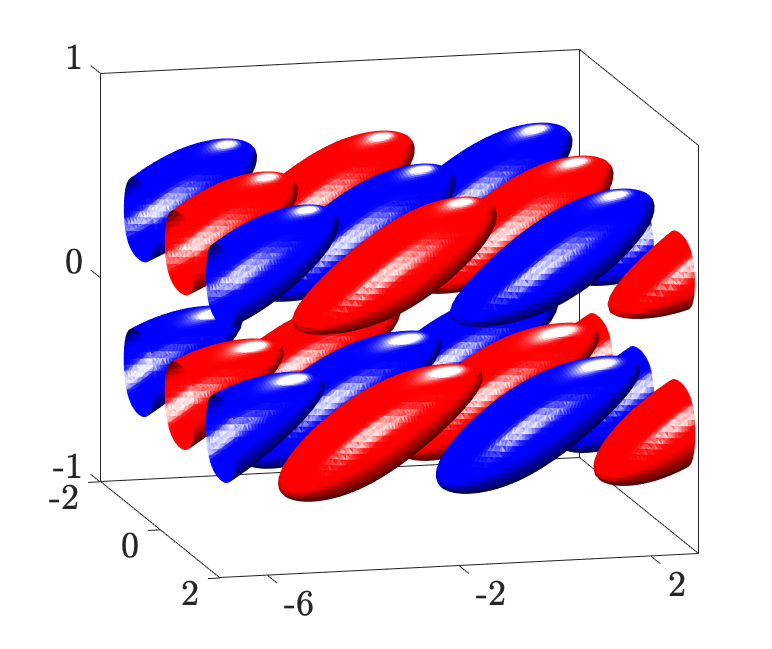}
        &&
	\includegraphics[width=5.4cm]{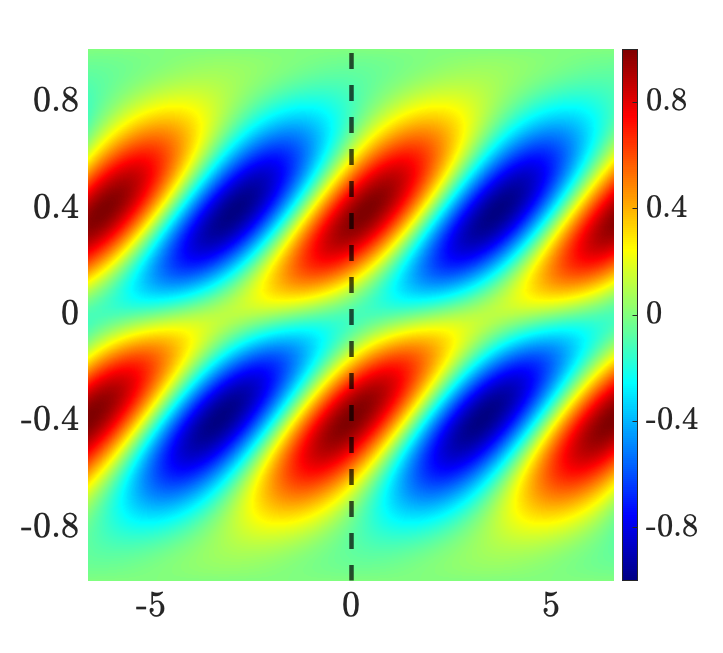}
        &&
	 \includegraphics[width=5.4cm]{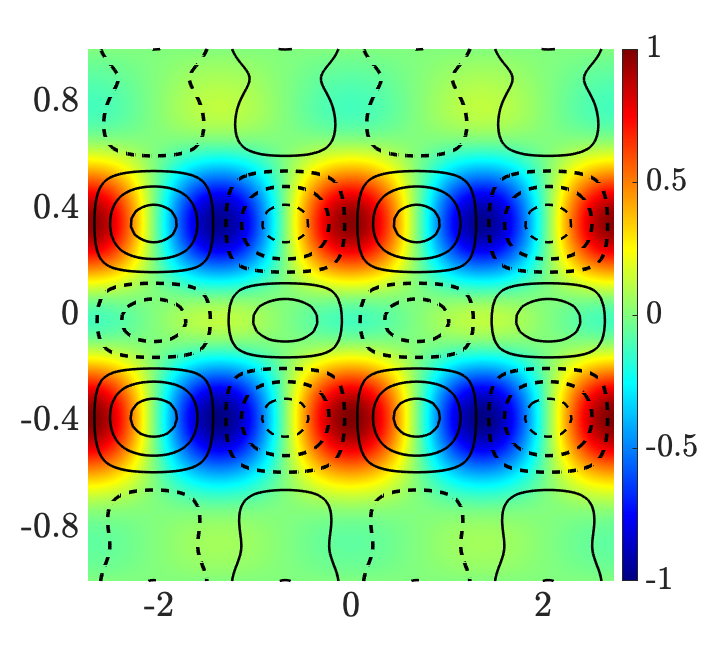}
	 \\[-2.4cm]
      &
        \\[-.15cm]
        \hspace{-.2cm}
        \begin{tabular}{c}
                \vspace{4.5cm}
                {\normalsize \rotatebox{90}{$y$}}
        \end{tabular}
        &
        \hspace{-.2cm}
	\includegraphics[width=5.4cm]{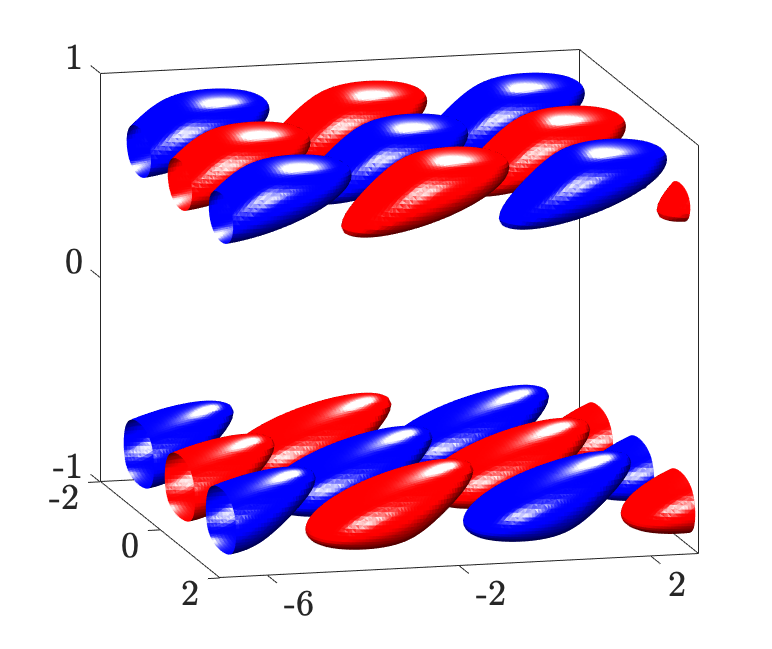}
        &&
	\includegraphics[width=5.4cm]{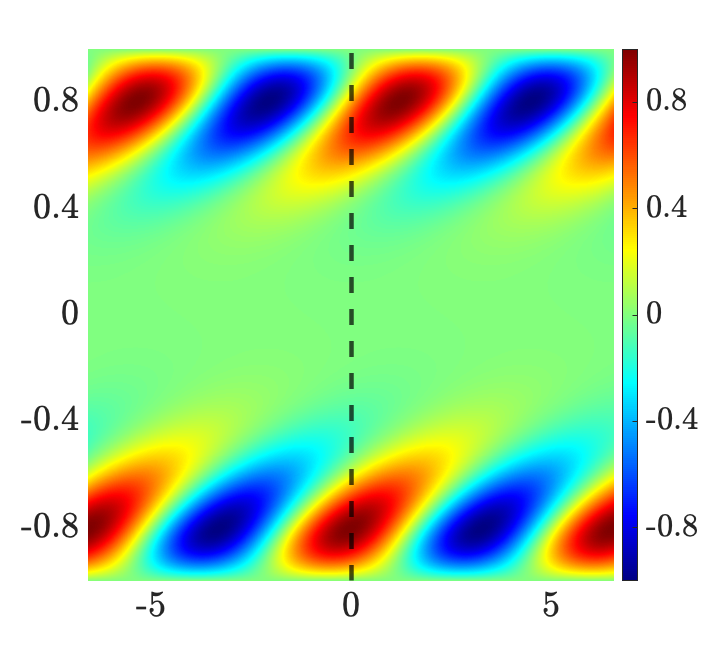}
        &&
	 \includegraphics[width=5.4cm]{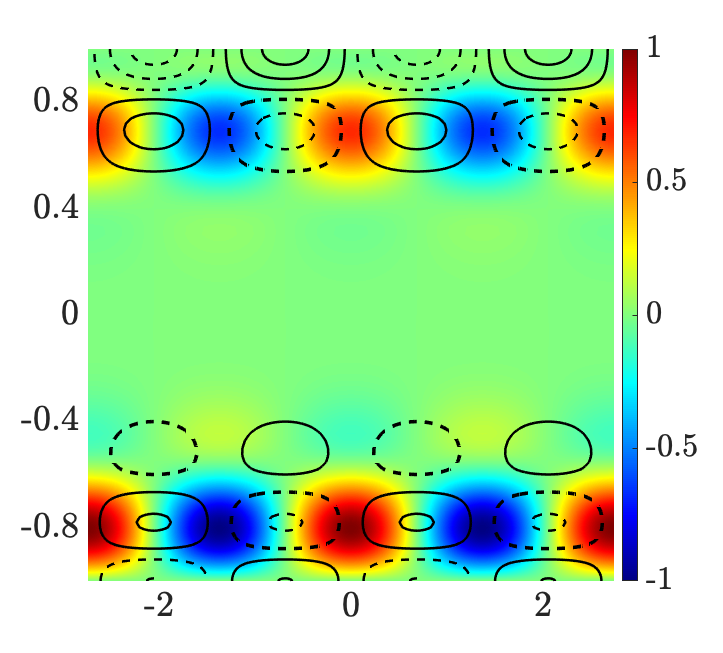}
	          \hspace{-.4cm}
        \\[-2.8cm]
        \hspace{-.4cm}
        \\[-.15cm]
        &
        \hspace{-.8cm}
        {\normalsize $z$} 
        \hspace{2.3cm}
        {\normalsize $x$}
        &&
        \normalsize {$x$}
	&&
        \normalsize {$z$}
        \end{tabular}
             \caption{{The streamwise component of the dominant flow structures of Couette flow with $R = 500$ and $\bk = (0.95, 2.29)$ in the absence (first rows) and presence (second rows) of stochastic base flow perturbations with {$f(y)=\bar{U}(y)/\max(|\bar{U}(y)|)$}, $\alpha=1$, and $\sigma^2_u=0.50$; (a) the principal modes; and (b) the second most energetic modes. The three columns correspond to: (left) the spatial structure of the eigenmodes with red and blue colors denoting regions of high and low velocity; (middle) the streamwise velocity at $z=0$; and (right) the $y-z$ slice of streamwise velocity (color plots) and vorticity (contour lines) at the streamwise location indicated by the dashed vertical lines in the middle panel.}}
        \label{fig.Flow-structures-Couette}
\end{figure}

\begin{figure}
	\begin{tabular}{cccccc}
	 \hspace{-.4cm}
	\subfigure[]{\label{fig.Poiseuille_Ubar_first}}
      &
        \\[-.5cm]
        \hspace{-.2cm}
        \begin{tabular}{c}
                \vspace{4.5cm}
                {\normalsize \rotatebox{90}{$y$}}
        \end{tabular}
        &
        \hspace{-.2cm}
	\includegraphics[width=5.4cm]{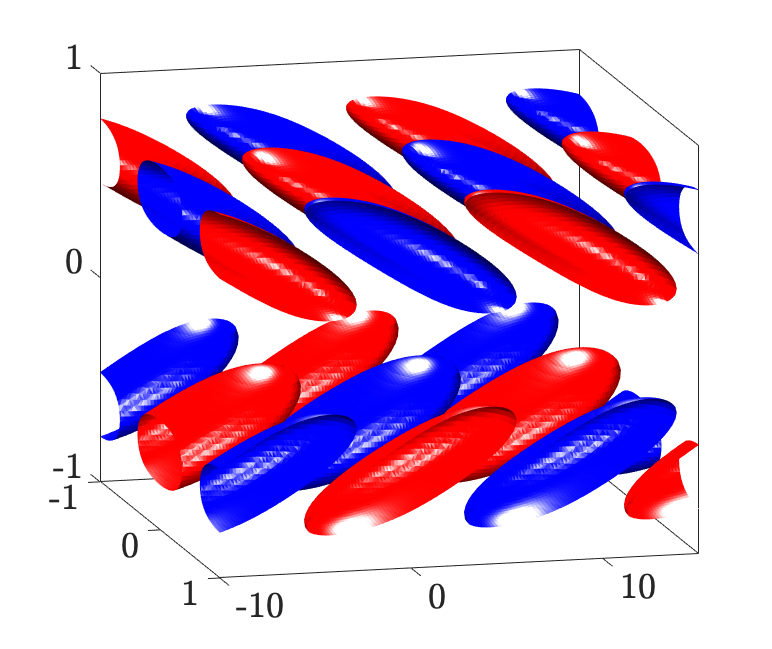}
        &&
	\includegraphics[width=5.4cm]{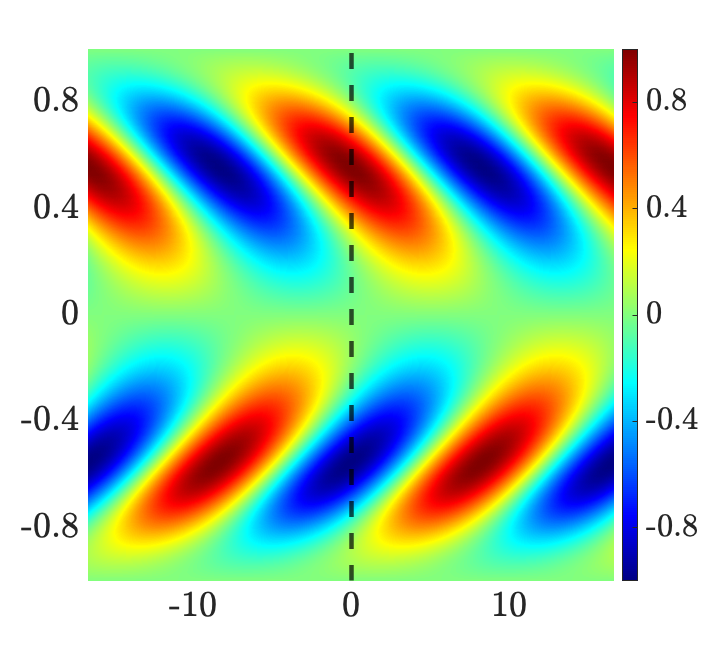}
        &&
	 \includegraphics[width=5.4cm]{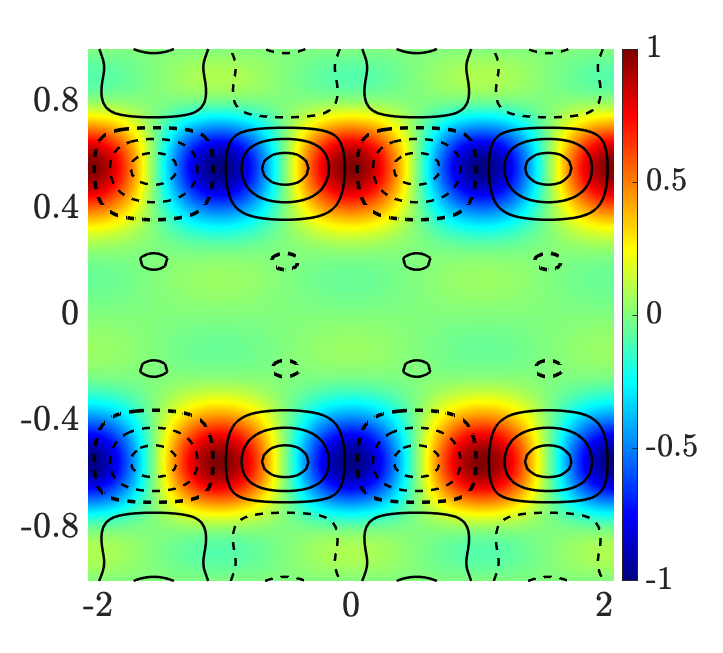}
        \\[-2.4cm]
      &
        \\[-.15cm]
        \hspace{-.2cm}
        \begin{tabular}{c}
                \vspace{4.5cm}
                {\normalsize \rotatebox{90}{$y$}}
        \end{tabular}
        &
        \hspace{-.2cm}
	\includegraphics[width=5.4cm]{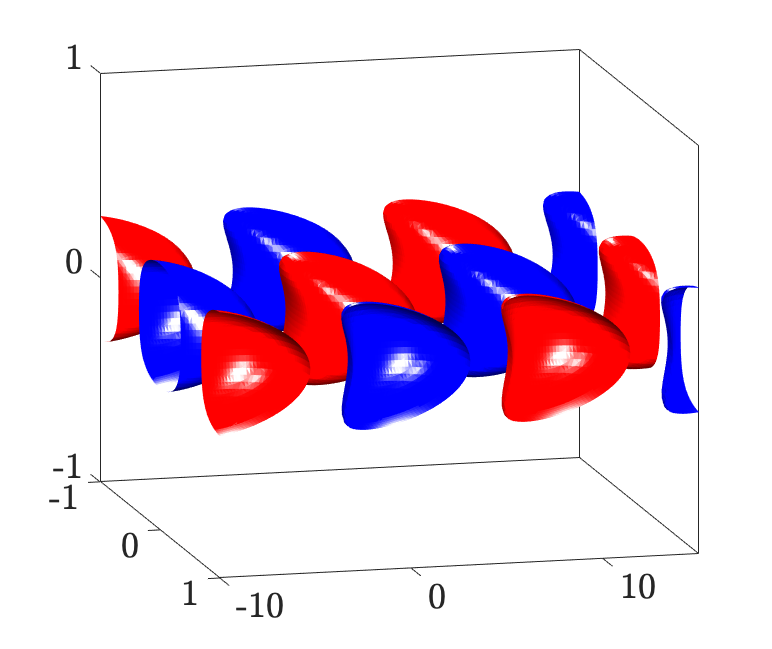}
        &&
	\includegraphics[width=5.4cm]{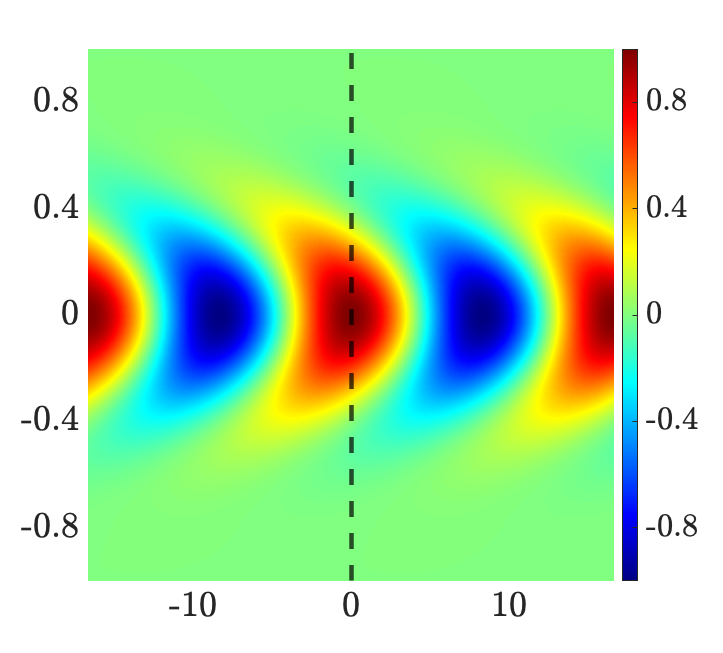}
        &&
	 \includegraphics[width=5.4cm]{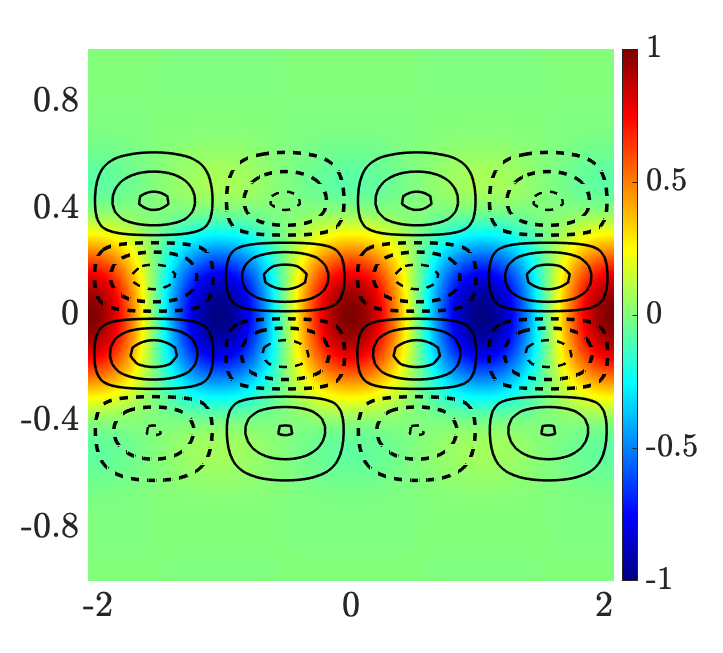}
        \\[-2.2cm]
         \hspace{-.4cm}
	\subfigure[]{\label{fig.Poiseuille_Ubar_second}}
      &
        \\[-.5cm]
        \hspace{-.2cm}
        \begin{tabular}{c}
                \vspace{4.5cm}
                {\normalsize \rotatebox{90}{$y$}}
        \end{tabular}
        &
        \hspace{-.2cm}
	\includegraphics[width=5.4cm]{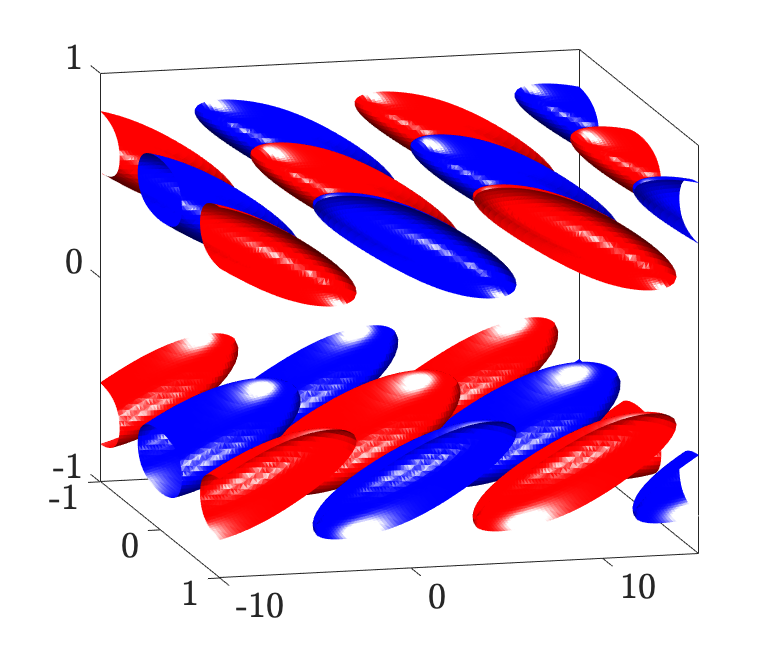}
        &&
	\includegraphics[width=5.4cm]{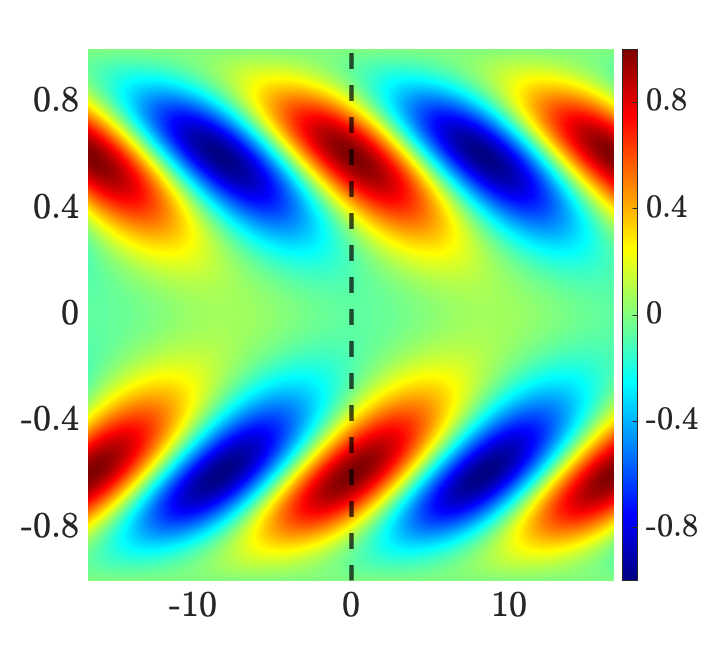}
        &&
	 \includegraphics[width=5.4cm]{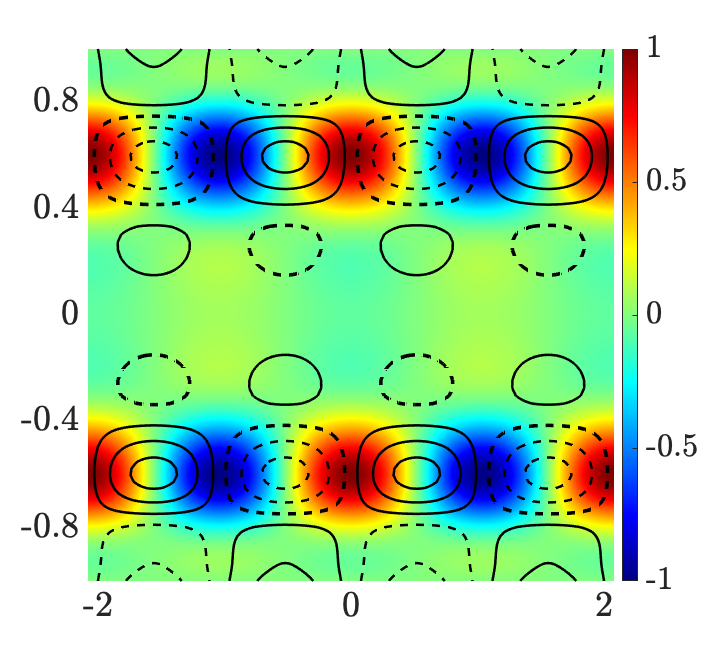}
	 \\[-2.4cm]
      &
        \\[-.15cm]
        \hspace{-.2cm}
        \begin{tabular}{c}
                \vspace{4.5cm}
                {\normalsize \rotatebox{90}{$y$}}
        \end{tabular}
        &
        \hspace{-.2cm}
	\includegraphics[width=5.4cm]{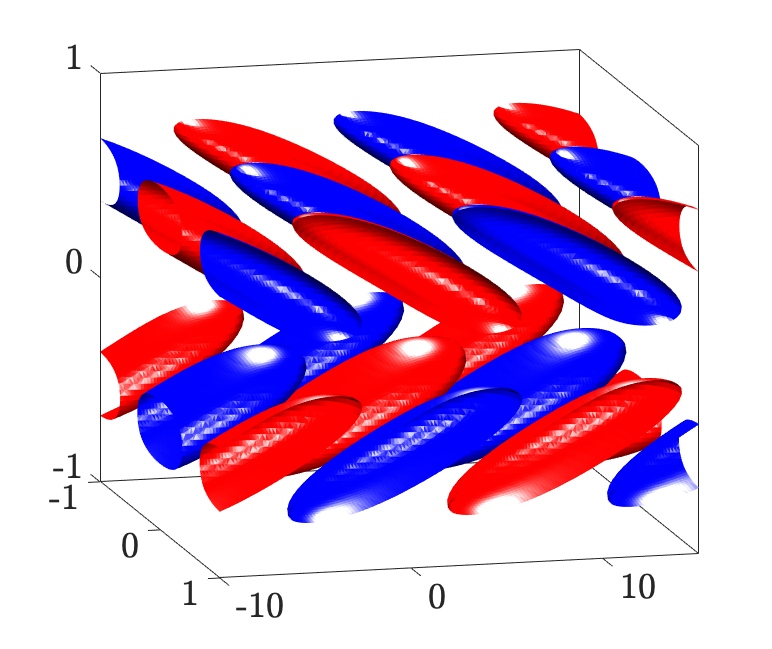}
        &&
	\includegraphics[width=5.4cm]{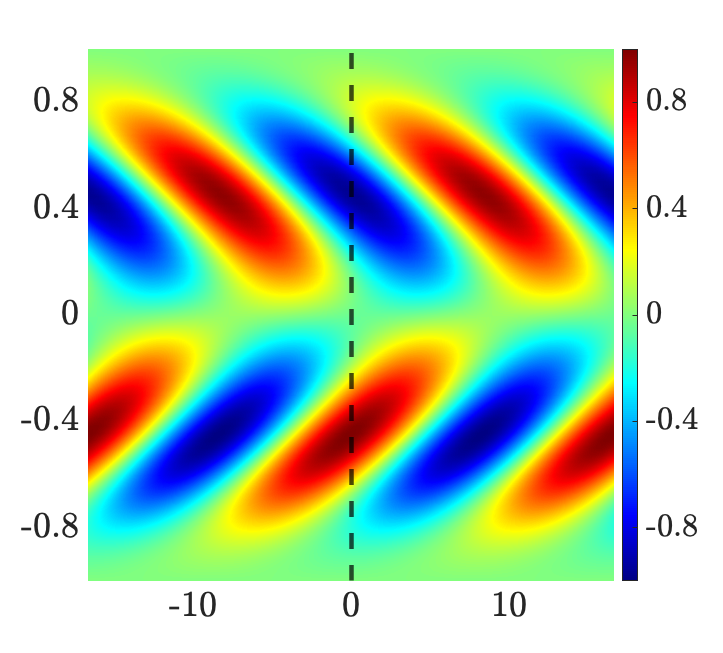}
        &&
	 \includegraphics[width=5.4cm]{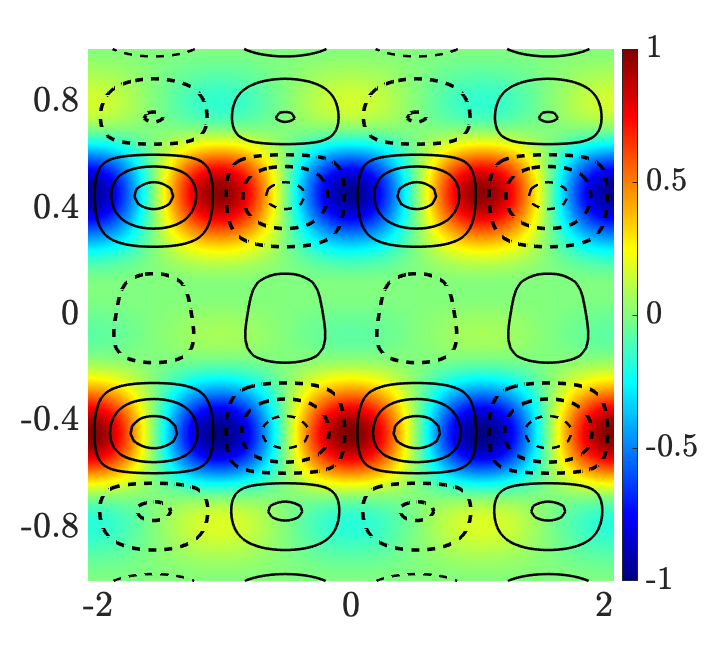}
	          \hspace{-.4cm}
        \\[-2.8cm]
        \hspace{-.4cm}
        \\[-.15cm]
        &
        \hspace{-.8cm}
        {\normalsize $z$} 
        \hspace{2.3cm}
        {\normalsize $x$}
        &&
        \normalsize {$x$}
	&&
        \normalsize {$z$}
        \end{tabular}
             \caption{{The streamwise component of the dominant flow structures of Poiseuille flow $R = 2000$ and $\bk = (0.38, 3.02)$ in the absence (first rows) and presence (second rows) of stochastic base flow perturbations with {$f(y)=\bar{U}(y)/\max(|\bar{U}(y)|)$}, $\alpha=1$, and $\sigma^2_u=0.21$; (a) the principal modes; and (b) the second most energetic modes. The three columns correspond to: (left) the spatial structure of the eigenmodes with red and blue colors denoting regions of high and low velocity; (middle) the streamwise velocity at $z=0$; and (right) the $y-z$ slice of streamwise velocity (color plots) and vorticity (contour lines) at the streamwise location indicated by the dashed vertical lines in the middle panel.}}
        \label{fig.Flow-structures-Poiseuille}
\end{figure}

As shown in the previous section, while streamwise elongated streaks represent the energetically dominant flow structures in the nominal flow, oblique modes become increasingly relevant as the amplitude and variance of base flow perturbations grow. To demonstrate the influence of base flow perturbations on the energy and spatial extent of dominant flow structures, we will focus on base flow perturbations with {$f(y)=\bar{U}(y)/\max(|\bar{U}(y)|)$} and the maximally affected oblique modes corresponding to the peaks in Figs.~\ref{fig.Energy_plots_laminar_alpha0p5}.
Figure~\ref{fig.principal-modes_laminar} shows the contribution of the first six eigenvalues of $\Phi(\bk)$ to the kinetic energy (sum of all eigenvalues) of Couette flow with $R=500$ and Poiseuille flow with $R = 2000$ at the wavenumber pairs that correspond to the maximum amplification in Figs.~\ref{fig.Ec_couette_withUbar} and~\ref{fig.Ec_poiseuille_withUbar}; $\bk=(0.95,2.29)$ in Couette flow and $\bk = (0.38,3.02)$ in Poiseuille flow. The base flow perturbations are chosen to be of unit amplitude ($\alpha=1$) with critically stable variance level identified in Sec.~\ref{sec.stability-laminar}. Based on Fig.~\ref{fig.eig_couette}, while streamwise base flow perturbations generally increase the energy of this oblique mode, they also concentrate the energy on the two most significant modes; the two most energetic modes in the perturbed flow contain $94\%$ of the total energy which is a significant increase relative to $75\%$ in the nominal flow. {Interestingly, in the presence of base flow perturbations that follow the Couette profile the two most significant eigenmodes will have the same energetic contributions which is also indicative of flow symmetries induced by such a multiplicative source of excitation (cf. Fig.~\ref{fig.Flow-structures-Couette}). This is in contrast to the role of base flow perturbations of Poiseuille flow that work to break structural symmetries (Fig.~\ref{fig.Flow-structures-Poiseuille}) by increasing the gap between the energy of the first and second most dominant eigenvalues (Fig.~\ref{fig.eig_poiseuille}); the perturbations of the Poiseuille flow profile} increase the energetic contribution of the principal eigenvalue from $48\%$ to $72\%$ and decrease the contribution of the second eigenvalue from $31\%$ to $16\%$. Nevertheless, the presence of base flow perturbations increases the energetic contribution of the first two modes by $9\%$ (from $79\%$ to $88\%$).

We next visualize the streamwise component of the two most significant modes identified in Fig.~\ref{fig.principal-modes_laminar} using Eq.~\eqref{eq.vel-components}. Figure~\ref{fig.Couette_Ubar_first} shows various views of the principal oblique structures corresponding to $\bk = (0.95,2.29)$ in nominal (first row) and perturbed (second row) Couette flow. While this wavenumber pair corresponds to nominally streamwise elongated mid-channel structures that are inclined to the walls, in the presence of base flow perturbations, it corresponds to near-wall streamwise elongated structures that are less inclined to the walls and exhibit an anti-symmetric arrangement with respect to the channel centerline. Similar to the energetically dominant streaks that are typically observed in such wall-bounded flows, the resulting oblique modes of the stochastically perturbed flow contain alternating regions of fast- and slow- moving fluid that are situated between counter-rotating vortical motion in the cross-stream plane (see third column of Fig.~\ref{fig.Couette_Ubar_first}). Figure~\ref{fig.Couette_Ubar_second} shows the spatial structure of the streamwise component of the second largest mode in Couette flow in the absence (first row) and presence (second row) of stochastic base flow perturbations. While under nominal conditions, the oblique mode corresponds to two rows of mid-channel streamwise elongated flow structures about the centerline, stochastic base flow perturbations give rise to near-wall flow structures that are similar to the principal modes (second row of Fig.~\ref{fig.Couette_Ubar_first}) but with a predominantly symmetric arrangement due to a phase shift. {As observed from Fig.~\ref{fig.eig_couette}, the equivalent energetic contribution of the two most significant eigenmodes is also reflective of symmetries in the structure of the corresponding flow structures shown in the second rows of Figs.~\ref{fig.Couette_Ubar_first} and~\ref{fig.Couette_Ubar_second}.}
A similar analysis of the flow structures corresponding to the oblique modes ($\bk = (0.38, 3.02)$) of Poiseuille flow is presented in Fig~\ref{fig.Flow-structures-Poiseuille}. Base flow perturbations cause the centerline conglomeration of the dominant flow structures that are nominally streamwise elongated and inclined to the wall (Fig.~\ref{fig.Poiseuille_Ubar_first}). On the other hand, apart from a slight wall-normal elevation, such perturbations do not influence the physical structure of the second most significant mode (Fig.~\ref{fig.Poiseuille_Ubar_second}). {Our observations suggest that the alignment of multiplicative uncertainties with the respective base flow profiles ($f(y) = \bar{U}(y)/\max(|\bar{U}(y)|)$) cause the predominant amplification of flow structures in regions that are most affected by such perturbations, i.e., in the vicinity of channel walls in Couette flow and in the middle of the channel in Poiseuille flow. We note that if $f(y)$ and $\bar{U}(y)$ were not aligned, e.g., $f(y)$ corresponding to Fig.~\ref{fig.fvec} in Poiseuille flow, the dominant flow structures would remain structurally unchanged albeit more energetic. These results are not reported here for brevity.}

\section{Effect of base flow variations on turbulent flow dynamics}
\label{sec.turbulent-flows}

{In this section, we examine the dynamics of stochastically forced turbulent channel flow in the presence of zero-mean white-in-time stochastic base flow perturbations. To this end, we augment the molecular viscosity in the NS equations~\eqref{eq.NS-eqn} with the turbulent viscosity $\nu_T$ of channel flow
\begin{align}
	\label{eq.NS-eqn}
	\ba{rcl}
		\tilde{\bu}_t 
		&\;=\;& 
		-\left( \tilde{\bu} \cdot \nabla \right) \tilde{\bu} \,-\, \nabla \tilde{P} \,+\,\dfrac{1}{R_\tau} \, \nabla \cdot\left((1\,+\,\nu_T)\left(\nabla \tilde{\bu} \,+\, (\nabla \tilde{\bu})^T\right)\right)
		\\[.15cm]
		0 
		&\;=\;& 
		\nabla \cdot \tilde{\bu}
	\ea
\end{align}
and linearize around the long-time averaged turbulent mean flow profile $\bu = [\,U(y)\,~0\,~0\,]^T$ provided by DNS of channel flow~\cite{deljim03,deljimzanmos04,hoyjim06,hoyjim08} (Fig.~\ref{fig.turbulent}) to obtain the linearized NS equations
\begin{align}
	\label{eq.dyn-fluc-turbulent-turb}
	\ba{rcl}
		\bv_t 
		&\;=\;& 
		-\left( \nabla \cdot \bu  \right) \bv \,-\, \left( \nabla \cdot \bv  \right) \bu \,-\, \nabla p \,+\,\dfrac{1}{R_\tau} \nabla \cdot \left((1\,+\,\nu_T)\left(\nabla v \,+\, (\nabla v)^T\right)\right)
		\\[.15cm]
		0 
		&\;=\;& 
		\nabla \cdot \bv
	\ea
\end{align}
which govern the dynamics of velocity, $\bv$, and pressure, $p$, fluctuations. Here, the Reynolds number $R_\tau = u_\tau h/\nu$ is defined in terms of the channel's half-height $h$ and the friction velocity $u_\tau=\sqrt{\tau_w/\rho}$, where $\tau_w$ is the wall-shear stress (averaged over horizontal directions and time), $\rho$ is fluid density, and $\nu$ is kinematic viscosity. For turbulent viscosity, we use the Reynolds and Tiederman~\cite{reytie67} turbulent viscosity profile
\begin{align}
	\label{eq.turbulent-viscosity}
    {\nu_{T}(y) 
    \;=\;
    \dfrac{1}{2} \left(\left( 1\,+\,\left(\,\dfrac{c_2}{3}\,R_\tau \,(\,1\,-\,y^2\,)(\,1\,+\,2\,y^2\,)(\,1-\mre^{-(1-|y|)R_\tau / c_1}\,)\right)^{2}\right)^{1/2} -\,1 \right)}
\end{align}
where parameters $c_1$ and $c_2$ are selected to minimize the least squares deviation between the steady-state solution to~\eqref{eq.dyn-fluc-turbulent-turb} using the averaged wall-shear stress $\tilde{P}_x=-1$ and the mean streamwise velocity obtained in experiments or simulations. Application of this least-squares procedure in finding the best fit to the mean velocity in turbulent channel flow resulting from DNS~\cite{deljim03,deljimzanmos04,hoyjim06,hoyjim08} yields $\{c_1 = 46.2, c_2 = 0.61\}$ at $R_\tau = 186$, $\{c_1 = 29.4, c_2 = 0.45\}$ at $R_\tau = 547$, $\{c_1 = 27, c_2 = 0.43\}$ at $R_\tau = 934$, and $\{c_1 = 25.4, c_2 = 0.42\}$ at higher Reynolds numbers. 
}

{
We assume the streamwise component of the base flow $\bu$ to be contaminated with an additive source of uncertainty $\gamma_u(y,t) = \alpha \bar{\gamma}(t) f(y)$. As a result, the dynamic operator $\bA$ in the state-space representation~\eqref{eq.lnse} takes the form
\begin{align}
	\label{eq.Aform-turbulent}
        \bA(\bk,t)
        &\;\DefinedAs\;
        \tbt{\bA_{11}}{0}{\bA_{21}}{\bA_{22}}
        \\[.25cm]
        \non
        \bA_{11}(\bk,t)
        &\;\DefinedAs\;
        \Delta^{-1}\!\Big(\dfrac{1}{R_\tau}\left( (1\,+\,\nu_T)\, \Delta^2 \,+\, 2\, \nu'_T \,\Delta \partial_y \,+\, \nu''_T (\,\partial^2_y\,+\,k^2\,) \right) +\,  
        \mri k_x \left(\bar{U}''(y) + \gamma_u''(y,t) \,-\, (\bar{U}(y) + \gamma_u(y,t))\,\Delta \right) \Big)
         \\[.15cm]
         \non
        \bA_{21}(\bk,t)
        &\;\DefinedAs\;
        -\mri k_z \left(\bar{U}'(y) + \gamma'_u(y,t)\right)
         \\[.15cm]
         \non
        \bA_{22}(\bk,t)
        &\;\DefinedAs\
        \dfrac{1}{R_\tau}\,\Big( (1\,+\,\nu_T)\Delta \,+\, \nu'_T \partial_y \Big)\,-\, \mri k_x\, \left(\bar{U}(y) +\gamma_u(y,t) \right)
\end{align}
where $\bar{U}(y)$ is the streamwise component of the nominal base flow profile $\bar{\bu}(y)$. In a similar manner as Eq.~\eqref{eq.A-decomp-original}, the operator-valued matrix $\bA$ can be decomposed into its nominal and perturbed components, i.e.,
\begin{align}
	\label{eq.A-decomp-turbulent}
    	\bA(\bk,t)
    	\;=\;
	    \bar{\bA}(\bk) \;+\; \alpha \,\bar{\gamma}_u(t)\,\bA_u(\bk)
\end{align}
where expressions for $\bar{\bA}$ and $\bA_u$ are given in Appendix~\ref{app.A0-Au-Aw-turb}.
}

{
We next discretize the differential operators in the linearized equations using $N = 151$ Chebyshev collocation points in the wall-normal direction and study the MSS and frequency response of the flow fluctuations in the presence of both additive stochastic forcing ${\bf f}$ and stochastic base flow perturbations $\gamma_u$. We will assume that perturbations $\gamma_u$ enter the dynamics through the same shape functions considered in the prior section.}

\subsection{{Stability analysis}}
\label{sec.stability-turb}

\begin{figure}
                \begin{tabular}{cccc}
      \subfigure[]{\label{fig.MSS-turb}}
        &&
        \hspace{.4cm}
       \subfigure[]{\label{fig.sigma2_critical_turbulent_Ubar}}
        &
        \\[-.6cm]
        \hspace{.2cm}
        \begin{tabular}{c}
                \vspace{.1cm}
                {\normalsize \rotatebox{90}{$\sigma^2_u$}}
        \end{tabular}
        &
        \begin{tabular}{c}
              \includegraphics[width=6.8cm]{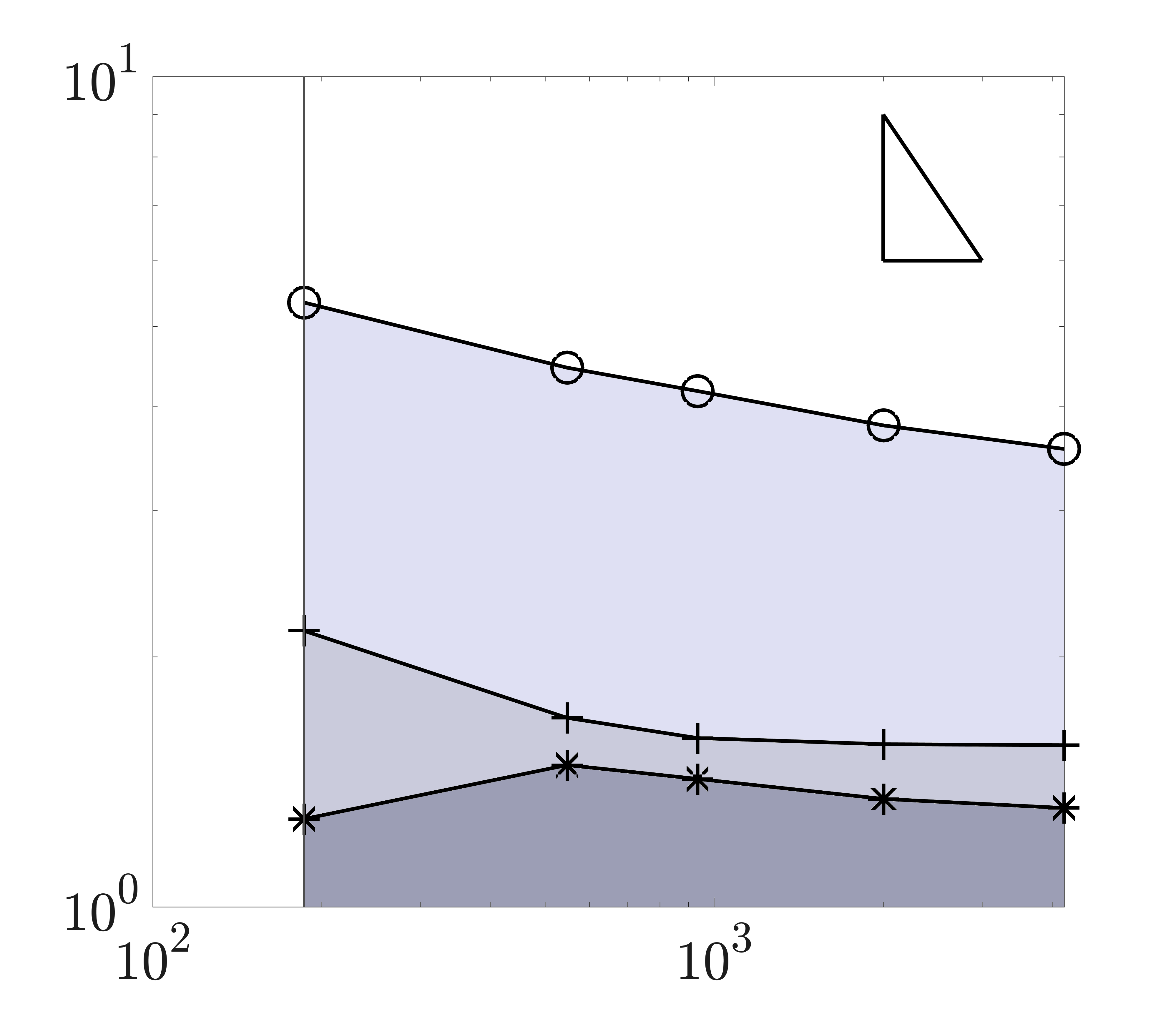}
        \end{tabular}
        &
        \hspace{.8cm}
        \begin{tabular}{c}
        		\vspace{.4cm}
        		{\normalsize \rotatebox{90}{$k_x$}}
        \end{tabular}
        &
        \begin{tabular}{c}
             \includegraphics[width=7.0cm]{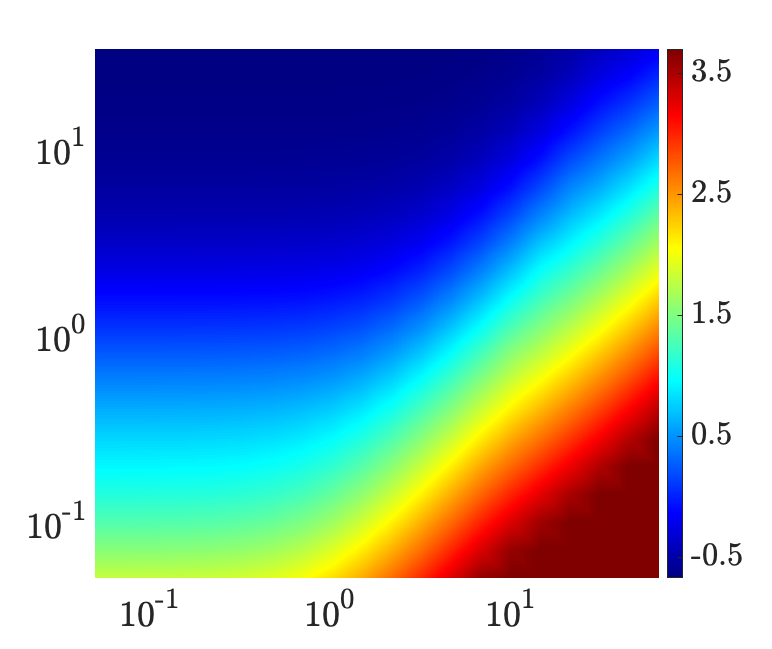}
        \end{tabular}
        \\[.2cm]
        &
        {\normalsize $R_\tau$}
        &&
        {\normalsize $k_z$}
        \end{tabular}
             \caption{{(a) Stability curves for fluctuation dynamics with $\bk = (2.5, 7)$ in turbulent channel flow subject to stochastic base flow perturbations following {shape functions $f(y)$ corresponding to Eq.~\eqref{eq.fshape-Ubar} ($+$) and those} shown in Figs.~\ref{fig.fshape1} ({\large$*$}) and~\ref{fig.fvec} ({\large$\circ$}). The shaded areas under the curves denote the variances of stochastic base flow uncertainty that do not violate MSS ($\rho(\mathbb{L})<1$ with $\alpha = 1$). The triangle in the upper right corner demonstrates an $R_\tau^{-1}$ slope. (b) Logarithmically scaled critical variance levels of stochastic multiplicative uncertainty $\bar{\gamma}_u$ with $\alpha=1$ and {$f(y)=\bar{U}(y)/\max(|\bar{U}(y)|)$} over the horizontal wavenumber spectrum of turbulent channel flow with $R_\tau=186$.}}
      \label{fig.stability-turbulent}
\end{figure}

For $\bk = (2.5,7)$, we analyze the MSS of the linearized NS equations around the DNS-generated mean velocity profile of turbulent channel flow $\bar{U}(y)$ at $R_\tau = 186$, $547$, $934$, $2003$, and $4179$~\cite{deljim03,deljimzanmos04,hoyjim06,hoyjim08}. While we focus on $\bk = (2.5,7)$, which is the horizontal wavenumber pair at which the premultiplied energy spectrum of channel flow at $R_\tau = 186$ peaks, similar stability trends were observed at other wavenumbers. The stability curves shown in Fig.~\ref{fig.MSS-turb} demonstrate the Reynolds number dependence of the critical variance $\sigma_u^2$ of stochastic base flow perturbation $\gamma_u(y,t)$. {As expected, the fragility of this mode to base flow perturbations increases as the Reynolds numbers grows apart from an initial increase observed in the case of full channel perturbations ($f(y)$ corresponding to Fig.~\ref{fig.fshape1}) from $R_\tau = 186$ to $547$. {The critical variance is found to approximately scale as $R_\tau^{-0.2}$ for the two extreme cases of shape functions in Fig.~\ref{fig.shape functions} and to approach $R_\tau$-invariance when perturbations follow the mean velocity profile (Eq.~\eqref{eq.fshape-Ubar}). As expected, the linearized dynamics demonstrate a higher tolerance} for stochastic perturbations entering in the near-wall regions, i.e., when $f(y)$ is given by Fig.~\ref{fig.fvec}. Similar to the findings of Sec.~\ref{sec.stability-laminar}, the critical variance of turbulent channel flow at any given Reynolds number $R_\tau$ is lowest for length-scales that are short in the streamwise dimension, but infinitely wide in the spanwise dimension (Fig.~\ref{fig.sigma2_critical_turbulent_Ubar}).}

The maximum tolerable variance of stochastic multiplicative uncertainty can provide guidelines for the number and quality of DNS-generated samples that should be involved in the statistical averaging process that leads to a stable equilibrium for linearized analysis. We note that for all Reynolds numbers studied here, and at all wall-normal locations, the reported variance of the numerically generated turbulent mean velocities is significantly lower than the critical values identified by our MSS analysis. Nevertheless, one implication of the uncovered Reynolds number dependence observed in Fig.~\ref{fig.MSS-turb} for statistical averaging is that the admissible variance in estimating the turbulent mean velocity $\bar{U}(y)$ reduces {at a much lower rate than $R_\tau^{-1}$. Furthermore, it is evident from Fig.~\ref{fig.MSS-turb} that the dynamics of turbulent fluctuations are generally more robust to base flow perturbations relative to that of laminar or transitional flow fluctuations.}

\subsection{{Energy spectrum of velocity fluctuations}}
\label{sec.energy-plots-turb}

{
We now analyze the effect of base flow perturbations on the energy spectrum of velocity fluctuations. We guarantee MSS by adjusting the variance of base flow perturbations to the maximum tolerable variance across all wavenumber pairs. For example, when {$f(y)=\bar{U}(y)/\max(|\bar{U}(y)|)$}, this critical variance corresponds to the shortest and widest length scales, and is identified as $0.13$; see Fig.~\ref{fig.sigma2_critical_turbulent_Ubar}. We note that the overall trend observed in Fig.~\ref{fig.sigma2_critical_turbulent_Ubar} along with the critical variance of the most sensitive mode is invariant to variations in the wall-normal extent of base flow perturbations dictated by the shape function $f(y)$. The steady-state covariance of velocity fluctuations in Eqs.~\eqref{eq.dyn-fluc-turbulent-turb} can be computed from solving Eq.~\eqref{eq.gen-lyap-original}. Following~\cite{moajovJFM12}, we select the covariance of white-in-time forcing to guarantee equivalence between the two-dimensional energy spectrum of turbulent channel flow and the flow obtained by the linearized NS equations in the absence of base flow perturbations ($\gamma_u=0$). This is achieved via the scaling
\begin{align*}
        \Omega(\bk)    
    	&\;=\;
	    \dfrac{\bar{E}(\bk)}{\bar{E_0}(\bk)}\,\Omega_0(\bk)
\end{align*}
where $\bar{E}(\bk)=\int_{-1}^{1} E(y,\bk) \,\mrd y$ is the two-dimensional energy spectrum of a turbulent channel flow obtained using the DNS-based energy spectrum $E(y,\bk)$~\cite{deljim03,deljimzanmos04}, and $\bar{E}_0(\bk)$ is the energy spectrum resulting from the linearized NS equations in the absence of base flow perturbations and subject to a white-in-time forcing ${\bf f}$ with covariance
\begin{align*}
    \Omega_0(\bk)
    \;=\;
	\tbt{\sqrt{E(y,\bk)}\,I}{0}{0}{\sqrt{E(y,\bk)}\,I}\,{\tbt{\sqrt{E(y,\bk)}\,I}{0}{0}{\sqrt{E(y,\bk)}\,I}}^*.
\end{align*}
}

\begin{figure}
\centering
	\begin{tabular}{cc}
	\subfigure[]{\label{fig.Enom_pre_turb_DNS}}
	&
	\\[-.6cm]
        \begin{tabular}{c}
        		\vspace{.1cm}
        		{\normalsize {$k_x$}}
        \end{tabular}
        &
        \hspace{-.2cm}
        \begin{tabular}{c}
                \includegraphics[width=6.0cm]{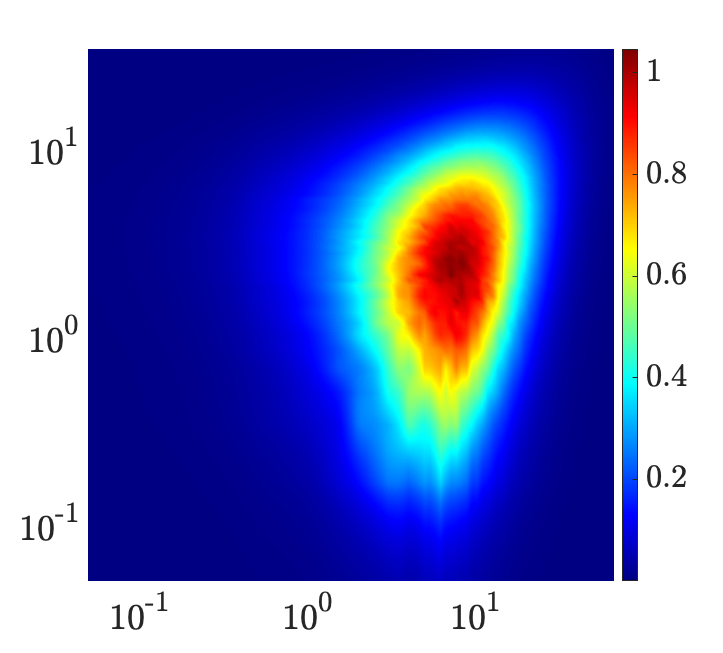}
	\end{tabular}
        \\[.2cm]
        &
        \hspace{.3cm}
        {\normalsize $k_z$}
        \end{tabular}
        \\
        \begin{tabular}{cccccc}
        \hspace{-.6cm}
      \subfigure[]{\label{fig.E2pre_turb_Ubar_alpha0p05}}
        &&
        \hspace{-.3cm}
       \subfigure[]{\label{fig.E2pre_turb_fshape1_alpha0p05}}
        &&
        \hspace{-.3cm}
       \subfigure[]{\label{fig.E2pre_turb_fvec_alpha0p05}}
        &
        \\[-.6cm]
        \begin{tabular}{c}
                \vspace{4.5cm}
                {\normalsize \rotatebox{90}{$k_x$}}
        \end{tabular}
        &
        \hspace{.2cm}
              \includegraphics[width=5cm]{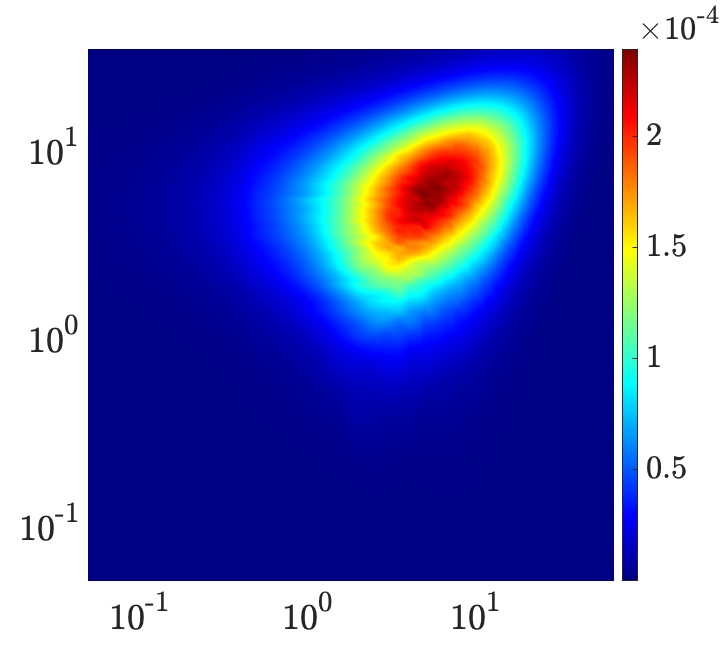}
        &&
        \hspace{.2cm}
              \includegraphics[width=5cm]{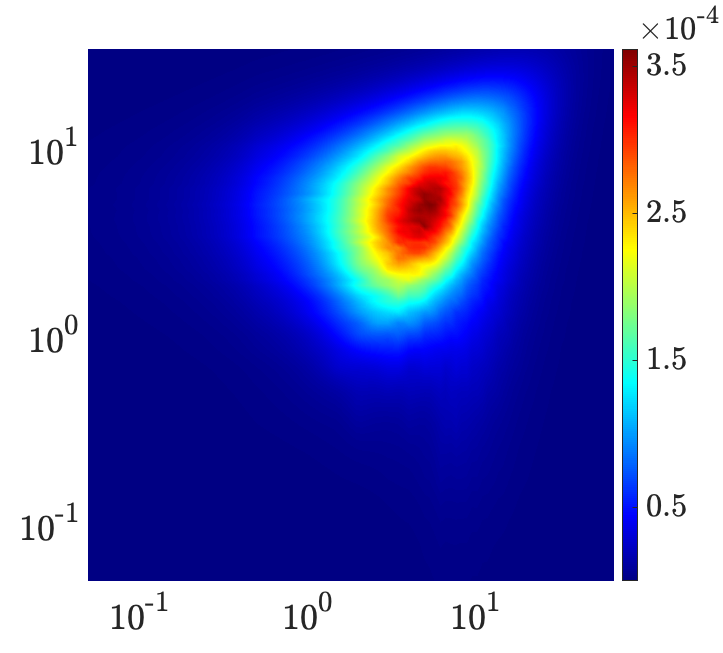}
        &&
        \hspace{.2cm}
             \includegraphics[width=5cm]{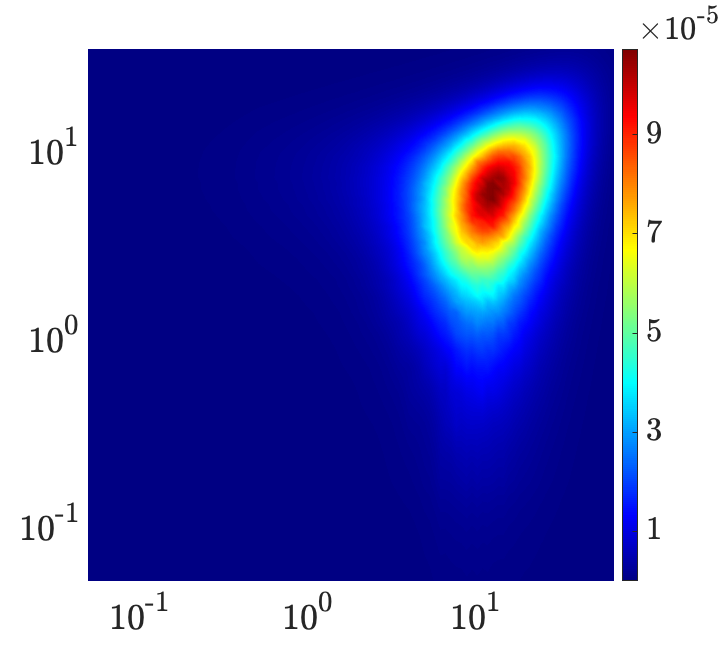}
        \end{tabular}
    \\[-2.5cm]
     \begin{tabular}{cccccc}
     \hspace{-.6cm}
      \subfigure[]{\label{fig.Ec_pre_turb_Ubar_alpha0p9}}
        &&
        \hspace{-.3cm}
      \subfigure[]{\label{fig.Ec_pre_turb_fshape1_alpha0p9}}
        &&
        \hspace{-.3cm}
      \subfigure[]{\label{fig.Ec_pre_turb_fvec_alpha0p9}}
        &
        \\[-.6cm]
        \begin{tabular}{c}
                \vspace{4.5cm}
                {\normalsize \rotatebox{90}{$k_x$}}
        \end{tabular}
        &
        \hspace{.2cm}
              \includegraphics[width=5cm]{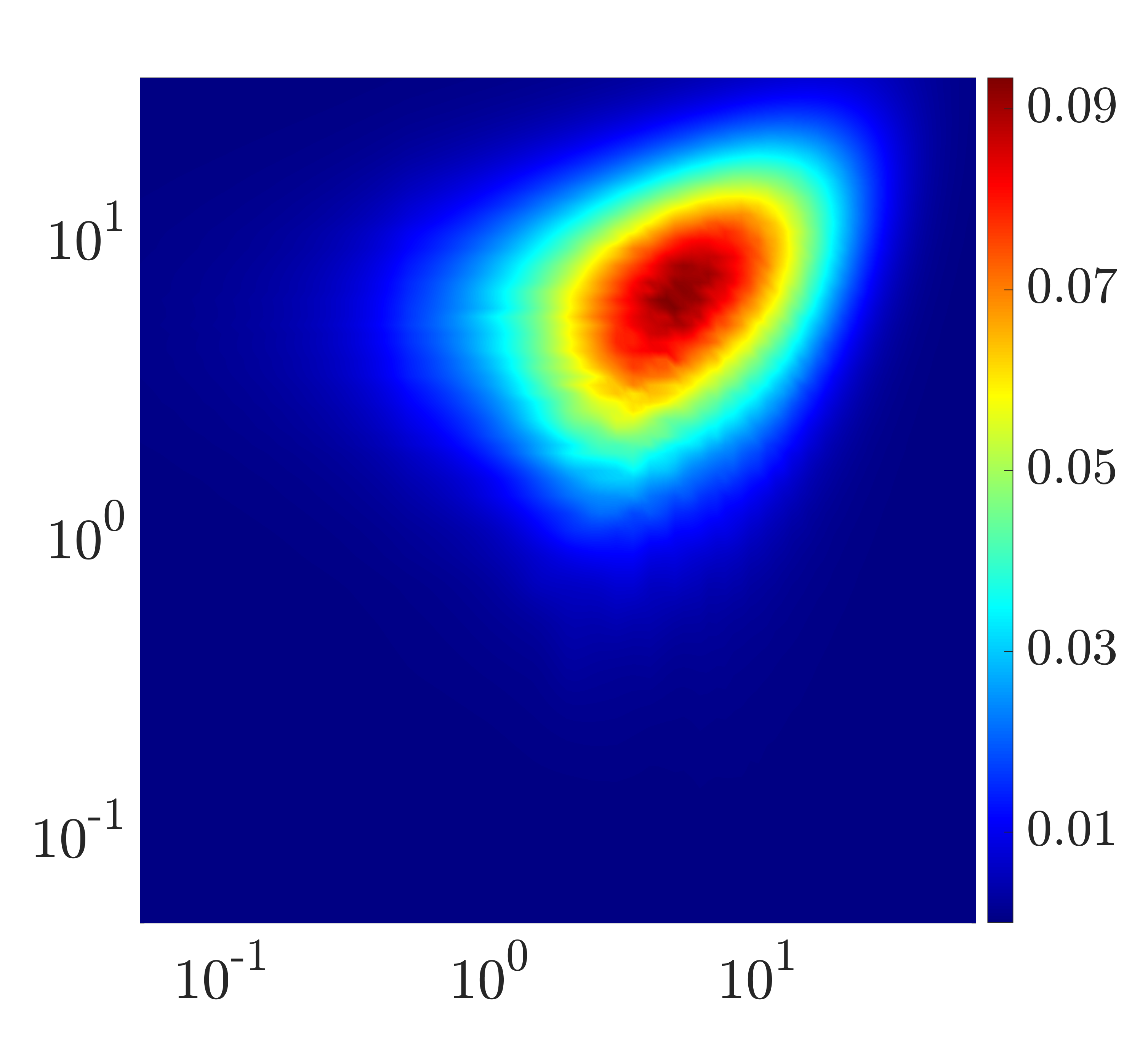}
        &&
        \hspace{.2cm}
              \includegraphics[width=5cm]{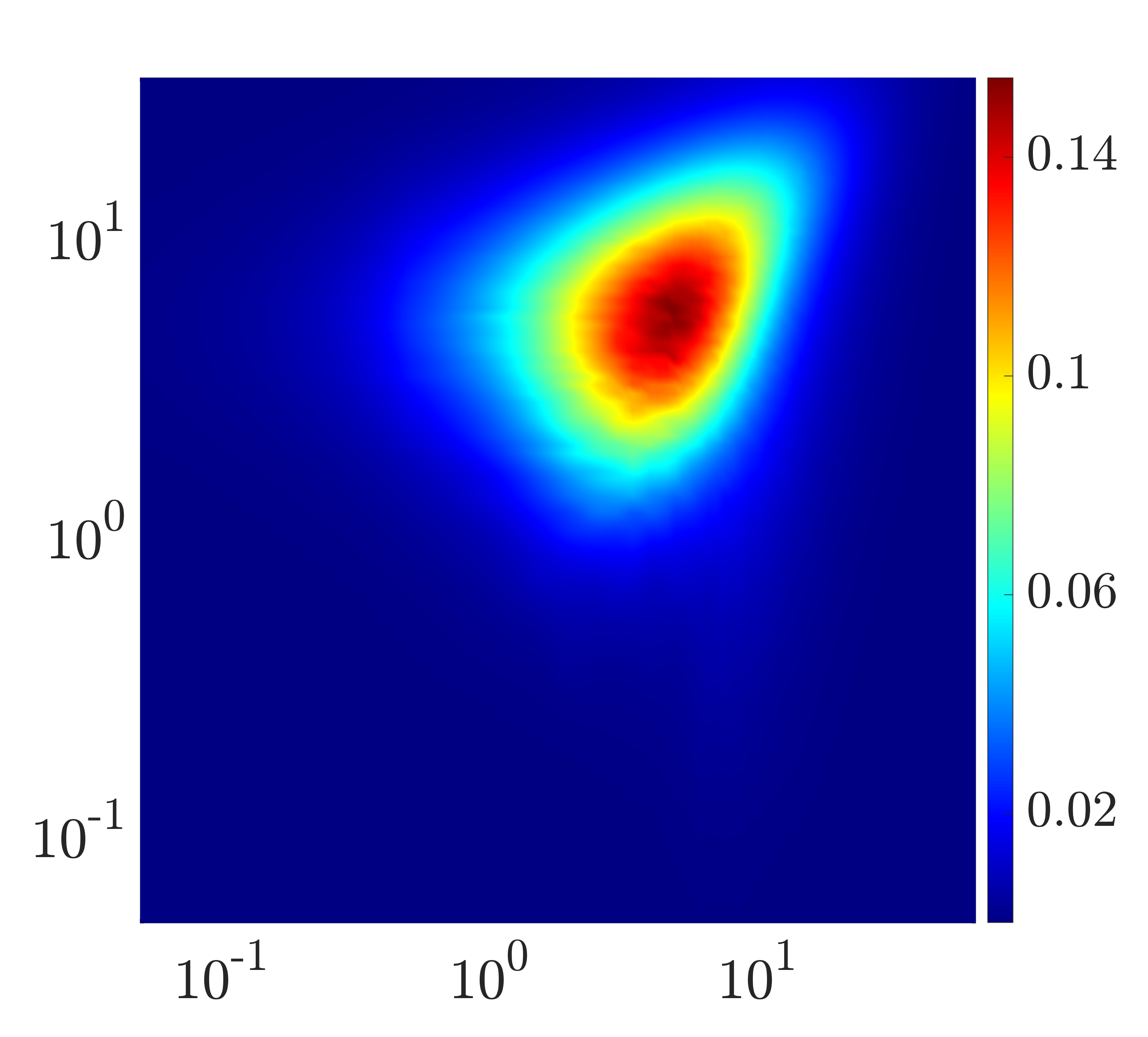}
        &&
        \hspace{.2cm}
             \includegraphics[width=5cm]{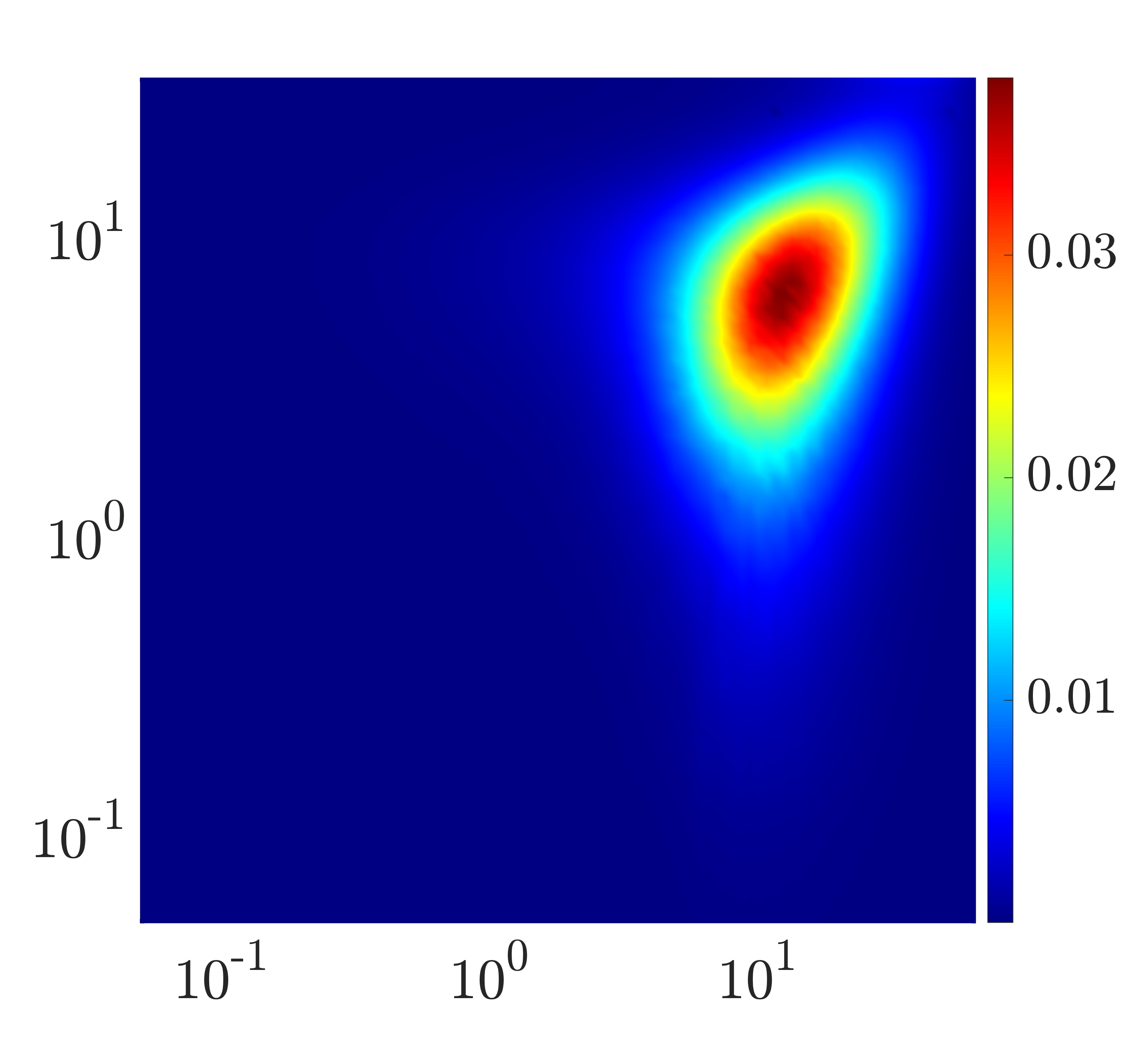}
        \\[-2.2cm]
        &
        {\normalsize $k_z$}
        &&
        {\normalsize $k_z$}
        &&
        {\normalsize $k_z$}
        \end{tabular}
                \caption{{(a) Premultiplied energy spectrum  of turbulent channel flow with $R_\tau=186$ in the absence of stochastic base flow perturbations ($E_0(\bk)$). {(b - g) Corrections} to the premultiplied energy spectra $E_c(\bk)$ of turbulent channel flow $R_\tau=186$ due to stochastic multiplicative uncertainty $\gamma_u$ with variance $\sigma^2_u = 0.13$ and perturbation amplitudes $\alpha = 0.05$ (second row) and $\alpha = 0.9$ (third row) that follow perturbations shapes {$f(y)$ corresponding to Eq.~\eqref{eq.fshape-Ubar} (b, e), Fig.~\ref{fig.fshape1} (c, f), and Fig.~\ref{fig.fvec} (d, g)}.}}
        \label{fig.Turbulent-energy-plots}
\end{figure}

\begin{figure}
\centering
	\begin{tabular}{cc}
        \begin{tabular}{c}
        		\vspace{.1cm}
        		{\normalsize \rotatebox{90}{$\int_{\bk} E_c(\bk) \,d\bk / \int_{\bk} E_0(\bk) \,d\bk$}}
        \end{tabular}
        &
        \begin{tabular}{c}
                \includegraphics[width=6.5cm]{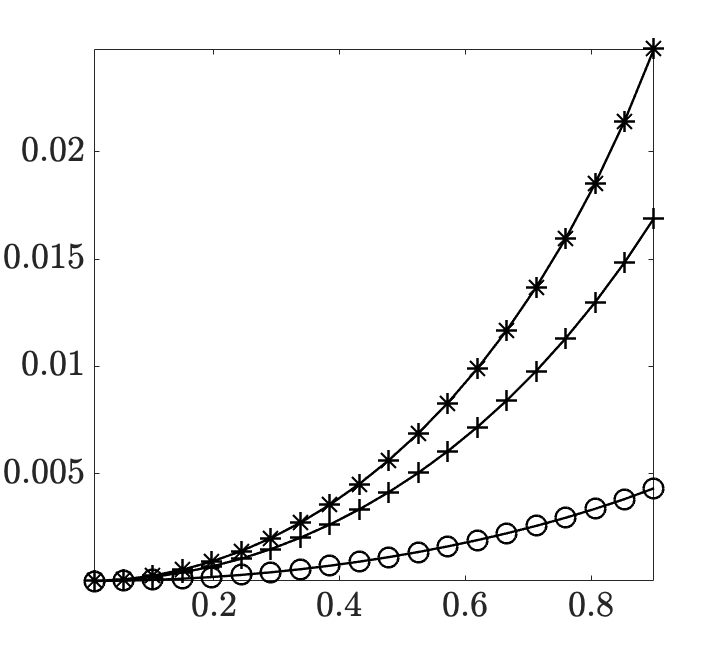}
	\end{tabular}
        \\[-.1cm]
        &
        \hspace{.3cm}
        {\normalsize $\alpha$}
        \end{tabular}
                \caption{The total effect of stochastic perturbations of amplitude $\alpha$ on the energy spectrum of turbulent channel flow with $R_\tau=186$. The curves demonstrate the $\alpha$ dependence of the energy correction due to base flow perturbations entering the dynamics through {the shape functions $f(y)$ corresponding to Eq.~\eqref{eq.fshape-Ubar} (+), in addition to those} depicted in Figs.~\ref{fig.fshape1} ({\large$*$}) and~\ref{fig.fvec} ({\large$\circ$}).}
        \label{fig.Ec-vs-alpha-turb}
\end{figure}

{
Figure~\ref{fig.Enom_pre_turb_DNS} shows the premultiplied energy spectrum of turbulent channel flow with $R_\tau=186$ in the absence of stochastic base flow perturbations in the linearized dynamics ($\gamma_u(t) = 0$). The changes to the premultiplied energy spectrum $k_x k_z E_c(\bk)$ due to stochastic multiplicative uncertainties with $\alpha=0.05$ entering through various wall-normal regions are shown in Figs.~\ref{fig.Turbulent-energy-plots}(b)-(d). Since the amplitude of base flow perturbations is small, the second-order correction to the perturbation series of energy provides a sufficient approximation of the change to the energy spectrum, i.e., $E_c(\bk) = \alpha^2 E_2$. Figures~\ref{fig.Turbulent-energy-plots}(e)-(g) consider the case of higher-amplitude base flow perturbations ($\alpha=0.9$) on the premultiplied energy spectrum by $k_x k_z E_c(\bk)$, where $E_c(\bk)$ is given in Eq.~\eqref{eq.Ec}. In computing $E_c(\bk)$, the limit was obtained using an $8$th-order perturbation series, i.e., $Ec = \alpha^2 E_2 + \alpha^4 E_4 + \alpha^6 E_6 + \alpha^8 E_8$, and verified using the Shanks transformation~\cite{sha55,van64}. As evident from the second and third rows of Fig.~\ref{fig.Turbulent-energy-plots}, the influence of base flow perturbations is concentrated at an energetically relevant region of the energy spectrum with a maximum at streamwise and spanwise wavenumbers that are slightly higher than those corresponding to the peak of the nominal energy spectrum (Fig.~\ref{fig.Enom_pre_turb_DNS}). Similar to the results presented in the previous subsection, stochastic base flow perturbation cannot influence streamwise streaks, which is because of the structure of $A_u(\bk)$ at $k_x=0$; see Appendix~\ref{app.A0-Au-Aw-turb}. Finally, as shown in Fig.~\ref{fig.Ec-vs-alpha-turb}, the total effect of stochastic base flow uncertainty of various amplitude, which can be quantified as $\int_{\bk} E_c(\bk) \,d\bk / \int_{\bk} E_0(\bk) \,d\bk$, follows a similar trend to what was observed for laminar flows (cf.~Fig.~\ref{fig.Ec_vs_alpha_laminar}).}

\subsection{{Maximally affected flow structures}}
\label{sec.principal flow structures-turbulent}

\begin{figure}
\centering
	\begin{tabular}{cc}
        \begin{tabular}{c}
        		\vspace{.1cm}
        		 {\normalsize \rotatebox{90}{$\lambda_j/\Sigma_i \lambda_i$}}
        \end{tabular}
        &
        \hspace{-.2cm}
        \begin{tabular}{c}
                \includegraphics[width=6.5cm]{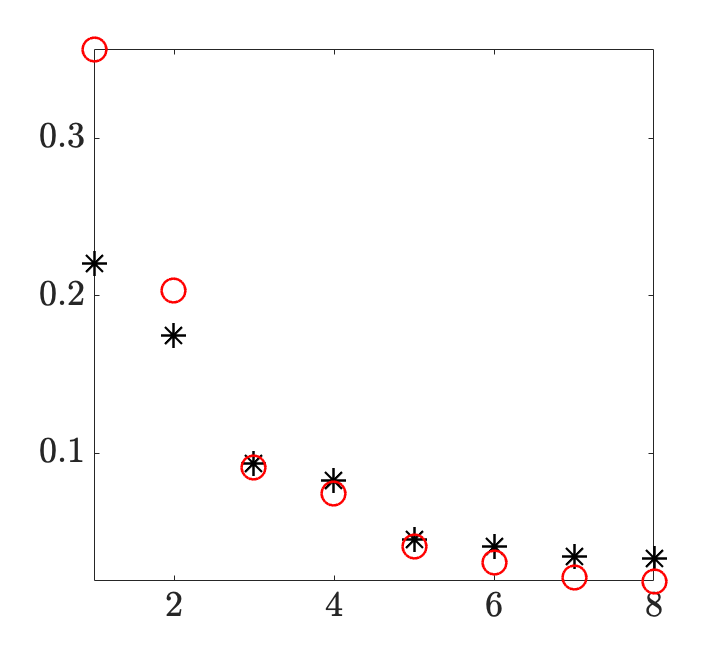}
	\end{tabular}
        \\[-.1cm]
        &
        \hspace{.3cm}
        {\normalsize $j$}
        \end{tabular}
                \caption{{Contribution of the first eight eigenvalues of the velocity covariance matrix $\Phi$ of channel flow in the absence (\tc{black}{\large $*$}), and presence (\tc{red}{\large $\circ$}) of base flow perturbations with {$f(y)=\bar{U}(y)/\max(|\bar{U}(y)|)$} and amplitude $\alpha = 1$ in turbulent channel flow with $R_\tau=186$ at $(k_x,k_z)=(1.86,1.94)$.}}
        \label{fig.eig-contribution-turbulent}
\end{figure}

\begin{figure}
\begin{tabular}{cccccc}
      \subfigure[]{\label{fig.turb_struct_nom}}
        \\[-.3cm]
        \hspace{-.2cm}
        \begin{tabular}{c}
                \vspace{.35cm}
                {\normalsize \rotatebox{90}{$y$}}
        \end{tabular}
        &
        \hspace{-.2cm}
        \begin{tabular}{c}
              \includegraphics[width=5cm]{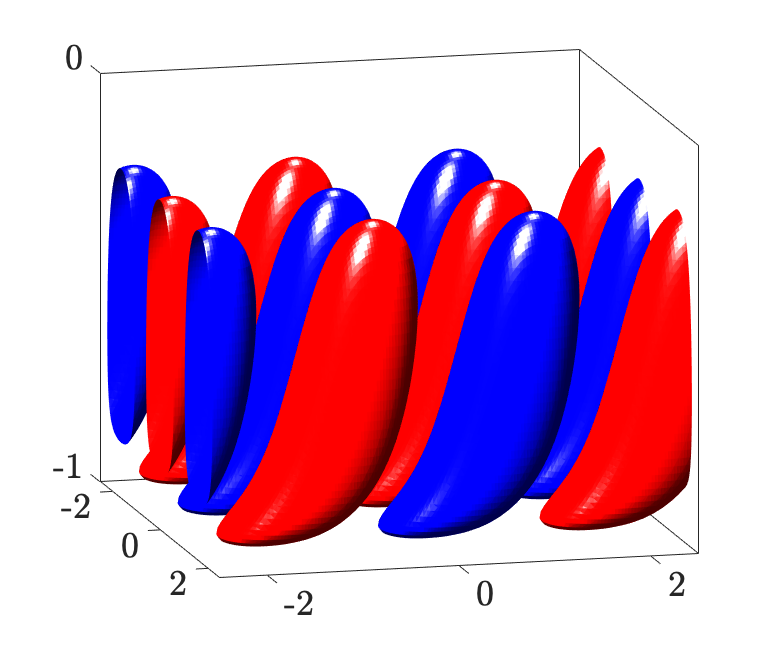}
        \end{tabular}
        &
          \hspace{.2cm}
        \begin{tabular}{c}
                \vspace{.1cm}
                {\normalsize \rotatebox{90}{$y$}}
        \end{tabular}
        &
        \begin{tabular}{c}
              \includegraphics[width=5cm]{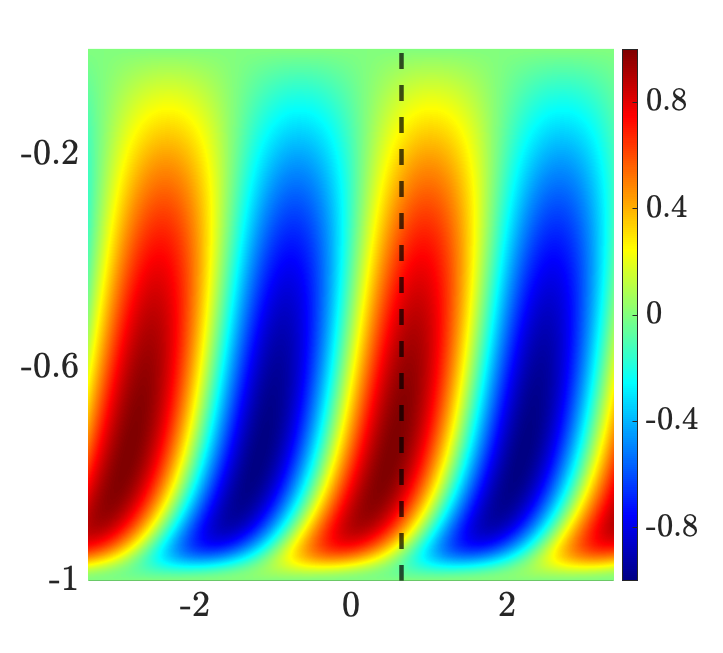}
        \end{tabular}
        &
        \hspace{-.2cm}
        \begin{tabular}{c}
        		\vspace{.4cm}
        		{\normalsize \rotatebox{90}{$y$}}
        \end{tabular}
        &
        \begin{tabular}{c}
             \includegraphics[width=5cm]{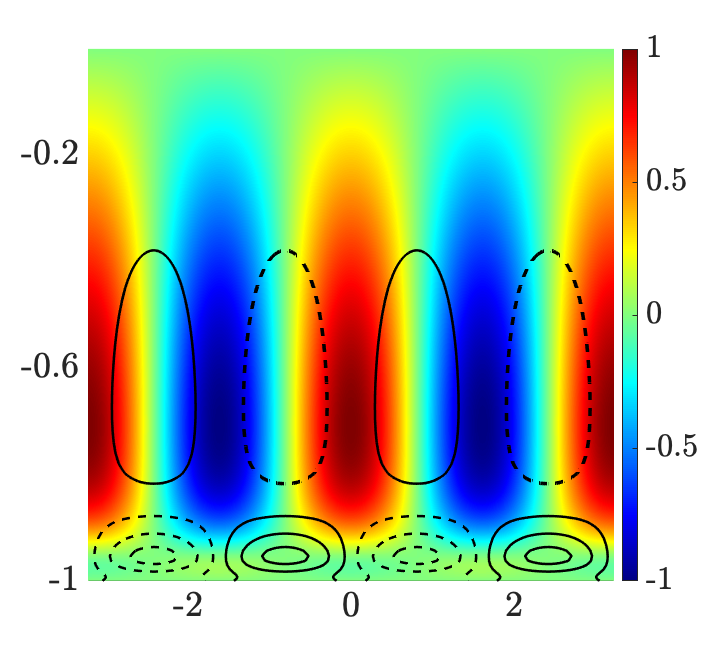}
        \end{tabular}
        \\[-.3cm]
        &
        \hspace{-1.4cm}
        {\normalsize $z$} 
        \hspace{2.3cm}
        {\normalsize $x$}
        &&
        \hspace{-.2cm}
        {\normalsize $z$}
        &&
        \hspace{-.2cm}
        {\normalsize $x$}
        \\
      \subfigure[]{\label{fig.turb_struct_pert}}
        \\[-.3cm]
        \hspace{-.2cm}
        \begin{tabular}{c}
                \vspace{.1cm}
                {\normalsize \rotatebox{90}{$y$}}
        \end{tabular}
        &
        \hspace{-.2cm}
        \begin{tabular}{c}
              \includegraphics[width=5cm]{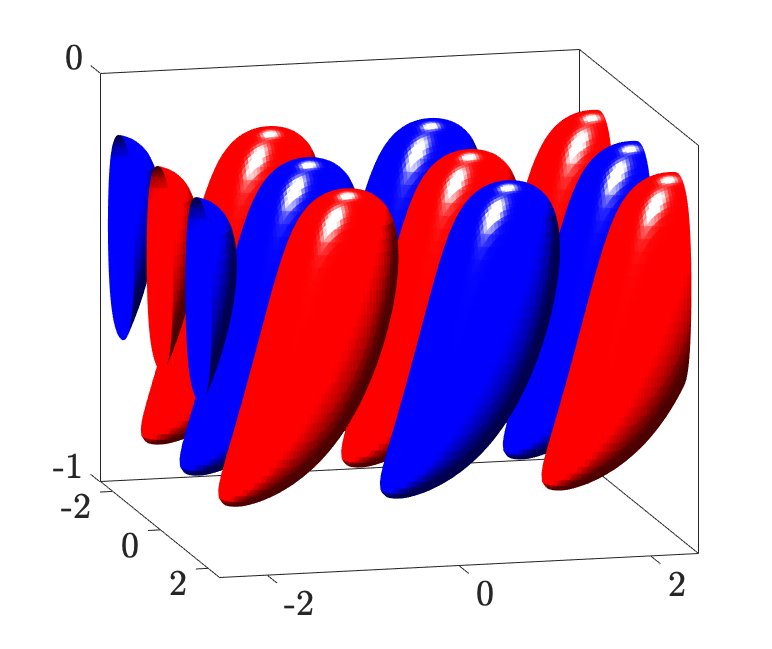}
        \end{tabular}
        &
          \hspace{.2cm}
        \begin{tabular}{c}
                \vspace{.1cm}
                {\normalsize \rotatebox{90}{$y$}}
        \end{tabular}
        &
        \begin{tabular}{c}
              \includegraphics[width=5cm]{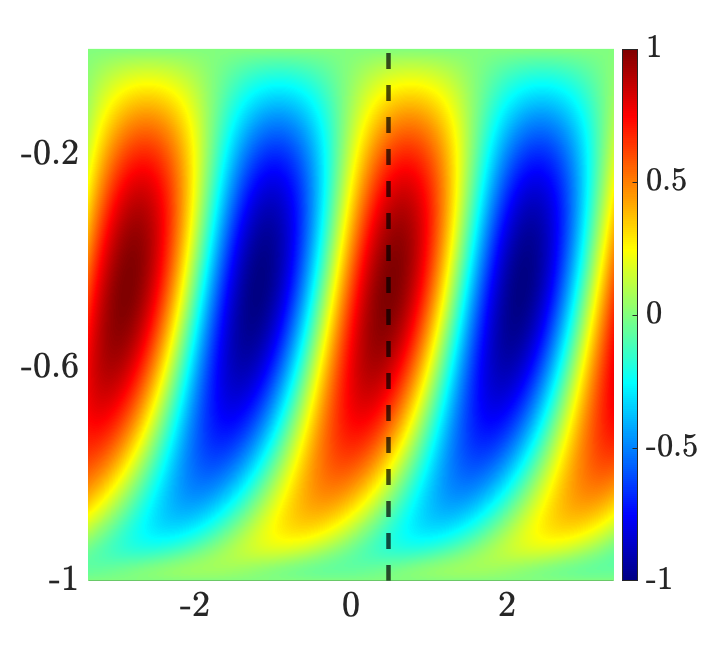}
        \end{tabular}
        &
        \hspace{-.2cm}
        \begin{tabular}{c}
        		\vspace{.4cm}
        		{\normalsize \rotatebox{90}{$y$}}
        \end{tabular}
        &
        \begin{tabular}{c}
             \includegraphics[width=5cm]{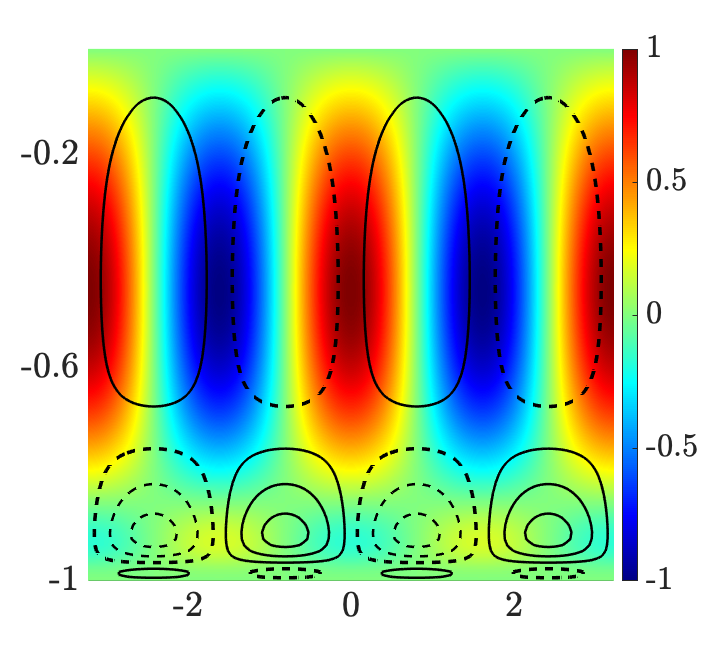}
        \end{tabular}
        \\[-.3cm]
        &
        \hspace{-1.4cm}
        {\normalsize $z$} 
        \hspace{2.3cm}
        {\normalsize $x$}
        &&
        \hspace{-.2cm}
        {\normalsize $z$}
        &&
        \hspace{-.2cm}
        {\normalsize $x$}
        \end{tabular}
             \caption{{The streamwise component of the dominant flow structures of turbulent channel flow with $R_\tau=186$ and $(k_x,k_z)=(1.86, 1.94)$ in the absence (a) and presence (b) of stochastic base flow perturbations of amplitude $\alpha=1$, shape {$f(y)=\bar{U}(y)/\max(|\bar{U}(y)|)$}, and variance $\sigma^2_u=0.49$. The three columns correspond to: (left) the spatial structure of the eigenmodes with red and blue colors denoting regions of high and low velocity; (middle) the streamwise velocity at $z=0$; and (right) the $y-z$ slice of streamwise velocity (color plots) and vorticity (contour lines) at the streamwise location indicated by the dashed vertical lines in the middle panel.}}
      \label{fig.flow_structures_turb}
\end{figure}

For a turbulent channel flow with $R_\tau=186$, we follow a similar procedure as Sec.~\ref{sec.principal flow structures} in analyzing the flow structures that are influenced by base flow perturbations with {$f(y)=\bar{U}(y)/\max(|\bar{U}(y)|)$}, $\alpha=1$, and the critically stable variance $\sigma^2_u=0.7$. Figure~\ref{fig.eig-contribution-turbulent} shows the contribution of the first eight eigenvalues of $\Phi(\bk)$ to the kinetic energy at the wavenumber pair corresponding to the maximum amplification in Fig.~\ref{fig.Ec_pre_turb_Ubar_alpha0p9}, i.e., $(k_x,k_z)=(1.86, 1.94)$. Base flow perturbations significantly increase the dominance of the principal eigenvalue (from $22\%$ of the total energy in the unperturbed state to approximately $36\%$). Figure~\ref{fig.flow_structures_turb} depicts the flow structures corresponding to the streamwise component of the most significant eigenmode in the absence and presence of streamwise base flow perturbations. It is evident that base flow perturbations shift the core of these energetic flow structures along with the counter-rotating vortical structures away from the wall while increasing the inclination of the flow structures to the wall.

\section{Reynolds number dependence}
\label{sec.Reynolds-number-dependence}

In this section, we analyze the Reynolds number dependence of the energy spectrum of streamwise constant velocity fluctuations ($k_x=0$) {in Poiseuille and Couette flows subject to streamwise base flow} perturbations $\bar{\gamma}_u$. For any finite $R$, we assume the dynamics of such fluctuations to be MSS. Theorem~\ref{Theorem1} establishes an explicit Reynolds number scaling for the energy spectrum $E(k_z)$ of streamwise constant fluctuations in channel flow subject to streamwise base flow uncertainty.
\begin{theorem}
\label{Theorem1}
    The variance amplification of streamwise constant velocity fluctuations in channel flow with nominal velocity $\bar{U}(y)$ subject to base flow perturbations is given by,
        \begin{align}
         \label{eq.energy_withR}
            E(k_z)
            \;=\; 
            f(k_z)\,R\,+\,g(k_z)\,R^2\,+\,h(k_z)\,R^3
        \end{align}
    where functions $f$, $g$, and $h$ are independent of $R$.
\end{theorem}
A proof for this theorem is provided in Appendix~\ref{app.Theorem1-proof} where it is shown that functions $f$, $g$ and $h$ represent traces of the solutions to Lyapunov equations which scale as $R$, $R^2$, and $R^3$, respectively. The function $f$ does not depend on $\bar{U}(y)$ and is thus the same for {both Poiseuille and Couette} flows. On the other hand, functions $g$ and $h$ depend on the underlying parallel base flow due to their dependence on the nominal shear $\bar{U}'(y)$. In nominal conditions, the energy spectrum of streamwise constant velocity fluctuations {of channel flow} can be decomposed into two components that scale with $R$ and $R^3$~\cite[Corollary 4]{jovbamJFM05}. The effect of base flow uncertainty is exclusively captured by the function $g$, which introduces a $R^2$ scaling to the energy spectrum of velocity fluctuations; see Appendix~\ref{app.Theorem1-proof} for details. In a similar manner, Theorem~\ref{Theorem2} uses a perturbation analysis to elucidate the Reynolds number dependence of changes to the energy content of streamwise elongated structures when the amplitude of base flow perturbations is small.
\begin{theorem}
\label{Theorem2}
    The variance amplification of streamwise constant velocity fluctuations in channel flow with nominal velocity $\bar{U}(y)$ subject to small-amplitude base flow perturbations is given by,
    \begin{align*}
        E(k_z)
        \;=\;
        E_0(k_z)\,+\,\alpha^2\,E_2(k_z) \,+\, O(\alpha^4),
    \end{align*}
    where
    \begin{align*}
        E_0(k_z)
        \;=\;
        f(k_z)\,R\,+\,h(k_z)\,R^3,
        \quad
        E_2(k_z)
        \;=\;
        g(k_z)\,R^2.
    \end{align*}
    The term $E_0$ denotes the nominal energy, $E_2$ captures the effect of base flow perturbations at the level of $\alpha^2$, and functions $f$, $g$, and $h$ are independent of $R$.
\end{theorem}
A proof for this theorem is provided in Appendix~\ref{app.Theorem2-proof}. Note that for $\alpha=1$, the functions $f$, $g$ and $h$ are the same in both Theorems. It is also evident that unless the amplitude $\alpha$ of base flow perturbation is sufficiently large, the energetic contribution of $h(z) R^3$ will dominate the energy of streamwise constant flow fluctuations especially at high Reynolds numbers.

\begin{figure}
	\begin{tabular}{cccccc}
        \subfigure[]{\label{fig.functionf_withR}}
        &
        &
        \hspace{-.6cm}
      \subfigure[]{\label{fig.functionh_withR3}}
        &
        &
        \hspace{-.6cm}
      \subfigure[]{\label{fig.functiong_withR2}}
        &
        \\[-.4cm]
        \hspace{.2cm}
        \begin{tabular}{c}
                \vspace{.1cm}
                {\normalsize \rotatebox{90}{$f$}}
        \end{tabular}
        &
        \begin{tabular}{c}
              \includegraphics[width=5cm]{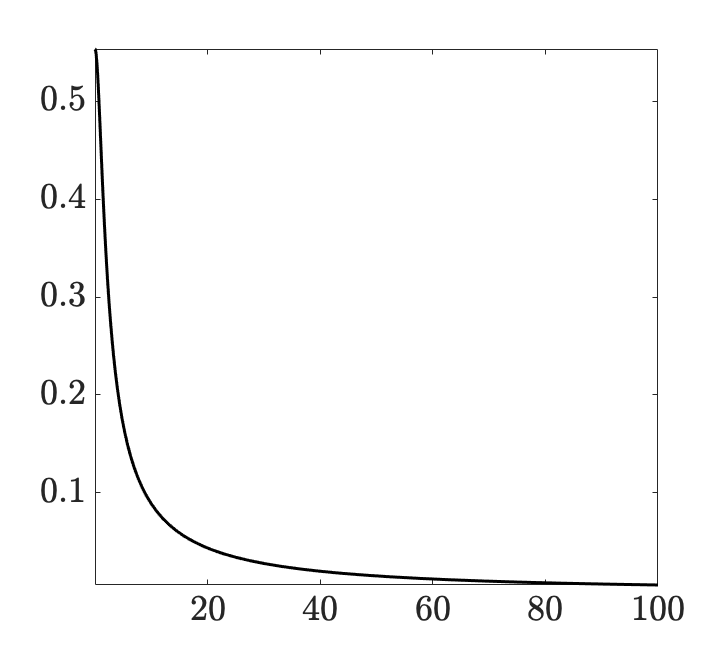}
        \end{tabular}
        &
          \hspace{-.2cm}
        \begin{tabular}{c}
                \vspace{.1cm}
                {\normalsize \rotatebox{90}{$h$}}
        \end{tabular}
        &
        \begin{tabular}{c}
              \includegraphics[width=5cm]{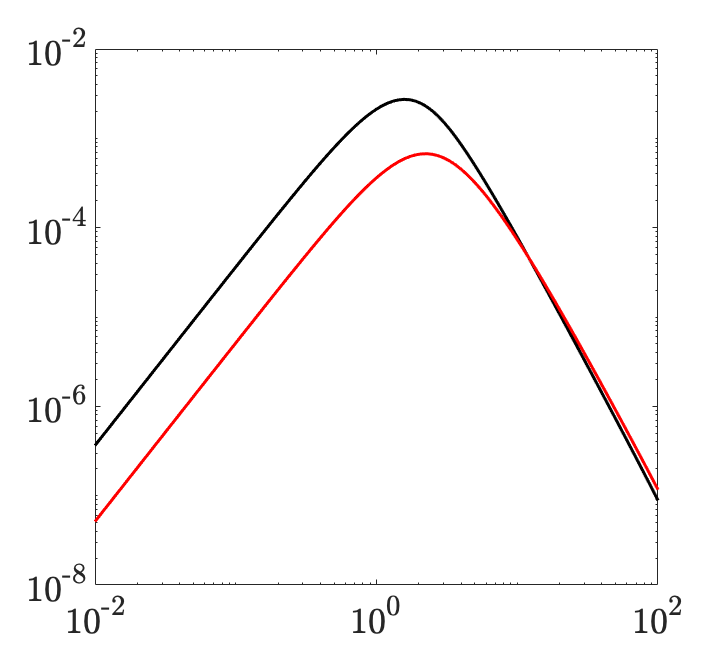}
        \end{tabular}
        &
        \hspace{-.2cm}
        \begin{tabular}{c}
        		\vspace{.4cm}
        		{\normalsize \rotatebox{90}{$g$}}
        \end{tabular}
        &
        \begin{tabular}{c}
             \includegraphics[width=5cm]{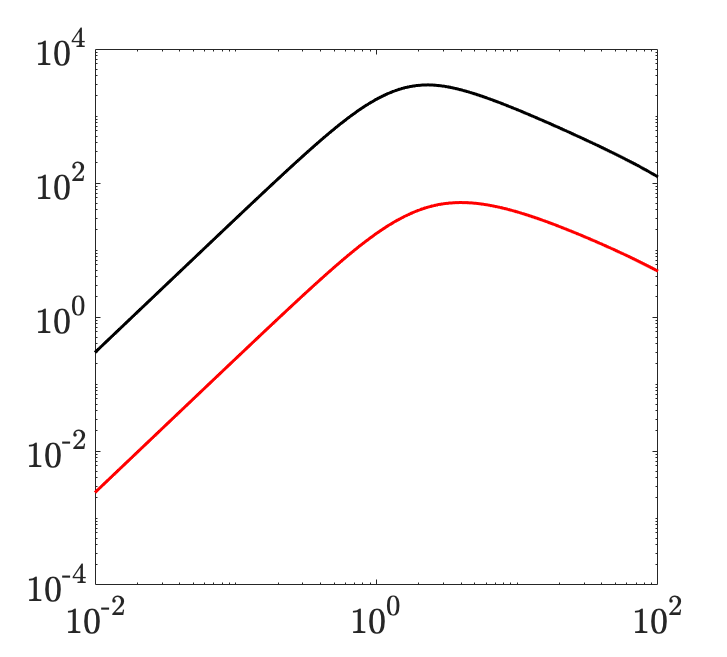}
        \end{tabular}
        \\[.2cm]
        &
        \hspace{.2cm}
        {\normalsize $k_z$}
        &&
        \hspace{.2cm}
        {\normalsize $k_z$}
        &&
        \hspace{.2cm}
        {\normalsize $k_z$}
        \end{tabular}
        \caption{{The $k_z$-dependence of functions (a) $f$, (b) $h$, and (c) $g$ in Eq.~\eqref{eq.energy_withR} for Couette flow with $R=500$ (\tc{black}{---}) and Poiseuille flow with $R=2000$ (\tc{red}{---}) subject to base flow perturbations of shape  {$f(y)=\bar{U}(y)/\max(|\bar{U}(y)|)$} of variance $\sigma^2_u = 1.13\times10^{5}$ and $\sigma^2_u = 3.22 \times 10^3$, respectively. The function $f$, which is responsible for the $O(R)$ energy amplification is the same for both Couette and Poiseuille flows.}}
        \label{fig.R_dependence_f-g-h}
\end{figure}

\begin{figure}
    \begin{tabular}{cccc}
      \subfigure[]{\label{fig.functiong_withR2_Couette_Ubar}}
        &
        &
        \hspace{.4cm}
      \subfigure[]{\label{fig.functiong_withR2_Poiseuille_Ubar}}
        &
        \\[-.4cm]
        \hspace{.2cm}
        \begin{tabular}{c}
                \vspace{.1cm}
                {\normalsize \rotatebox{90}{$\sigma^2_u$}}
        \end{tabular}
        &
        \begin{tabular}{c}
              \includegraphics[width=6.5cm]{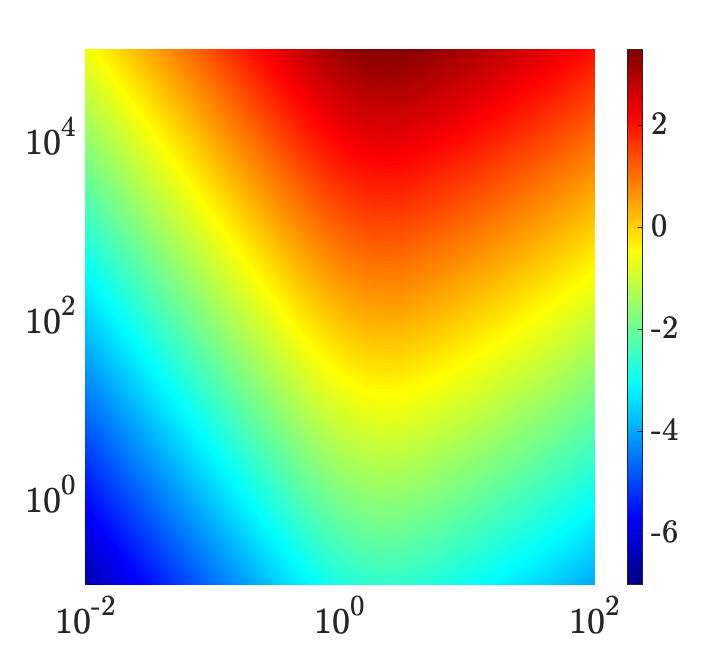}
        \end{tabular}
        &
        \hspace{.8cm}
        \begin{tabular}{c}
        		\vspace{.4cm}
        		{\normalsize \rotatebox{90}{$\sigma^2_u$}}
        \end{tabular}
        &
        \begin{tabular}{c}
             \includegraphics[width=6.5cm]{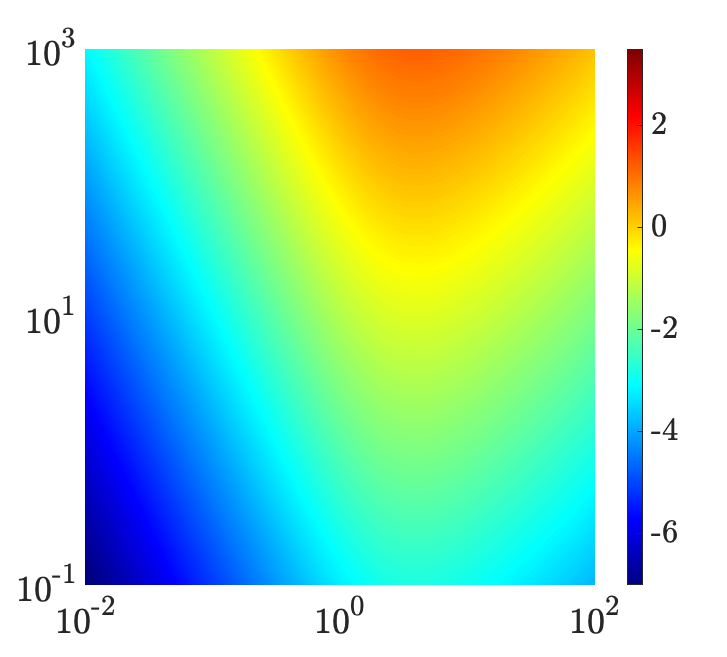}
        \end{tabular}
        \\[.2cm]
        &
        \hspace{-.2cm}
        {\normalsize $k_z$}
        &&
        \hspace{-.2cm}
        {\normalsize $k_z$}
        \end{tabular}
             \caption{{Logarithmically scaled terms that are responsible for the $O(R^2)$ energy amplification in Eq.~\eqref{eq.energy_withR} ($\log_{10}(g(k_z,\sigma^2_u))$) as a function of spanwise wavenumber $k_z$ and base flow perturbation variance $\sigma^2_u$. Perturbations to the base flow follow {$f(y)=\bar{U}(y)/\max(|\bar{U}(y)|)$}. (a) Couette flow with $R=500$; and (b) Poiseuille flow with $R=2000$.}}
      \label{fig.function_g}
\end{figure}

Figure~\ref{fig.R_dependence_f-g-h} illustrates the $k_z$ dependence of functions $f$, $g$, and $h$ for streamwise constant laminar channel flow subject to both white-in-time exogenous excitation and white-in-time base flow perturbations with {$f(y)=\bar{U}(y)/\max(|\bar{U}(y)|)$}. As explained above, and shown in Appendices~\ref{app.Theorem1-proof} and \ref{app.Theorem2-proof}, the function $f$ is independent of the choice of base flow, and is thus, identical for both Couette and Poiseuille flows; see Fig.~\ref{fig.functionf_withR}. Figure~\ref{fig.functionh_withR3} shows the dependence of $h$ on the spanwise wavenumber $k_z$ for Couette flow with $R=500$ and Poiseuille flow with $R=2000$. For both flows, the function $h$, which corresponds to the dominant Reynolds number scaling ($O(R^3)$) at high Reynolds numbers, peaks at around the same spanwise wavenumbers ($k_z=1.59$ and $2.09$ in Couette and Poiseuille flows, respectively) as their nominal spectral energy peak (cf.\ Fig.~\ref{fig.Energy_plots_laminar_nominal}). In the presence of streamwise base flow perturbations $\bar{\gamma}_u$ with $\alpha=1$ and variance levels corresponding to the critical variances obtained from Fig.~\ref{fig.sigma2_critical_kxkzgrd} at $k_x=0$, the energy of streamwise constant fluctuations is complemented with the additional term $g$, which scales as $R^2$. Figure~\ref{fig.functiong_withR2} shows the $k_z$ dependence of this function for Couette and Poiseuille flows. The spanwise wavenumbers at which the function $g$ peaks for these two flows ($k_z=1.91$ and $3.02$, in Couette and Poiseuille flows, respectively) is in agreement with the energy spectra in Figs.~\ref{fig.Ec_couette_withUbar} and~\ref{fig.Ec_poiseuille_withUbar} for $k_x \approx 0$. To further elucidate the dependence of $g$ on the variance of streamwise base flow perturbations, we compute this function for various spanwise wavenumbers $k_z$ and a range of variances $\sigma^2_u$ for which MSS is guaranteed; see Fig.~\ref{fig.function_g}. As shown in Fig.~\ref{fig.function_g}, for the range of considered variances, the dependence of $g$ on $k_z$ predominantly follows the trends observed in Fig.~\ref{fig.functiong_withR2}.

\vsp
{\color{blue}
\begin{remark}
We note that the generalization of Theorems~\ref{Theorem1} and~\ref{Theorem2} or the development of similar results for streamwise constant turbulent channel flows is inhibited by the involved Reynolds number dependence of the nominal mean velocity profile $\bar{U}(y)$, its wall-normal derivative $\bar{U}'(y)$, and the turbulent viscosity $\nu_T(y)$ per Eq.~\eqref{eq.turbulent-viscosity}.
\end{remark}
}


\section{Concluding remarks}
\label{sec.conclusion}

In the present study, we have developed an input-output framework for studying the influence of persistent stochastic base flow perturbations on the stability and energy content of velocity fluctuations in wall-bounded shear flows. We have provided verifiable conditions for the MSS of the linearized dynamics subject to stochastic base flow variations and have shown that the second-order statistics of fluctuations around the uncertain base state can be obtained as solutions to a generalized Lyapunov equation. We have used this framework to perform a thorough study of the effects of white-in-time structured stochastic base flow variations on transitional and turbulent channel flows. For transitional flows, the Reynolds number dependence of critical uncertainty variances uncovered by our method are in agreement with previously reported scaling laws for the magnitude of deterministic base flow variations. {We have shown that in both laminar and turbulent flows, shorter (in $x$) and wider (in $z$) wavelengths are least susceptible to base flow variations. Our results provide further evidence for the robustness of turbulent flows relative to their transitional counterparts. Furthermore, we observe a significantly weaker power-law dependence on the friction Reynolds number for the critical variance of stochastic perturbations to the DNS-based turbulent mean velocity relative to the power-law dependence extracted in the laminar case.}


{In laminar flow, channel-wide base flow perturbations predominantly affect the oblique modes, especially those with $k_x\approx O(1)$ and $k_z\sim O(1)$. On the other hand, near-wall perturbations result in the dominant amplification of TS waves. Our results show that the amplification of streamwise elongated structures is relatively robust to base flow perturbations, especially if such perturbations are confined to the near-wall region of the flow.} We have shown that the latter is due to the structure of dynamical perturbations induced by streamwise base flow variations at $k_x=0$ and that streaks would also become susceptible to such sources of uncertainty if variations were allowed to enter other components of the base state. {We demonstrate that large-amplitude base flow perturbations can influence the distribution of energy among various length scales and lead to the dominance of flow structures that are significantly different from those that dominate the nominal flow. Notably, perturbations of the Poiseuille flow result in an increase in the energetic dominance of the principal mode that is excited by persistent stochastic excitations. In turbulent channel flow, base flow variations influence both oblique and streamwise elongated structures to a greater extent than two-dimensional TS modes. They also increase the wall-normal separation of dominant flow structures as well as their inclination to the wall. Regardless of their wall-normal extent, however, the effect of base flow perturbations on turbulent flows is significantly less than what is observed for laminar flows.}

In addition to studying the dependence of the energy spectrum on spatial frequencies, we uncover the Reynolds number dependence of the energy of streamwise elongated fluctuations in the presence of streamwise perturbations of {the Poiseuille and Couette base flow profiles}. We show that the contribution of base flow perturbations to the amplification of streamwise elongated flow structures scales as $R^2$, which trails the $R^3$ scaling of energy amplification under nominal conditions. This scaling trend further explains the robust amplification of such flow structures, especially at high Reynolds numbers.

The utility of the proposed input-output framework goes beyond the the analysis of streamwise base flow perturbations in laminar and turbulent flows. The framework allows for structured stochastic uncertainty to enter other components of the velocity field that may originate from exogenous sources that influence the long-time behavior of the flow, e.g., surface-mounted actuators or roughness elements. Given the influence of base flow variations on the second-order statistics of the linearized NS equations, it is also anticipated that the statistics of this source of multiplicative uncertainty can be shaped to model turbulence in wall-bounded flows. The development of a systematic framework for modeling turbulent flow statistics via stochastic base flow variations is a topic for future research that would directly uncover essential dynamical perturbations that account for the absence of nonlinear interactions in linearized models.

\appendix

\section{Operators $\bar{\bA}$, $\bA_u$, and $\bA_w$ in Eq.~\eqref{eq.A-decomp-original}}
\label{app.A0-Au-Aw}

Operators $\bar{\bA}$, $\bA_u$, and $\bA_w$ in Eq~\eqref{eq.A-decomp-original} are given by:
\begin{align*}
    \bar{\bA}(\bk)
    &\;=\;
    \tbt{\;\bA_{11}\;}{0}{\;\bA_{21}\;}{\;\bA_{22}\;}, \quad\quad
    \bA_{11}(\bk)
    \;=\;
   \Delta^{-1}\Big(\dfrac{1}{R}\,\Delta^2 \,+\, \mri k_x \left(\bar{U}'' - \bar{U}\Delta\right) \,+\,  \mri k_z \left(\bar{W}'' \,-\, \bar{W}\Delta\right) \Big) 
    \\[.25cm]
    \bA_{21}(\bk)
    &\;=\;
     -\mri k_z\, \bar{U}' \,+\,  \mri k_x\, \bar{W}', \quad\quad
    \bA_{22}(\bk)
    \;=\;
    \dfrac{1}{R}\, \Delta \,-\, \mri k_x\, \bar{U}  \,-\, \mri k_z\, \bar{W}
    \\[.35cm]
    \bA_u(y,t)
     &\;=\;
    \tbt{\;\Delta^{-1}\,\mri k_x \left(f_u'' \,-\, f_u\,\Delta\right)\;}{\;0\;}{\; -\mri k_z\, f_u'\;}{\;-\mri k_x\, f_u\;}
      \\[.45cm]
    \bA_w(y,t)
     &\;=\;
    \tbt{\;\Delta^{-1}\,\mri k_z\left(f_w'' \,-\, f_w\,\Delta\right)\;}{\;0\;}{\; -\mri k_x\, f_w'\;}{\;-\mri k_z\, f_w\;}.
\end{align*}

\section{Perturbation analysis for solving the generalized Lyapunov Eq.~\eqref{eq.gen-lyap1}}
\label{app.pert-anal}

The solution to Eq.~\eqref{eq.gen-lyap1} can be efficiently computed using a perturbation analysis in $\alpha$. Following the form of the perturbed dynamical matrix $A$ in Eq.~\eqref{eq.A-decomp-original}, for sufficiently small $\alpha$, the solution $X$ can be expanded using the perturbation series,
\begin{align}
\label{eq.X-pert}
    X(\bk) 
    ~=~ 
    X_0(\bk) \;+\; \alpha\, X_1(\bk) \;+\; \alpha^2\, X_2(\bk) \;+\; \dots.
\end{align}
Substituting~\eqref{eq.X-pert} into Eq.~\eqref{eq.gen-lyap1} and collecting powers of $\alpha$ yields the sequence of standard algebraic Lyapunov equations,
\begin{align}
\label{eq.lyap-pert}
    \ba{c}
        \alpha^0:~ \bar{A}\,X_0 \;+\; X_0\,\bar{A}^* \;=\; -B\,\Omega\, B^*
        \\[.35cm]
        \alpha^{n}:~ \bar{A}\,X_{n} \;+\; X_{n}\,\bar{A}^* ~=~ -[\delta(n-1) \,-\, 1] \left(\sigma_u^2\, A_u\,X_{n-2}\,A_u^* \;+\; \sigma_w^2\, A_w\,X_{n-2}\,A_w^*\right)
    \ea
\end{align}
where $\delta(n)$ is the discrete delta function. Based on this perturbation expansion, $X_n = 0$ for odd values of $n$. This is because the right-hand-side of the algebraic Lyapunov equation is $0$ for odd $n$. As a result, the structure identified for the steady-state covariance matrix $X$ follows the perturbation series given in~\eqref{eq.X-pert-anal}. For small-size perturbations similar to those considered in Sec.~\ref{sec.laminar-flows}, the limit of the perturbation series~\eqref{eq.E-pert} can be obtained with one or two perturbation terms to $E_0$. As mentioned in Sec.~\ref{sec.energy-plots-laminar}, the Shanks transformation can be used to overcome the problem of slow convergence or even divergence of the sequence when the perturbation amplitude $\alpha$ is large.

\section{Operators $\bar{\bA}$ and $\bA_u$ in Eq.~\eqref{eq.A-decomp-turbulent}}
\label{app.A0-Au-Aw-turb}

Operators $\bar{\bA}$ and $\bA_u$ in Eq.~\eqref{eq.A-decomp-turbulent} are given by:
\begin{align*}
\bA(\bk,t)
        &\;\DefinedAs\;
        \tbt{\bA_{11}}{0}{\bA_{21}}{\bA_{22}}
        \\[.25cm]
        \non
        \bA_{11}(\bk,t)
        &\;=\;
        \Delta^{-1}\,\Big(\dfrac{1}{R_\tau} \left( \,(1\,+\,\nu_T)\, \Delta^2 \,+\, 2\, \nu'_T \,\Delta \partial_y \,+\, \nu''_T (\,\partial^2_y\,+\,k^2\,) \,\right) \,+\,
        \mri k_x \left(\bar{U}'' \,-\, \bar{U}\,\Delta \right) \Big)
         \\[.15cm]
         \non
        \bA_{21}(\bk,t)
        &\;=\;
        -\mri k_z\, \bar{U}',\quad \quad
        \bA_{22}(\bk,t)
        \;=\;
        \dfrac{1}{R_\tau} \,\left( \,(1\,+\,\nu_T)\,\Delta \,+\, \nu'_T\, \partial_y\, \right)\,-\, \mri k_x\, \bar{U}
        \\[.25cm]
    \bA_u(y,t)
     &\;=\;
    \tbt{\;\Delta^{-1}\,\mri k_x \left(f_u'' \,-\, f_u\,\Delta\right)\;}{\;0\;}{\; -\mri k_z\, f_u'\;}{\;-\mri k_x\, f_u\;}.
\end{align*}


\section{Proof of Theorem~\ref{Theorem1}}
\label{app.Theorem1-proof}

For streamwise constant channel flow ($k_x=0$), the dynamic operators $\bar{A}$ and $A_u$ in Eq.~\eqref{eq.gen-lyap1} are given by,
\begin{align}
	\label{eq.Aform-kx0}
        \bar{A}
        \,=\,
        \tbt{(1/R)\mathscr{L}}{0}{\mathscr{C}_p}{(1/R) \mathscr{I}},
        \quad
        A_u
        \,=\,
        \tbt{0}{0}{-\mri k_z \gamma'_u(y,t)}{0}
\end{align}
where the operators $\mathscr{L}$, $\mathscr{C}_p$ and $\mathscr{I}$ are parametrized by the spanwise wavenumber $k_z$ and the Reynolds number $R$. Moreover, assuming a solenoidal white-in-time exogenous excitation ${\bf f}$ with covariance $\Omega=I$, we will have $B \Omega B^* = I$. Let the state covariance $X$ take the form
\begin{align}
\label{eq.Xform}
    X
    \;=\;
    \tbt{X_{1}}{X_{2}}{X^*_{2}}{X_{3}}.
\end{align}
Substituting this matrix together with those in~\eqref{eq.Aform-kx0} into Eq.~\eqref{eq.gen-lyap1} yields the set of coupled Sylvester equations:
\begin{align*} 
        \mathscr{L}\,X_{1} \,+ \,X_{1}\, \mathscr{L}^*
        &\;=\;
        -R\,I
        \\[.15cm]
        \non
        \mathscr{L}\, X_{2}\,+\,X_{2}\,\mathscr{I}^*
        &\;=\;
        -R\,X_{1}\,\mathscr{C}^*_p
        \\[.15cm]
        \mathscr{I}X_{3}\,+\,X_{3}\,\mathscr{I}^*
        &\;=\;
        -R\, \left(\mathscr{C}_p\, X_{2} \,+\, X^*_{2}\,\mathscr{C}^*_p \,-\, \alpha^2\, \sigma^2\, k^2_z\, \gamma'_u\, X_{1}\, \gamma'_u  \,+\, I \right).
\end{align*}
From these equations it is evident that $X_1$ and $X_2$ scale as $R$ and $R^2$, respectively, and as a result, $X_3$ will contain terms that scale with all orders $O(R)$, $O(R^2)$, and $O(R^3)$. Let $X_1 \,=\, R\, \tilde{X}_1$ and $X_3 \,=\, R\, \tilde{X}_{3,1}\,+\,R^2 \tilde{X}_{3,2}\,+\,R^3 \tilde{X}_{3,3}$. Thus, the variance amplification {$E = \trace(CXC^*)$} of streamwise constant fluctuations can be decomposed as
\begin{align*}
    E
    \;=\;
    R\,f\,+\,R^2\,g\,+\,R^3\,h
\end{align*}
where {$f \,\DefinedAs\, \trace(C\,(\tilde{X}_1+\tilde{X}_{3,1})\,C^*)$, $g\,\DefinedAs\, \trace(C\, \tilde{X}_{3,2}\, C^*)$, and $h \,\DefinedAs\, \trace(C\, \tilde{X}_{3,3}\, C^*)$}.


\section{Proof of Theorem~\ref{Theorem2}}
\label{app.Theorem2-proof}

For streamwise constant channel flow ($k_x=0$), substituting the dynamic operators $\bar{A}$ and $A_u$ from Eq.~\eqref{eq.Aform-kx0} together with the perturbation series~\eqref{eq.X-pert-anal} for covariance matrix $X$ in its block operator form (cf.~Eq.~\eqref{eq.Xform}) into Eq.~\eqref{eq.gen-lyap1} yields the set of coupled Sylvester equations:
\begin{align*}
    \mathscr{L}\,X_{1,0} \;+\; X_{1,0}\,\mathscr{L}^*
        &\;=\;
        -R\,I
        \\[.15cm]
        \mathscr{L}\,X_{2,0} \;+\; X_{2,0}\,\mathscr{I}^*
        &\;=\;
        -R\,X_{1,0}\,\mathscr{C}^*_p
        \\[.15cm]
        \mathscr{I}\,X_{3,0}\;+\;X_{3,0}\,\mathscr{I}^*
        &\;=\;
        -R\,\left(\,\mathscr{C}_p\,X_{2,0}\;+\;X^*_{2,0}\,\mathscr{C}^*_p\;+\; I\,\right).
\end{align*}
at the level of $\alpha^0$, and 
\begin{align*}
        \mathscr{I}\,X_{3,2}\,+\,X_{3,2}\,\mathscr{I}^*
        &\;=\;
        \sigma^2\,k^2_z \,R\,\gamma'_u\, X_{1,0} \,\gamma'_u.
\end{align*}
at the level of $\alpha^2$. Here, we have assumed solenoidal white-in-time exogenous excitation ${\bf f}$ with covariance $\Omega=I$, which yields $B \Omega B^* = I$, and $X_{i,j}$ denote the $i$th block (cf.~\eqref{eq.Xform}) of the $j$th term in the perturbation series~\eqref{eq.X-pert-anal}. The nominal variance amplification can be computed as $E_0 = {\trace(CX_0C^*)} = R\,f\,+\,R^3\,h$ with functions $f$ and $h$ following the forms described in Appendix~\ref{app.Theorem1-proof}, i.e., {$f \,\DefinedAs\, \trace(C\,(\tilde{X}_{1,0} + \tilde{X}_{3,0,1})\,C^*)$ and $h \,\DefinedAs\, \trace(C\,\tilde{X}_{3,0,3}\,C^*)$, where $X_{1,0} \,=\, R\, \tilde{X}_{1,0}$ and $X_{3,0} = R\, \tilde{X}_{3,0,1} \,+\, R^3\, \tilde{X}_{3,0,3}$. On the other hand, since the exogenous forcing does not include a contribution at the level of $\alpha^2$, $X_{1,2} = X_{2,2} = 0$, $X_{3,2} = R^2\, \tilde{X}_{3,2}$, and $E_2 = \trace(C\,X_2\,C^*) = R^2\,g$ with $g \,\DefinedAs\, \trace(C\,\tilde{X}_{3,2}\,C^*)$}.

\vspace*{-1.6ex}


%

\end{document}

%% file: Fig3.tex
%
%
%
%
%

%
%
%
%
\typeout{Reading Tikz styles:  Tikz_common_styles}
%
\tikzstyle{block} = [draw,rectangle,thick,minimum height=2em,minimum width=1.0cm,
                                top color=blue!10, bottom color=blue!10]%
\tikzstyle{sum} = [draw,circle,inner sep=0mm,minimum size=2mm]%
\tikzstyle{connector} = [->,thick]%
\tikzstyle{line} = [thick]%
\tikzstyle{branch} = [circle,inner sep=0pt,minimum size=1mm,fill=black,draw=black]%
\tikzstyle{guide} = []%
%
\tikzstyle{deltablock} = [block, top color=red!10, bottom color=red!10]%
\tikzstyle{controlblock} = [block, top color=green!10, bottom color=green!10]%
\tikzstyle{weightblock} = [block, top color=orange!10, bottom color=orange!10]%
\tikzstyle{clpblock} = [block, top color=cyan!10, bottom color=cyan!10]%
%
\tikzstyle{block_dim} = [draw,rectangle,thick,minimum height=2em,minimum width=2em,
                                color=black!15]%
\tikzstyle{sum_dim} = [draw,circle,inner sep=0mm,minimum size=2mm,color=black!15]%
\tikzstyle{connector_dim} = [->,thick,color=black!15]%
%
\tikzstyle{smalllabel} = [font=\footnotesize]%
\tikzstyle{axiswidth}=[semithick]%
\tikzstyle{axiscolor}=[color=black!50]%
\tikzstyle{help lines} =[color=blue!40,very thin]%
\tikzstyle{axes} = [axiswidth,axiscolor,<->,smalllabel]%
\tikzstyle{axis} = [axiswidth,axiscolor,->,smalllabel]%
\tikzstyle{tickmark} = [thin,smalllabel]%
\tikzstyle{plain_axes} = [axiswidth,smalllabel]%
\tikzstyle{w_axes} = [axiswidth,->,smalllabel]%
\tikzstyle{m_axes} = [axiswidth,smalllabel]%
\tikzstyle{dataplot} = [thick]%
%

%
\large

\noindent
\begin{tikzpicture}[scale=.8, auto, >=stealth']
    
     \node[block, minimum height = 1.cm, top color=RoyalBlue!15, bottom color=RoyalBlue!15] (sys1) {\;\LARGE${\cal{M}}$\,};
     
     \node[block, minimum height = 2.0cm, top color=red!10, bottom color=red!10] (sys2) at ($(sys1.south) - (0cm,1.8cm)$) {$\,{\alpha} \begin{bmatrix}\\[-.25cm] \;\mrd \tilde{\gamma}_{u} I &\\[.1cm]&\,\mrd \tilde{\gamma}_{w} I\; \;\\[.05cm]\end{bmatrix}$};
                          
     \node[] (R1) at ($(sys1.east) + (2.6cm,-0.3cm)$){};
     
%
     
    
    \draw [connector] ($(sys1.west) + (-2.4cm,.3cm)$) -- node [midway, above] {$\mrd \tilde{\bf f}$} ($(sys1.west) + (0cm,.3cm)$);
                    	
    
    \draw [line] (sys2.west) -|  ($(sys1.west) + (-2.4cm,-.3cm)$);
    
    \draw [connector] ($(sys1.west) + (-2.4cm,-.3cm)$) -- node [midway, below] {$\mrd \tilde{\br}$} ($(sys1.west) + (0cm,-.3cm)$);
    
    \draw [connector] ($(sys1.east) + (0cm,.3cm)$) -- node [midway, above] {$\bv$} ($(sys1.east) + (2.5cm,.3cm)$);
    
    \draw [line] ($(sys1.east) + (0cm,-.3cm)$) -- node[midway, below] {$\bz$} (R1.west);
    
    \draw [connector] (R1.west) |- (sys2.east);
    
                                       
\end{tikzpicture}